\documentclass[aps,prb,10pt,twocolumn,superscriptaddress,balancelastpage,showpacs,letterpaper]{revtex4-2}
\usepackage[utf8]{inputenc}
\usepackage[T1]{fontenc}
\usepackage{amsmath,amssymb}
\usepackage{graphicx}
\usepackage{bm}
\usepackage{mathrsfs}
\usepackage{braket}
\usepackage{xcolor}
\usepackage{csquotes}
\usepackage{siunitx}
\usepackage{enumitem}
\usepackage{hyperref}
\usepackage{bookmark}
\hypersetup{colorlinks=true, linkcolor=purple, citecolor=teal}

\newcommand{\beq}{\begin{equation}}
\newcommand{\eeq}{\end{equation}}

\makeatletter
\let\@orig@make@capt@title\@make@capt@title
\def\@make@capt@title#1#2{\@orig@make@capt@title{{\bf #1}}{#2}}
\makeatother

\begin{document}
\title{Dynamical phase transitions in the nonreciprocal Ising model}
\author{Yael Avni}
\affiliation{University of Chicago, James Franck Institute, 929 E 57th Street, Chicago, IL 60637}

\author{Michel Fruchart}
\affiliation{Gulliver, ESPCI Paris, Université PSL, CNRS, 75005 Paris, France}

\author{David Martin}
\affiliation{University of Chicago, Kadanoff Center for Theoretical Physics and Enrico Fermi Institute, 933 E 56th St, Chicago, IL 60637}

\author{Daniel Seara}
\affiliation{University of Chicago, James Franck Institute, 929 E 57th Street, Chicago, IL 60637}

\author{Vincenzo Vitelli}
\affiliation{University of Chicago, James Franck Institute, 929 E 57th Street, Chicago, IL 60637}
\affiliation{University of Chicago, Kadanoff Center for Theoretical Physics, 933 E 56th St, Chicago, IL 60637}

\begin{abstract}
Nonreciprocal interactions in many-body systems lead to time-dependent states, commonly observed in biological, chemical, and ecological systems. The stability of these states in the thermodynamic limit and the critical behavior of the phase transition from static to time-dependent states are not yet fully understood. To address these questions, we study a minimalistic system endowed with nonreciprocal interactions: an Ising model with two spin species having opposing goals. The mean-field equation predicts three stable phases: disorder, static order, and a time-dependent swap phase. Large scale numerical simulations support the following: (i) in 2D, the swap phase is destabilized by defects; (ii) in 3D, the swap phase is stable, and has the properties of a time crystal; (iii) the transition from disorder to swap in 3D is characterized by the critical exponents of the 3D XY model, and corresponds to the breaking of a continuous symmetry, time translation invariance; (iv) when the two species have fully anti-symmetric couplings, the static-order phase is unstable in any finite dimension due to droplet growth; (v) in the general case of asymmetric couplings, static order can be restored by a droplet-capture mechanism preventing the droplets from growing indefinitely. We provide details on the full phase diagram which includes first- and second-order-like phase transitions and study how the system coarsens into swap and static-order states. 
\end{abstract}
\maketitle

\section{Introduction}

Many-body systems with nonreciprocal interactions emerge in contexts ranging from neuroscience~\cite{sompolinsky1986temporal,Derrida1987,Parisi1986} and social networks~\cite{hong2011kuramoto,hong2011conformists,garnier2024unlearnable} to ecology~\cite{ros2023generalized,Bascompte2006,Loreau2013,May2007} and open quantum systems~\cite{Chiacchio2023,Metelmann2015,Clerk2022}. 
These systems exhibit rich collective behavior including time-dependent states that can arise because the dynamics does not follow from the minimization of a single potential function~\cite{saha2020scalar,you2020nonreciprocity,fruchart2021non,frohoff2023nonreciprocal,rana2024defect,brauns2023non,liu2023non,Kreienkamp2022}.
However, it remains unknown whether these time-dependent states can be thought of as proper non-equilibrium phases of matter, in the sense of the term used in statistical physics; and if so, whether they exhibit well-defined phase transitions and critical behavior.
In particular, we are interested in the fate of these time-dependent states in systems made of locally coupled units subject to thermal noise in finite spatial dimensions when they are taken to the thermodynamic limit.

In order to address this question, we consider a nonreciprocal version of the Ising model, in which two species of spins have incompatible goals: spins of species A tend to align with spins of species B, while spins B tend to anti-align with spins A.
This leads to a cyclic dynamics where spins of each species tend to flip in turn. 
We study this system through analytical arguments and extensive Monte-Carlo simulations.
A short account of our results is presented in Letter form in Ref.~\cite{Avni_Letter}; we summarize here the most salient points.
First, numerical simulations indicate that in our model, a time-dependent phase with long-range spatial and temporal order indeed exists in three spatial dimensions. 
In this phase, the magnetization oscillates for arbitrarily long times. The system acts as a perfect clock in the thermodynamic limit, in the sense that the coherence time of the oscillations diverges with the size of the system. 
Second, a continuous phase transition between this oscillatory phase and a disordered ({\it i.e.} paramagnetic) phase, corresponding to a Hopf bifurcation in mean-field, is observed. Numerical evidence shows that it is associated with scale invariance, and the corresponding static critical exponents are compatible with the 3D XY universality class.
Third, we found that a static ordered phase (similar to a ferromagnetic phase) can exist when nonreciprocal interactions are present, but only if there is some symmetric coupling between the species.
However, we found no evidence of a direct transition between the oscillatory phase and the static-order phase, in contrast with mean-field predictions.

\section{Background}
The following section provides background and context on different aspects related to our work, as well as relations to existing literature. It can be skipped on first reading. 

\subsection{Nonreciprocal interactions between groups}

There is a rich literature on nonequilibrium lattice models, including variants of the Ising model with asymmetric couplings exhibiting a variety of complex phenomena~\cite{odor2004universality,Mukamel2000,henkel2008non,Godreche2009,tauber2014critical}.
A distinguishing feature of the nonreciprocal Ising model introduced in this paper is the presence of macroscopic nonreciprocal interactions between two groups of spins of comparable sizes. 
This is a rather generic feature that arises in a variety of physical contexts, some of which can be captured by Ising or Ising-like minimal models.
Examples include:
\begin{itemize}[wide=0pt,nosep]
\item \textit{Nonlinear parametric resonators}, that can be mapped to Ising spins. Networks of coupled parametric oscillators have been used as platforms to realize (reciprocal) \enquote{Ising machines} that solve combinatorial optimization problems~\cite{Honjo2021,Mohseni2022,Goto2019,Inagaki2016,Lucas2014,Mohseni2022,Goto2016,Alvarez2024}. 
This is essentially because a single of these resonators implements a double-well potential, and a linear coupling effectively produces a (reciprocal) Ising-like coupling.
Reference~\cite{han2023controlled} shows experimentally and analytically that driven non-identical parametric resonators can be mapped into nonreciprocally coupled Ising spins. 
A network of such resonators could directly realize the nonreciprocal Ising model discussed in the present work, as well as implement models of learning with nonreciprocal couplings between the artifical neurons~\cite{Amit1989,Parisi1986,Derrida1987,Mezard2011,Aguilera2021,Pourret2008,Engelken2023}.
\item \textit{Models of collective opinion dynamics}, such as in the voter model~\cite{FernandezGracia2014,odor2004universality}, where the values of the spins represent different opinions. Different communities may have asymmetric interactions: for instance, this is the case of conformists and contrarians (or alternatively trend-setters)~\cite{castellano2009statistical,Jusup2022,FernandezGracia2014,masuda2013voter,Touboul2019,Yi2013,Hong2011,Bagnoli2015,Caulkins2007}. Similarly, spin models have been used to model game theory and decision making of economic agents~\cite{garnier2024unlearnable}.
\item \textit{Models of the collective response of transmembrane receptors}, that have been put forward for the high sensitivity of cells to external chemicals in chemotaxis~\cite{Shi1998,Duke1999,Mello2003,Hu2010,Tu2013}. It has been argued that nonreciprocal couplings between the receptors and the kinases that serve as an output signal is required to account for experimental observations~\cite{Hathcock2023b,Hathcock2023,Sherry2024}.
\item \textit{Models of neurons} based on networks of spins, such as the Hopfield model~\cite{Hopfield1982,Krotov2019}, can contain asymmetric interactions~\cite{sompolinsky1986temporal,Xu1996,Sompolinsky1987,Crisanti1987,Tianping2001,Amit1989,Parisi1986,Derrida1987}, that account not only for existence of asymmetric neural connections but also for the presence of different groups of neurons called excitatory and inhibitory, which correspond to different signs of the couplings~\cite{Izhikevich2007,Amit1989}.
This approach has also been extended to larger brain structures like different brain regions: active/non-active regions are represented by spin variables while their interaction is mimicked by the spin alignment interaction~\cite{lynn2021broken}.
\item \textit{Models of spinor Bose-Einstein condensates} based on two populations of Ising spins in a cavity (a Dicke model) that effectively interact nonreciprocally have been considered~\cite{Chiacchio2023,Buca2019b,Chiacchio2019} to account for non-stationary dynamics observed in experiments \cite{Dogra2019}. 
\item \textit{Model inference} in which symmetric and asymmetric Ising models have also been used as a variational ansatz~\cite{Nguyen2017,Zdeborova2016,Dettmer2016,Mezard2011,Cocco2009}.
\end{itemize}

In several of these classes of systems, it has been conjectured that criticality 
may play a nontrivial role \cite{Mora2011,Munoz2018,Beggs2012,Buice2007,Cowan2013,Cowan2016,Borile2013}.
Our nonreciprocal Ising model can be seen as a stripped down version of these situations.
In realistic cases, (i) there may be more than two populations (ii) the spins may take more than two possible states and (iii) the network topology may be more complex than $\mathbb{Z}^d$ (see {\it e.g.} Refs.~\cite{Watts1998,Barthelemy2011}). 
While these differences are likely to induce additional complexities and phenomena, we expect that the basic features of the nonreciprocal Ising model will persist for systems in the same universality class~\cite{odor2004universality,Munoz2018}.
In particular, we expect that locally-coupled many-body systems undergoing 
a Hopf bifurcation in mean-field will enter this universality class even if they are not originally composed of coupled spin-like degrees of freedom.

\subsection{Phase transitions to time-dependent states}

An underlying motivation for this work is to construct a bridge between bifurcations in non-variational systems and phase transitions out of equilibrium. 
In the equilibrium Ising model, for instance, the ferromagnet/paramagnet transition corresponds to a pitchfork bifurcation in mean-field.
This means that the scalar order parameter $\phi$ ({\it e.g.} the magnetization) follows the dynamics $({\rm d}/{\rm d}t) \phi = \alpha \phi - \phi^3$ in which $\alpha \propto T - T_{\text{c}}$ is proportional to the distance from the critical temperature.
Conversely, the thermal field theory corresponding to locally-coupled pitchfork bifurcations where \begin{equation}
    \partial_t \phi = \alpha \phi - \phi^3 + D \Delta \phi + \eta(t,r),
\end{equation} in which $\eta$ is a thermal noise, leads upon renormalization to the Ising critical point~\cite{tauber2014critical}.
This dynamics can be seen as the model A dynamics (in the formalism of Hohenberg and Halperin~\cite{Hohenberg1967}) associated to a $\phi^4$ free energy $\mathcal{F}$, namely
\begin{equation}
    \Gamma \partial_t \phi = - \frac{\delta \mathcal{F}}{\delta \phi}
\end{equation}
where $\Gamma$ is a kinetic coefficient.

In nonreciprocal and other non-equilibrium systems, the dynamics of the (not necessarily scalar) order parameter $\vec{\phi}$ is typically non-variational,
\begin{equation}
    \Gamma \partial_t \vec{\phi} = \vec{f}[\vec{\phi}] \neq - \frac{\delta \mathcal{F}}{\delta \vec{\phi}}
\end{equation}
so bifurcations with no analogue at equilibrium can occur, in particular those leading to limit cycles. The simplest instance is the Hopf bifurcation, which describes the appearance of a limit cycle from a fixed point, see Refs.~\cite{Strogatz2018,Kuznetsov2023}. Unless otherwise specified, \enquote{fixed points} in this paper refer to fixed points of the physical dynamics, not fixed points of a renormalization group (RG) flow.
The Hopf bifurcation describes a variety of spatially extended systems~\cite{kuramoto1984chemical} ranging from oscillatory chemical reactions and extended lasers~\cite{aranson2002world} to models of mammalian hearing~\cite{camalet2000auditory,hudspeth2010critique}.
This bifurcation is described by the equation
\begin{equation}
\dot{z} = (\alpha + i \omega) z - |z|^2 z   
\end{equation}
for the complex numbers $z$ and takes place at $\alpha = 0$.
When $z$ is replaced by a complex field $\psi$ with appropriate diffusive couplings, the corresponding equation is known as the complex Ginzburg-Landau equation~\cite{aranson2002world}.
The fate of this bifurcation has been analyzed using perturbative RG in Refs.~\cite{risler2004universal,risler2005universal}, and later reproduced and generalized in Refs.~\cite{daviet2024nonequilibrium,Zelle2024}.
In these works, a phase transition generalizing the Hopf bifurcation is predicted based on a $d=4-\epsilon$ expansion and formally related to the RG fixed point of model A dynamics with $O(2)$ symmetry (i.e., to the XY universality class).
Using this analogy Refs.~\cite{risler2004universal,risler2005universal} predict that this phase transition, which leads to an oscillating phase with long-range order, should exist for space dimension $d>2$, 
while no coherent oscillations are expected in $d<2$, and quasi-long-range order in conjectured in $d=2$~\footnote{
References~\cite{risler2004universal,risler2005universal} conjecture that quasi-long-range may be present at $d=2$, but more recent studies based on a mapping to a version of the Kardar-Parisi-Zhang (KPZ) equation with compact variables suggest that the phase with algebraic order does not persist in the limit of large systems (nor an equivalent of the Berezinskii-Kosterlitz-Thouless (BKT) transition) but are replaced with a stretched exponential decay of correlations corresponding to the rough phase of the KPZ equation~\cite{altman2015two,wachtel2016electrodynamic}.
In addition, Refs.~\cite{aranson1998spiral} argue that the modified effective interaction between spiral defects (compared to that between XY vortices) allows for a finite density of defects at any temperature, potentially suppressing the algebraic order.}.
To the best of our knowledge, these predictions have only been tested qualitatively, in Refs.~\cite{wood2006critical,wood2006universality}, where numerical simulations of coupled three-state oscillators were performed for lattices of size $L^d$ going up to $L=80$ in $d=3$. A comparison of the rescaled finite-size scaling data of Refs.~\cite{wood2006critical,wood2006universality} with Ising and XY exponents suggested a better collapse with XY exponents.
The results of our large-scale simulations with lattices of sizes up to $L=320$ in $d=3$ are in agreement with the predictions of Refs.~\cite{risler2004universal,risler2005universal} and with the qualitative observations of Refs.~\cite{wood2006critical,wood2006universality}. 
In particular, we obtain values of the critical exponents that are in very good agreement with XY exponents but not with Ising exponents.

The Hopf bifurcation is not the only way to produce a limit cycle. 
Another relevant bifurcation is the infinite-period bifurcation (or SNIC bifurcation) that describes the creation of a limit cycle by a saddle-node bifurcation that connects heteroclinic orbits together~\cite{Strogatz2018}. 
Here, multiple SNIC bifurcations (related by symmetry) arise at the same time in the mean-field transition from a static ordered (symmetry-breaking) state to a limit cycle. To the best of our knowledge, the field-theoretical version of this class of bifurcations has not been studied.
It has been conjectured that it does not exist in spatially extended systems of arbitrary dimension in Ref.~\cite{assis2011infinite} based on the instability of the static ordered state in the three-state oscillator model of Refs.~\cite{wood2006critical,wood2006universality}. 
Our simulations suggest that a static ordered state can exist despite the presence of nonreciprocal coupling (provided that they are not fully anti-symmetric, at least in our model), but even in this case we did not find evidence of any direct transition between the static-order and oscillating phases.

\subsection{Stability of spatially coherent oscillations}

Our work is also directly related to the question of the stability and spatial coherence of time-periodic phases in noisy locally-coupled classical many-body systems. By considering the nucleation and growth of droplets from fluctuations, Ref.~\cite{Bohr1987,Grinstein1988,bennett1990stability,Grinstein1994} argued that spatially coherent oscillations are unstable in the thermodynamic limit if they arise from the spontaneous breaking of a discrete time-translation invariance (in particular when the period of oscillation is commensurate with the discrete time unit characterizing the evolution) \footnote{Note that these arguments require that the system is isotropic and that the noise is unbounded so that it can create arbitrarily large droplets~\cite{bennett1990stability,Grinstein1994}.}.
This is not the case in our model, where the oscillating phase arises from the spontaneous breaking of a continuous time-translation invariance (and the period is generically incommensurate with the discrete time unit). 
The case of spontaneously broken continuous time-translation invariance is considered in Refs.~\cite{bennett1990stability,grinstein1993temporally,Grinstein1994} using a mapping between the desynchronization of local oscillators and the roughening of surfaces described by the Kardar-Parisi-Zhang (KPZ) equation~\cite{Kardar1986}.
In the KPZ universality class, surfaces are always rough in $d \leq 2$, while two phases (rough or smooth) are possible for $d>2$, see {\it e.g.} \S~5.7.4 of Ref.~\cite{Livi2017}.
Based on this mapping, it is argued that temporal order can exist in $d>2$ but not in $d\leq 2$ in isotropic systems.
Note that in the mapping to the KPZ equation, the angular nature of phase variables (defined on a compact space) is neglected, which amounts to ignoring topological defects, expected not to affect the large-scale physics~\cite{grinstein1993temporally}. 
A compact version of the KPZ equation has been studied in Refs.~\cite{Chen2013,altman2015two,wachtel2016electrodynamic,Aranson1998}, but not in isotropic systems in $d=3$.
Similarly, Refs.~\cite{chan2015limit,daviet2024nonequilibrium} suggested an analogy with the Mermin-Wagner theorem~\cite{Mermin1966,Hohenberg1967} in the time domain, according to which the fluctuations of the Goldstone modes associated with broken time-translation invariance destroy order in $d \leq 2$.
 
References~\cite{Chate1991,Chate1992,Chate1995,Chate1997,Gallas1992,Binder1992,Hemmingsson1993,Losson1995} performed numerical simulations of cellular automata and coupled map lattices (see Ref.~\cite{Chate1992} for an introduction) in various space dimensions, from which it emerged that periodic oscillations are possible for $d \geq 3$ within this class of sytems~\cite{Chate1995,Pomeau1993,Chaté1996}.
Similar behaviors were reported in simulations of predator-prey models on lattice~\cite{Lipowski1999,Lipowski2000,Mobilia2006,Antal2001b,Antal2001,Tauber2024,odor2004universality,tauber2014critical}. These results are in agreement with the theoretical arguments reviewed above, and are are also consistent with our numerical observations.
Similar questions arise in the study of the synchronization of locally coupled oscillators with randomly distributed frequencies (like in the Kuramoto model), see \S~IV.A of Ref.~\cite{acebron2005kuramoto} and references therein, but the physics is different because of the random frequencies.

\subsection{Classical continuous time crystals}

The works reviewed in the previous paragraphs are primarily concerned with the \emph{spatial} coherence of the oscillations.
In a nutshell, they ask whether there can be macroscopic quantities averaged over the entire system (such as the magnetization; we call these order parameters) that oscillate forever.
When this is the case, we can ask another question: is the oscillation of the macroscopic order parameter temporally coherent? In other words, is it acting as a good clock?
Addressing this question is another of the key motivations of our work.
To illustrate the difference, let us consider a simple noisy phase oscillator
\begin{equation}
    \dot{\phi} = \omega + \sqrt{2 D_\phi} \eta(t)
\end{equation}
where $\eta(t)$ is a white noise with $\langle\eta\rangle = 0$ and $\langle\eta(t) \eta(t')\rangle = \delta(t-t')$. 
Here, $\phi$ can be seen as the argument of a complex number $z(t) = A e^{i \phi(t)}$ whose amplitude $A$ is very quickly relaxing to its steady-state value.
In this simple system, we observe that the average value $\braket{\phi}$ always oscillates (averaging the equation of motion leads to $\braket{\phi(t)} = \phi(0) + \omega t$). 
However, this oscillator only keeps track of time ({\it i.e.} of its initial phase) for so long. 
To see that, consider the correlation function $C(t) = \langle e^{i [\phi(t) - \phi(0)]} \rangle$ which describes the average phase that an assembly of independent clocks would have at time $t$, provided that they were synchronized at time $t=0$. 
Here, it behaves as \footnote{
This can be seen by integrating $d\phi_t = \omega dt + \sqrt{2 D_\phi} dW_t$ where $W_t$ is a standard Wiener process, leading to $\phi(t) - \phi(0) = \omega t + \sqrt{2 D_\phi} W_t$. We find $C(t) = e^{i \omega t} \langle e^{i \sqrt{2 D_\phi} W_t}\rangle$. As $W_t$ follows a normal distribution $\mathcal{N}(\mu=0,\sigma^2=t)$, its moment generating function is $M_{W_t}(x) \equiv \mathbb{E}(e^{x W_t}) = e^{\mu x + \sigma^2 x^2/2}$ from which the result is found.
}
\begin{equation}
    C(t) \propto \cos(\omega t) e^{- t/\tau}
\end{equation}
where $\tau = 1/D_\phi$, meaning that the oscillator loses track of its original phase after $\mathcal{Q} = \omega \tau$ oscillations ($\mathcal{Q}$ is known as the quality factor of the noisy oscillator).
In fact, any noisy oscillator behaves in this way at long times~\cite{Gaspard2002b}.

Thermodynamic bounds on the number $\mathcal{Q}$ of coherent oscillations have been considered~\cite{Cao2015,Oberreiter2022,Shiraishi2023,Ohga2023}, and suggest that an infinite amount of energy (more precisely of entropy production) is needed to obtain infinite coherence.
It is however not excluded that a macroscopic system in the thermodynamic limit can have an infinite coherence time with a finite cost per unit volume. Our numerical results suggest that this indeed happens in our model: we evaluate the coherence time of oscillations, which scales as $\tau(L) \sim L^{3}$ (in $d=3$). This points to the existence of perfectly coherent oscillations in the thermodynamic limit. Additionally, we measure the entropy production rate per unit volume and find that it converges to a finite value, independent of system size.
In this sense, our model can be seen as a classical continuous time crystal~\cite{Khemani2019,Sacha2017,Iemini2018,Buca2019,Kessler2019,Kongkhambut2022,Yao2020,Zaletel2023,winfree1980geometry,Wilczek2012,Shapere2012,Yao2018,kongkhambut2022observation,wu2024dissipative}.
We conjecture that this property may extend to other systems with locally coupled oscillators that spontaneously break a continuous time-translation symmetry.

\section{Outline and summary}

The outline of the paper is as follows: we present the nonreciprocal Ising model with general asymmetric coupling between the two species and derive its mean-field equation in Sec.~\ref{Sec_model}. In Secs.~\ref{fully_anti} and ~\ref{Sec_2D}--\ref{summary_purely_antisymmetric} we restrict ourselves to the fully anti-symmetric case in which the system exhibits $C_4$ symmetry. In Sec.~\ref{Sec_model} we show that the mean-field equation predicts three distinct phases: disorder, time-dependent order that we dub ``swap”, and static order, and analyze the bifurcations that separate the phases. Details on the Monte-Carlo simulations are given in~Sec.~\ref{MonteCarlo_App}.
Our numerical simulations in 2D are presented in Sec.~\ref{Sec_2D}, suggesting that the swap phase is unstable in 2D due to spiral defects. In Sec.~\ref{Sec_3D} we show that the swap phase is stable in 3D, and calculate the static critical exponents of the phase transition from disorder to swap. These critical exponents are in excellent agreement with the equilibrium XY critical exponents, which is compatible with renormalization-group studies and with the breaking of a continuous time-translation symmetry. We calculate the coherence time of the magnetization and show that it grows with system size, suggesting a time-crystal behavior in the thermodynamic limit. In Sec.~\ref{Sec_static}, we provide numerical evidence and theoretical arguments that suggest that the static-order phase is destroyed in any finite dimension due to droplet growth. In Sec.~\ref{coarsening} we discuss how the system coarsens into swap and static-order states. The phase diagram of the fully anti-symmetric case is summarized in Sec.~\ref{summary_purely_antisymmetric}. In Sec.~\ref{Sec_K_+} we allow the presence of inter-species reciprocal couplings, which reduces the symmetry to $\mathbb{Z}_2$, and accounts for a general asymmetric interaction. We present mean-field analysis and 2D and 3D simulation results. These results reveal
that the swap phase stability is unchanged with respect to the fully anti-symmetric case ({\it i.e.} it is stable in 3D but not in 2D). However, static order can be restored by an intriguing droplet-capture mechanism. Contrary to the mean-field prediction, we find no evidence of a direct transition between the swap and the static-order phase. We summarize our results and conclusions in Sec.~\ref{Sec_conclusions}.


\section{The model} \label{Sec_model}

In the equilibrium Ising model, adjacent on-lattice spins (${\sigma=\pm 1}$) stochastically align with one another through an energy function. Its dynamics is given by kinetic Ising models which, while lacking uniqueness, always operate by minimizing the global free energy. In what follows, we generalize these kinetic Ising models to the case of nonreciprocal interactions where such global potential function does not exist.
To this aim, we define a Glauber dynamics~\cite{glauber1963time} where each spin tends to minimize its own {\it selfish energy}.
Similar formulations were used among the physical sciences~\cite{lynn2021broken,Mello2003,han2023controlled,Loos2023,guislain2023nonequilibrium,guislain2023discontinuous,liu2023non,Lima2006,Sanchez2002,Lipowski2015,garnier2024unlearnable}.

\subsection{The nonreciprocal (NR) Ising model}
\begin{figure}[ht]
\centering
{\includegraphics[width=0.44\textwidth,draft=false]{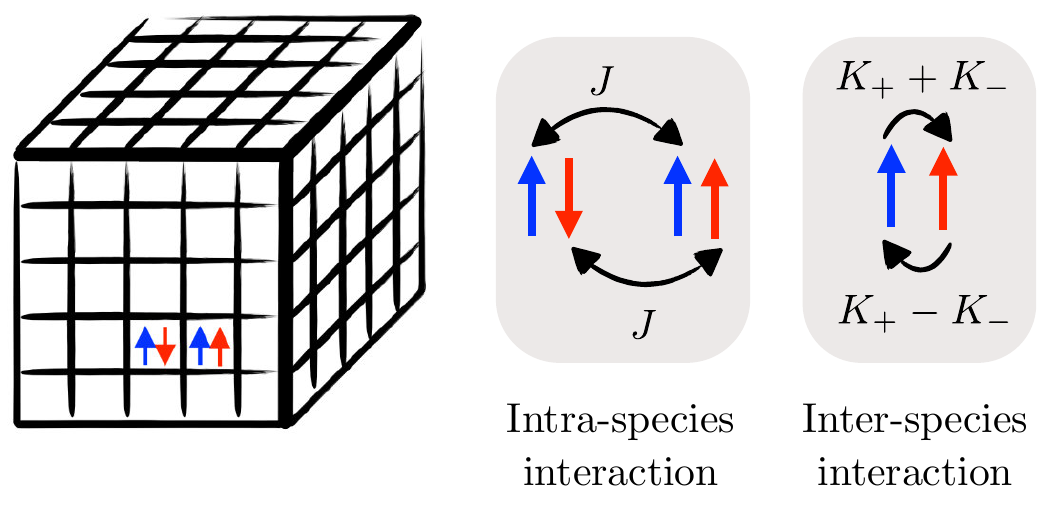}}
\caption{
\textbf{Nonreciprocal (NR) Ising model.}
The model includes two species per site with two interaction types: (i) reciprocal nearest-neighbors interaction of strength $J$ between same-species spins and (ii) nonreciprocal same-site interaction of strength $K_++K_-$ for species A (blue) 
 and $K_+-K_-$ for species B (red), between spins of different species.}
\label{Model}
\end{figure}

We consider two spin species, labeled by Greek indices $\alpha=A,B$, located on a $d$-dimensional cubic lattice of linear size $L$. Each site $i\in \{1,...,L^d\}$ contains one A-spin and one B-spin (Fig.~\ref{Model}, left). 
The state of the system is described by a vector 
\begin{equation}
    \vec{\sigma}=\{\sigma^A_1,...,\sigma^A_{L^d},\sigma^B_1,...,\sigma^B_{L^d}\}
\end{equation}
in which each spin $\sigma^{\alpha}_i = \pm 1$.

We consider transition rates given by the generalized Glauber dynamics~\cite{glauber1963time} in which the transition rate $w(\vec{\sigma}|\vec{\sigma}')$ from configuration $\vec{\sigma}$ to $\vec{\sigma}'$ is given by
\begin{equation}
\label{rates}
w(F^{\alpha}_{i}\vec{\sigma}|\vec{\sigma})=\frac{1}{2\tau}\left[1-\tanh\left(\frac{\Delta E^{\alpha}_i}{2k_B T}\right)\right]
\end{equation}
where $T$ is the temperature, $k_B$ the Boltzmann constant, and $\tau$ is a characteristic time for single spin flips. In addition,
\begin{equation}
    \Delta E^{\alpha}_i =  E^{\alpha}_i(F^{\alpha}_{i}\vec{\sigma}) - E^{\alpha}_i(\vec{\sigma})
\end{equation}
is the difference in selfish energy between the states.
Here $F^{\alpha}_{i}$ is a single-spin-flipping operator, {\it i.e.}, when acting on $\vec{\sigma}$ it flips the $\alpha$-species spin on site $i$ from $\sigma^{\alpha}_{i}$ to $-\sigma^{\alpha}_{i}$ while keeping all the other spins unchanged. For instance,
\begin{equation}
\begin{split}
 &F_{i}^{A}\{\sigma_{1}^{A},...,\sigma_{i}^{A},...,\sigma_{L^{d}}^{A},\sigma_{1}^{B},...,\sigma_{L^{d}}^{B}\} \\
&\quad =\{\sigma_{1}^{A},...,-\sigma_{i}^{A}...,\sigma_{L^{d}}^{A},\sigma_{1}^{B},...,\sigma_{L^{d}}^{B}\}
 \end{split}
\end{equation}

For the selfish energy, we choose
\begin{equation}
    \label{SelfishEnergy}
    E^{\alpha}_i = -J\sum_{j\,{\rm nn\,of}\,i} \sigma_i^{\alpha}\sigma_j^{\alpha} - K _{\alpha\beta} \sigma_i^\alpha\sigma_i^\beta,
\end{equation}
where
\begin{equation} \label{K_matrix}
K_{\alpha\beta}=\left(\begin{array}{cc}
0 & K_{+}+K_{-}\\
K_{+}-K_{-} & 0
\end{array}\right)
\end{equation}
and $J,K_+,K_->0$. The sum in Eq.~(\ref{SelfishEnergy}) runs over nearest neighbors of $i$ and summation over spin indices $\beta$ is implied.
The selfish energy indicates that spins of the same species tend to align with their neighbors, while spins of different species interact nonreciprocally (see Fig.~\ref{Model}, right): spins $A$ tend to align with spins $B$, whereas spins $B$ either tend to align with spins $B$ but to a lesser extent (if $K_+>K_-$), or tend to anti-align with spins $A$ (if $K_+<K_-$). The latter case will be the main focus of this paper as it leads to a rich phenomenology related to limit cycles. 
Note that in our model, inter-species interactions only occur between spins located on the same site.

In the continuous-time limit, the probability $P(\vec{\sigma}, t)$ to observe a specific configuration of spins $\vec{\sigma}$ at time $t$ evolves according to the master equation
\begin{align}
\label{MasterEquation}
 \partial_{t}P(\vec{\sigma},t)&=-\sum_{i,\alpha}w(F^{\alpha}_{i}\vec{\sigma}|\vec{\sigma})P(\vec{\sigma},t)\\
&+\sum_{i,\alpha} w(\vec{\sigma}|F^{\alpha}_{i}\vec{\sigma})P(F^{\alpha}_{i}\vec{\sigma},t). \nonumber
\end{align} 
As detailed in Sec.~\ref{MonteCarlo_App}, we perform Monte-Carlo simulations that sample trajectories obtained from a discrete-time version of Eq.~\eqref{MasterEquation}.

\begin{figure}[ht]
\centering
{\includegraphics[width=0.25\textwidth,draft=false]{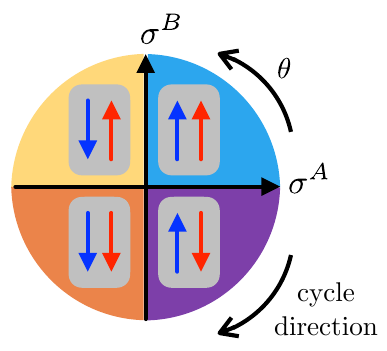}}
\caption{
\textbf{Mapping between the state of two spins located on the same site ($\sigma^A$ and $\sigma^B$) and their corresponding angle variable $\theta$.} The color associated to each quadrant is used throughout the paper to indicate the corresponding spin state. The direction of the cycle is determined by the sign of $K_-$ (throughout the paper $K_- > 0$ which implies clockwise direction).
}
\label{theta}
\end{figure}

We note that the Glauber dynamics of the nonreciprocal Ising model can be written in terms of a new variable $\theta_i$ (instead of $\sigma^{\alpha}_i$), which is defined as the angle on the $\left(\sigma^A_i,\sigma^B_i\right)$ plane.
This angle $\theta_i$ can take one of the four values respectively corresponding to the four possible states per site: $\pi/4$ ($\uparrow \uparrow$), $3\pi/4$ ($\downarrow\uparrow$), $5\pi/4$ ($\downarrow \downarrow$), and $7\pi/4$ ($\uparrow \downarrow$) (see Fig.~\ref{theta} and an explicit mapping for the fully anti-symmetric case in Sec.~\ref{locally_coupled_oscillators}).
The color code given in Fig.~\ref{theta} is used throughout the paper.

\subsection{Broken detailed balance} \label{sec:BDB}
We now show that the nonreciprocal Ising model described in the previous section is out of equilibrium, in the sense that it breaks detailed balance.

A Markov chain with rates $w$ satisfies detailed balance if it has a stationary distribution $P^{\text{ss}}$ such that
\begin{equation}
    \frac{w(\vec{\sigma}\to \vec{\sigma}')}{w(\vec{\sigma}' \to \vec{\sigma})}
    =
    \frac{P^{\text{ss}}(\vec{\sigma}')}{P^{\text{ss}}(\vec{\sigma})}.
\end{equation}
Let us assume that our kinetic model possesses an (unknown) stationary distribution $P^{\text{ss}}(\vec{\sigma})$. Considering a loop $\vec{\sigma}_1 \to \vec{\sigma}_2 \to \cdots \to \vec{\sigma}_n$ in state space, as well as the same loop travelled in opposite direction, detailed balance implies~\cite{kalpazidou2007cycle}
\begin{equation}
    \frac{w(\vec{\sigma}_1\to \vec{\sigma}_2)}{w(\vec{\sigma}_1 \to \vec{\sigma}_n)}
    \cdot
    \frac{w(\vec{\sigma}_2\to \vec{\sigma}_3)}{w(\vec{\sigma}_n \to \vec{\sigma}_{n-1})}
    \cdots
    \frac{w(\vec{\sigma}_{n}\to \vec{\sigma}_1)}{w(\vec{\sigma}_2 \to \vec{\sigma}_1)}
    = 1.
\end{equation}
If we find a cycle such that this ratio is not unity, then detailed balance cannot be satisfied. This is known as the Kolmogorov criterion.
Here, a relevant cycle to consider consists in flipping the spins $(\sigma_i^A, \sigma_i^B)$ of a fixed site $i$ according to $\uparrow \uparrow \to \uparrow \downarrow \to \downarrow \downarrow \to \downarrow \uparrow \to \uparrow \uparrow$, while keeping all the other spins unchanged. Using Eqs.~(\ref{rates}), ~(\ref{SelfishEnergy}) and assuming all the neighboring spins are up, we obtain
\begin{equation}
\begin{split}w\left(\uparrow\uparrow\to\uparrow\downarrow\right) & =f(\tilde{J}+\tilde{K}_{+}-\tilde{K}_{-})\\
w\left(\uparrow\downarrow\to\downarrow\downarrow\right) & =f(\tilde{J}-\tilde{K}_{+}-\tilde{K}_{-})\\
w\left(\downarrow\downarrow\to\downarrow\uparrow\right) & =f(-\tilde{J}+\tilde{K}_{+}-\tilde{K}_{-})\\
w\left(\downarrow\uparrow\to\uparrow\uparrow\right) & =f(-\tilde{J}-\tilde{K}_{+}-\tilde{K}_{-})\\
w\left(\uparrow\uparrow\to\downarrow\uparrow\right) & =f(\tilde{J}+\tilde{K}_{+}+\tilde{K}_{-})\\
w\left(\downarrow\uparrow\to\downarrow\downarrow\right) & =f(\tilde{J}-\tilde{K}_{+}+\tilde{K}_{-})\\
w\left(\downarrow\downarrow\to\uparrow\downarrow\right) & =f(-\tilde{J}+\tilde{K}_{+}+\tilde{K}_{-})\\
w\left(\uparrow\downarrow\to\uparrow\uparrow\right) & =f(-\tilde{J}-\tilde{K}_{+}+\tilde{K}_{-})
\end{split}
\end{equation}
where $f(x)\equiv\left(1-\tanh (x)\right)/(2\tau)$ and we defined the normalized coupling coefficients $\tilde{K}_-=K_-/(k_B T)$, $\tilde{K}_+=K_+/(k_B T)$ and $\tilde{J} \equiv 2 d J/(k_B T)$.
Using the identity $(1+\tanh(x))/(1-\tanh(x))={\rm e}^{2x}$ we find that the forward and reverse cycles satisfy
\begin{equation}
\label{eq:BDBCycle}
\!\!\!
\frac{
w\left(\uparrow\uparrow\to\uparrow\downarrow\right) 
w\left(\uparrow\downarrow\to\downarrow\downarrow\right) 
w\left(\downarrow\downarrow\to\downarrow\uparrow\right) 
w\left(\downarrow\uparrow\to\uparrow\uparrow\right)
}{
w\left(\uparrow\uparrow\to\downarrow\uparrow\right)
w\left(\downarrow\uparrow\to\downarrow\downarrow\right)
w\left(\downarrow\downarrow\to\uparrow\downarrow\right)
w\left(\uparrow\downarrow\to\uparrow\uparrow\right)
}
=e^{8\tilde{K}_-}
\end{equation}
which means that detailed balance is broken in our model whenever $\tilde{K}_- \neq 0$.
The breaking of detailed balance implies the system is driven away from equilibrium.
Under certain assumptions on the physical realization of the model that go by the name of \textit{local detailed balance}~\cite{Seifert2012,Horowitz2019,Maes2021}, the cycle produces entropy 
$\Delta S/k_B = 8 \tilde{K}_-$ (see Appendix~\ref{EntProd} for details).
We refer to Refs.~\cite{Martynec2020,Alston2023,Suchanek2023,Nardini2017,Seara2021} for discussions about irreversibility and entropy production in nonequilibrium phase transitions.

\subsection{The mean-field (MF) equation} \label{Sec_MF}

We now derive mean-field (MF) equations of motion for the magnetizations of each species from the master equation \eqref{MasterEquation} (see Refs.~\cite{Suzuki1968,Forgacs1984,Leung2000}). 
The mean-field equations are a useful guide, but it is not expected that they accurately describe the behavior of the system in any finite dimension.

The average magnetization of a single spin at time $t$ is
\begin{equation}
\langle \sigma^{\alpha}_i(t) \rangle = \sum_{\vec \sigma} \sigma^{\alpha}_{i} P({\vec \sigma},t). 
\end{equation}
Using Eq.~(\ref{MasterEquation}) and changing summation variables, we get
\beq
\label{glauber_rate_se}
\frac{\rm{d}}{{\rm d}t}\langle\sigma^{\alpha}_{i}\rangle=-2\langle\sigma^{\alpha}_{i}w(F^{\alpha}_{i}\vec{\sigma}|\vec{\sigma})\rangle.
\eeq
Substituting the rates (Eq.~(\ref{rates})) and the selfish energy (Eq.~(\ref{SelfishEnergy})) we obtain
\begin{equation}
\begin{split}
&\tau\frac{{\rm d}}{{\rm d}t}\langle\sigma_{i}^{\alpha}\rangle=-\langle\sigma_{i}^{\alpha}\rangle\\
&+\left\langle\tanh\left[\frac{J}{k_B T}\sum_{j\, {\rm nn\,of}\, i }\sigma_{j}^{\alpha}+\frac{K_{\alpha\beta}}{k_B T}  \sigma_{i}^{\beta}\right]\right\rangle
\end{split}
\end{equation}
where we have used the identity ${x \tanh{x b} = \tanh{b}}$ for $x=\pm1$. 
Applying the mean-field approximation, $\langle g(\vec{\sigma})\rangle=g(\langle \vec{\sigma}\rangle)$ and defining the average magnetization as $\langle\sigma^{\alpha}_{i}\rangle\equiv m^{\alpha}_i$ yields
\beq \label{MF_eq0}
\frac{{\rm d} m_i^{\alpha}}{{\rm d}t}=-\frac{m_i^{\alpha}}{\tau}+\frac{1}{\tau}\tanh\left(\frac{J\sum_{j\, {\rm nn\,of}\, i }m_j^{\alpha}+K_{\alpha\beta}m_i^{\beta}}{k_{B}T}\right).
\eeq
We now assume that the average magnetization changes over length scales much larger than the lattice spacing and write it as a continuum variable $m_i^{\alpha}=m_{\alpha}({\vec r},t)$. We thus get
\beq \label{MF_eq1}
\partial_{t}m_{\alpha}=-\frac{m_{\alpha}}{\tau}+\frac{1}{\tau}\tanh\left(\frac{2dJm_{\alpha}+K_{\alpha\beta}m_{\beta}+J a^{2}\nabla^{2}m_{\alpha}}{k_{B}T}\right)\;,
\eeq
where $a$ is the lattice spacing. Finally, upon rescaling time by $\tau$ and space by $a$ we obtain
\begin{equation}
    \label{mean_field}
    \partial_t m_{\alpha}=-m_{\alpha}+\tanh\left[\tilde{J} m_{\alpha}+\tilde{K}_{\alpha\beta}m_{\beta} + D \nabla^{2} m_{\alpha} \right]
\end{equation}
where $D \equiv J / (k_BT)$ and $\tilde{K}_{\alpha \beta} \equiv K_{\alpha \beta} / (k_BT)$.
Equation~(\ref{mean_field}) is the mean-field equation of the general nonreciprocal Ising model. 
Note that it is invariant under the diagonal up-down $\mathbb{Z}_2$ symmetry
\begin{equation}
    \mathscr{S}_1 : (m_A, m_B) \mapsto (-m_A, -m_B).
    \label{symmetrygen1}
\end{equation}

\section{The fully anti-symmetric case} \label{fully_anti}
In this section as well as in Secs.~\ref{Sec_2D}--\ref{summary_purely_antisymmetric}, we will study the {\it fully anti-symmetric case}, in which $K_+=0$. 
To simplify the notations, we write $K_- \equiv K$ (and $\tilde{K}_- \equiv \tilde{K}$). In this case, the mean-field equation reads
\begin{equation}
    \label{mean_field_pure}
    \partial_t m_{\alpha}=-m_{\alpha}+\tanh\left[\tilde{J} m_{\alpha}+\tilde{K}\varepsilon_{\alpha\beta}m_{\beta} + D \nabla^{2} m_{\alpha} \right]
\end{equation}
where $\varepsilon_{\alpha \beta}$ is the Levi-Civita symbol.
Equation \eqref{mean_field_pure} has the symmetry 
\begin{equation}
    \mathscr{S}_2 : (m_A, m_B) \mapsto (-m_B, m_A)
\end{equation}
in addition to that in Eq.~\eqref{symmetrygen1}, so overall it is invariant under the cyclic group $C_4$.

\begin{figure*}[ht]
\centering
{\includegraphics[width=0.8\textwidth,draft=false]{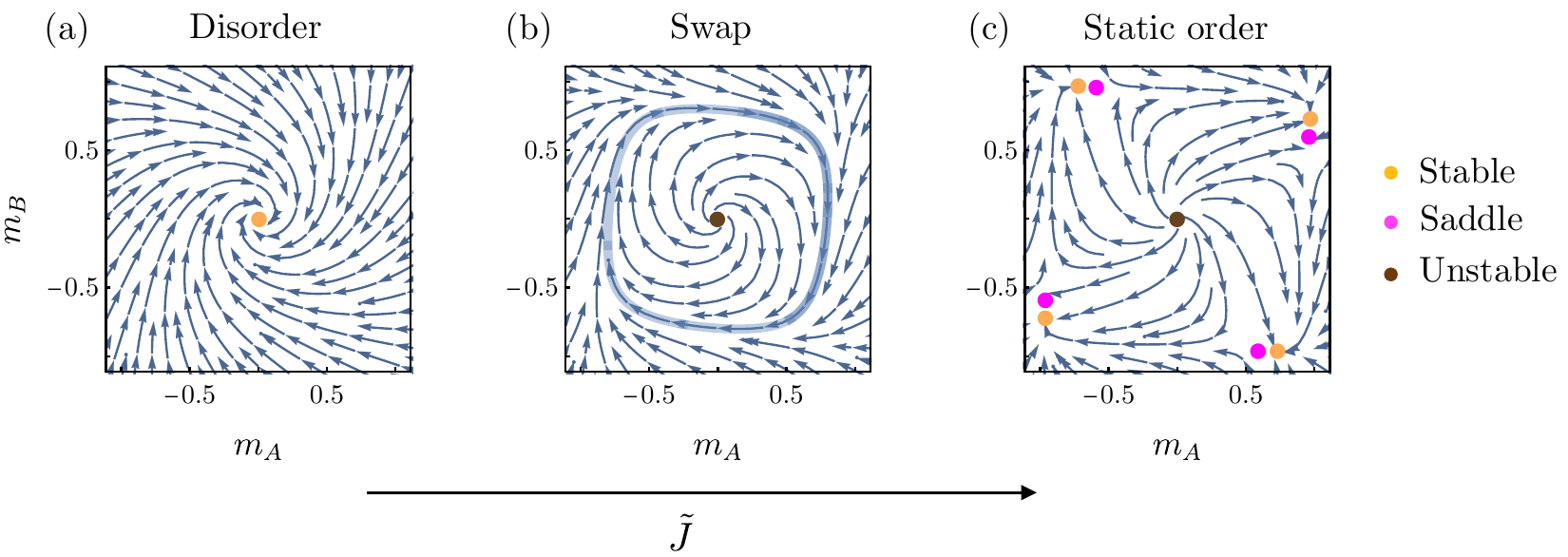}}
\caption{
\textbf{Mean-field (MF) flow in $\left(m_A,m_B\right)$ space.} Arrows correspond to the vector $\vec{v}=(\partial_t m_A,\partial_t m_B)$ calculated from the spatially homogeneous full mean-field equation, Eq.~(\ref{mean_field_pure}). Fixed points are represented by dots. The limit cycle in the swap phase is highlighted by a semi-transparent blue line. Nonreciprocal inter-species coupling is $\tilde{K} = 0.4$, and reciprocal coupling between same-species nearest neighbors is (a) $\tilde{J} = 0.8$ , (b), $\tilde{J} = 1.3$, and (c) $\tilde{J} = 1.8$, showing the disorder, swap, and static-order phases, respectively.}
\label{MF_Flow}
\end{figure*}
\subsection{Connection with locally-coupled oscillators} \label{locally_coupled_oscillators}
In the fully anti-symmetric case, the dynamics of $\theta_i$ is described by the forward and backward moves,
\begin{subequations}\label{ForwardBackward}
\begin{align}
& \theta_{i}\to\theta_{i}+\frac{\pi}{2}    \,\,\,\, {\rm with\,\,rate} \,\,\,\, w^+ \\
& \theta_{i}\to\theta_{i}-\frac{\pi}{2}    \,\,\,\, {\rm with\,\,rate} \,\,\,\, w^-
\end{align}
\end{subequations}
where the rates are
\beq
w^{\pm}=\frac{1}{2\tau}\left[1-\tanh\left(\frac{\Delta E(\vec{\theta})}{2 K_B T}\pm\tilde{K} \right)\right]
\eeq
and
\beq \label{thetaDynamics2}
E(\vec{\theta})=-2J\sum_{\langle i,j\rangle}\cos\left(\theta_{i}-\theta_{j}\right).
\eeq
with the sum running over nearest-neighbor pairs. Equations~(\ref{ForwardBackward}--\ref{thetaDynamics2}) can be viewed as describing locally-coupled oscillators with internal frequency and discrete states. In this sense, they share similarities with the Wood {\it et al.} cyclic model~\cite{wood2006universality,wood2006critical,assis2011infinite} although oscillators are restricted to forward moves (backward moves are prohibited) in these references. For $K=0$, our model reaches an equilibrium state that is equivalent to the clock model (vector Potts model) for $q=4$~\cite{suzuki1967solution,tobochnik1982properties}.

\subsection{Analysis of the bifurcations} \label{Bifurcations}
In the ${m_\alpha\ll1}$ limit and for small gradients, the mean-field equation in the fully anti-symmetric case (Eq.~(\ref{mean_field_pure})) can be approximated by
\begin{equation}
\label{MeanFieldExpanded}
\begin{aligned} 
\partial_t m_{\alpha}  =&-(1-\tilde{J})m_{\alpha}+\tilde{K} \varepsilon_{\alpha\beta}m_{\beta}+D \nabla^{2}m_{\alpha}\\
 & -\frac{1}{3}\left(\tilde{J}m_{\alpha}+\tilde{K} \varepsilon_{\alpha\beta} m_{\beta}\right)^{3}. 
\end{aligned}
\end{equation}
For $\tilde{K}=0$, Eq.~(\ref{MeanFieldExpanded}) reduces to two uncoupled mean-field equations that derive from a $\phi^4$ Ginzburg-Landau potential.
For any $\tilde{K}\neq 0$, Eq.~(\ref{MeanFieldExpanded}) does not derive from a potential, and describes a non-equilibrium system. Note that due to the lack of rotational symmetry in the cubic term, Eq.~(\ref{MeanFieldExpanded}) differs from the Complex Ginzburg-Landau (CGL) equation~\cite{aranson2002world,garcia2012complex}, which is given by
\beq \label{CGL_eq}
\partial_t \psi  =a \psi+b \nabla^{2}\psi
  + c|\psi|^2 \psi
\eeq
where $\psi({\vec r},t)$ is a complex field and $a$, $b$ and $c$ are complex numbers. However, Eq.~(\ref{MeanFieldExpanded}) becomes CGL-like in limits discussed below.

Here, we restrict our analysis to spatially homogeneous solutions, $m_\alpha({\vec r},t)\equiv m_\alpha(t)$.
The linear stability of such solutions to Eq.~\eqref{MeanFieldExpanded} is unaffected by the Laplacian term, and we therefore analyze the spatially homogeneous mean-field equation given by
\begin{equation}
\begin{aligned} \label{MeanFieldHomog}
\partial_t m_{\alpha}  =&-(1-\tilde{J})m_{\alpha}+\tilde{K} \varepsilon_{\alpha\beta}m_{\beta}\\
 & -\frac{1}{3}\left(\tilde{J}m_{\alpha}+\tilde{K} \varepsilon_{\alpha\beta} m_{\beta}\right)^{3}.
\end{aligned}
\end{equation}

Figure~\ref{MF_Flow} shows the phase-space velocity field $\vec{v}=(\partial_t m_A,\partial_t m_B)$ from the spatially homogeneous full mean-field equation, Eq.~(\ref{mean_field_pure}) along with the fixed points which are obtained by numerically solving the $\partial_t m_A = \partial_t m_B = 0$ equations.
The stability of the fixed points (shown in the figure as orange, pink, and brown for stable, saddle, and unstable, respectively) is determined by analyzing the eigenvalues of the Jacobian matrix $\mathcal{J}$ of the linearized equation $\partial_t \delta \vec{m}=\mathcal{J} \delta \vec{m} $ where linearization is performed around each fixed point. 

Three distinct phases are found: a disordered phase in which ${m_A=m_B=0}$ is the only stable fixed point (Fig.~\ref{MF_Flow}a), a limit-cycle phase that we dub \enquote{swap phase} as both species repeatedly flip their magnetizations
(Fig.~\ref{MF_Flow}b), and a static-order (ferromagnetic) phase in which, depending on initial conditions, the system ends up in one out of four stable fixed points with non-zero $m_A$ and $m_B$ (Fig.~\ref{MF_Flow}c).

\begin{figure}[ht]
\centering
{\includegraphics[width=0.28\textwidth,draft=false]{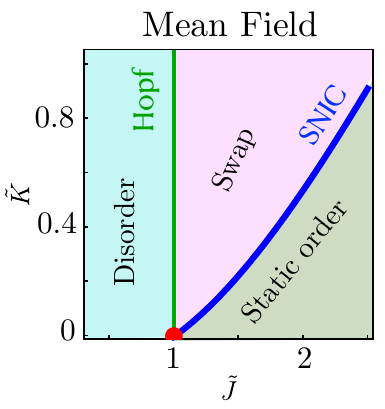}}
\caption{
{\bf Mean-field phase diagram}. The three phases shown on the diagram are (i) disorder, (ii) swap and (iii) static order. The lines separating the phases are Hopf bifurcation (thin green) and SNIC bifurcation (thick blue). Red dot is the Pitchfork bifurcation. Obtained from numerical analysis of the spatially homogeneous Eq.~(\ref{mean_field_pure}).}
\label{MF_Phase_diagram}
\end{figure}

The mean-field phase diagram obtained from the spatially homogeneous full mean-field equation is shown in Fig.~\ref{MF_Phase_diagram}, as a function of the couplings $\tilde{J}$ and $\tilde{K}$. The phases are determined by the fixed points and their stability as described above. When nonreciprocal interactions are turned off (${\tilde{K}=0}$), a pitchfork bifurcation at ${\tilde{J} = 1}$ (red point in Fig.~\ref{MF_Phase_diagram}) separates the disordered phase (in blue) from the static-order phase (in green), as in the usual Ising model.
When nonreciprocal interactions are present (${\tilde{K}\neq 0}$), the swap phase arises (in pink) and is separated from the disordered state by a Hopf bifurcation at $\tilde{J}=1$ (green line), and from the static-order phase by another bifurcation known as a ($C_4$-symmetric) saddle-node on an invariant circle (SNIC) bifurcation~\cite{Strogatz2018,Izhikevich2007} (blue line). We analyze each of these bifurcations below using Eq.~(\ref{MeanFieldHomog}).

\subsubsection{Pitchfork bifurcation}
For ${\tilde {K}=0}$, Eq.~(\ref{MeanFieldHomog}) leads to the standard pitchfork bifurcation at ${\tilde{ J} = 1}$. The ${m_A=m_B=0}$ point becomes unstable and four stable fixed points 
\beq
m_A=\pm m_B=\pm \sqrt{\frac{3(\tilde{J}-1)}{\tilde{J}^3}}
\eeq
emerge, corresponding to the ordered phase.

\subsubsection{Hopf bifurcation} \label{sec_Hopf}
For $\tilde{K}\neq 0$, the eigenvalues of $\mathcal{J}$ at $m_A=m_B=0$ are complex,
\beq
\lambda_{1,2}=-1+\tilde{J}\pm i\tilde{K}.
\eeq
This means that there is a Hopf bifurcation at ${\tilde{J}=1}$: the ${m_A=m_B=0}$ point become unstable for ${\tilde J}>1$  and a stable limit cycle emerges, corresponding to the swap phase.

The limit cycle is more easily analyzed in polar coordinates, ${m_A = r\cos \theta}$ and ${m_B = r\sin \theta}$ where Eq.~(\ref{MeanFieldHomog}) becomes
\begin{align}\label{MF_polar}
\partial_{t}r&=(\tilde{J}-1)\,r-\frac{1}{4}\tilde{J}^{3}\bigg[1+\kappa^{2}\nonumber\\
&+\left(\frac{1}{3}-\kappa^{2}\right)\cos(4\theta)+\left(\kappa-\frac{1}{3}\kappa^{3}\right)\sin(4\theta)\bigg]r^{3}
\\\nonumber
\partial_{t}\theta&=-\tilde{K}+\frac{1}{4}\tilde{J}^{3}\bigg[\kappa+\kappa^{3}\\
\label{MF_radial}&-\left(\kappa-\frac{\kappa^{3}}{3}\right)\cos(4\theta)+\left(\frac{1}{3}-\kappa^{2}\right)\sin(4\theta)\bigg]r^{2}\; ,
\end{align}
with $\kappa \equiv \tilde{K}/\tilde{J}$.

The limit cycle describes anharmonic oscillations, where both $r$ and $\dot{\theta}$ vary with $\theta$. 
Its shape has a rotational symmetry of order 4 due to the $C_4$ symmetry of the mean-field equation. However, very close to the bifurcation, {\it i.e.} $\tilde{J}-1\ll1$ and any fixed $\tilde{K}$, the oscillations are harmonic~\cite{kuramoto1984chemical}. This can be seen as follows.

\begin{figure*}[ht]
\centering
{\includegraphics[width=0.6\textwidth,draft=false]{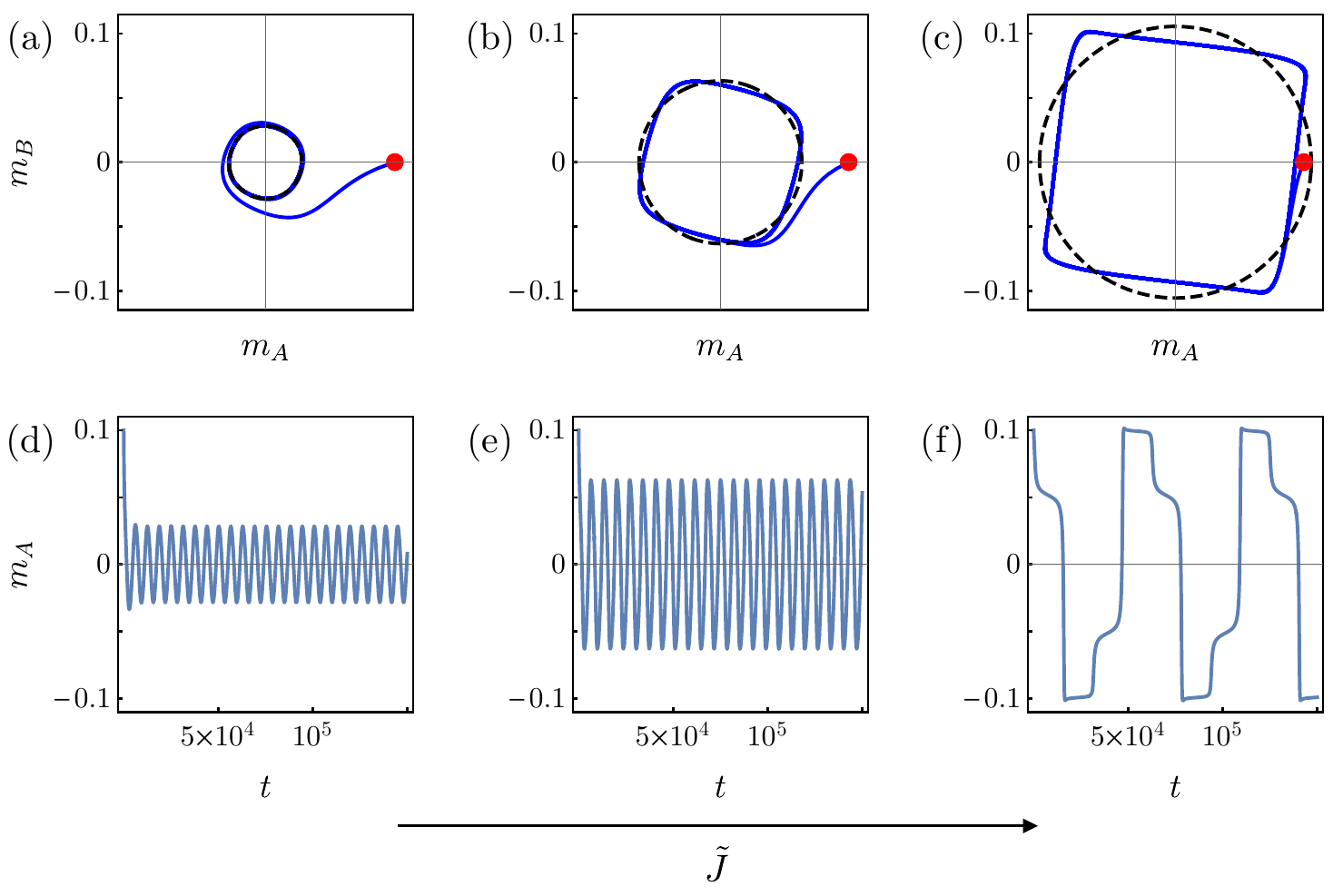}}
\caption{
\textbf{Mean-field swap phase at different distances from the Hopf and SNIC bifurcations.} 
Numerical solutions of the spatially homogeneous mean-field dynamics, Eq.~(\ref{MeanFieldHomog}). (a-c) $m_B$ vs. $m_A$. (d-f) $m_A$ vs. $t$. From left to right: $\tilde{J}=1.0002$, $1.001$, and $1.0028$. In all plots $\tilde{K}=0.001$. Red dot is the initial state and dashed black line is the steady state predicted from Eq.~(\ref{r_of_theta2}), valid near the Hopf bifurcation.}
\label{FigMF2}
\end{figure*}

The equation for $\theta$ in the ${\tilde{J}-1\ll1}$ limit is
 \beq \label{thetat_Hopf}
\partial_{t}\theta=-\tilde{K} + \mathcal{O}(\tilde{J}-1)\; ,
 \eeq
 where we used the fact that ${r^{2}\sim \tilde{J}-1
}$. The angle $\theta$ then has the solution ${\theta = \theta_0 - \tilde{K} t}$, and the equation for $r$ becomes
\begin{align} \label{rt_Hopf}
\partial_{t}r & =(\tilde{J}-1)\,r-\frac{1}{4}\bigg[1+\tilde{K}^{2}+\left(\frac{1}{3}-\tilde{K}^{2}\right)\cos(4\tilde{K}t)\\
 & -\left(\tilde{K}-\frac{1}{3}\tilde{K}^{3}\right)\sin(4\tilde{K}t)\bigg]r^{3}+\mathcal{O}\left((\tilde{J}-1)^{5/2}\right)
\nonumber
\end{align}
where we have set $\theta_0$ to zero. Equation~(\ref{rt_Hopf}) can be solved analytically. In steady state ($t\to\infty$ limit), its solution reads
\begin{equation}
r\left(t\right)=2\sqrt{\frac{\tilde{J}-1}{1+\tilde{K}^{2}}}\times\frac{1}{\sqrt{1-(\tilde{J}-1)f(\tilde{J},\tilde{K},t)}}
\end{equation}
with
\begin{align}
f(\tilde{J},\tilde{K},t) & =\frac{1-\tilde{J}+3(\tilde{J}-3)\tilde{K}^{2}+2\tilde{K}^{4}}{3(1+\tilde{K}^{2})\left[(\tilde{J}-1)^{2}+4\tilde{K}^{2}\right]}\cos(4\tilde{K}t)\\
 & -\frac{\tilde{K}\left(5-3\tilde{J}+(\tilde{J}-7)\tilde{K}^{2}\right)}{3(1+\tilde{K}^{2})\left[(\tilde{J}-1)^{2}+4\tilde{K}^{2}\right]}\sin(4\tilde{K}t) \nonumber
\end{align}
Taking the limit $\tilde{J}-1\ll1$ once more $(\tilde{J}-1)f(\tilde{J},\tilde{K},t)
$ goes to zero and we arrive at
\beq \label{r_of_theta2}
r\left(t\right)=2\sqrt{\frac{\tilde{J}-1}{1+\tilde{K}^{2}}}
\eeq
which is a circular motion of constant angular velocity in phase space. This harmonic solution can be rationalized by looking at the two times scales corresponding respectively to the angular and radial motion near the bifurcation (described by Eqs.~(\ref{thetat_Hopf}) and~(\ref{rt_Hopf}), respectively). The angular motion time scale is $T_{\theta} \sim \tilde{K}^{-1}$ while the radial motion time scale is determined by the relaxation time of $r(t)$ to its new minimum when $\theta$ changes, which scales as $T_r\sim(\tilde{J}-1)^{-1}$. 
The two time scales $T_{\theta}$ and $T_r$ can be interpreted as \enquote{oscillation time-scale} and \enquote{alignment time-scale}, respectively. 
Since $T_r\gg T_{\theta}$, the angular motion occurs much more rapidly, leaving insufficient time for $r$ to reach a new minimum when $\theta$ changes. Thus, only the $\theta$-averaged value of the right-hand side of Eq.~(\ref{MF_polar}) matters. 
This is supported by Fig.~\ref{FigMF2} which shows typical phase-space trajectories in the swap phase: as $\tilde{J}-1$ increases, the motion indeed loses its rotational symmetry and become more \enquote{square-like}.

The above result is a consequence of a more general theorem. 
A reaction-diffusion equation that undergoes a Hopf bifurcation such as the mean-field evolution Eq.~(\ref{MeanFieldExpanded}) can be reduced, close to the bifurcation, to the CGL equation ({\it i.e.}, the CGL equation is its normal form). 
The transformation to the CGL equation guarantees that the steady-state solution of the non-transformed set of equations is an elliptical limit cycle (or in special cases like here, circular)~\cite{kuramoto1984chemical}. 
Because of this reduction, we expect the disorder-to-swap phase transition in our model to be in the same universality class as the Hopf bifurcation in the CGL equation, which has been studied in Refs.~\cite{risler2004universal,risler2005universal}.

\subsubsection{Infinite-period (SNIC) bifurcation}
Let us now start from the limit-cycle swap phase. 
Upon increasing $\tilde{J}$ or decreasing $\tilde{K}$, the system goes through a ($C_4$-symmetric) saddle-node on an invariant circle (SNIC) bifurcation, where four pairs of stable and saddle fixed points emerge from the limit cycle (pink and orange points in Fig.~\ref{MF_Flow}c). 
The SNIC bifurcation is also known under several other names: infinite-period bifurcation (because the period of oscillation diverges at the bifurcation point), saddle-node on a limit cycle (SNLC), saddle-node inﬁnite-period (SNIPER), etc.
As the dynamics is invariant under the action of the group $C_4$ (see Sec.~\ref{fully_anti}), four copies of the SNIC bifurcation happen at the same time. 
Bifurcations in such a dynamical system with symmetries are known as equivariant bifurcations, see Refs.~\cite{Crawford1991,Golubitsky1988,Chossat2000} for an introduction. 
Once the stable fixed points emerge, the system is in a static-order phase. 

Noticing that the SNIC bifurcation occurs when $\tilde{K}$ and $\tilde{J}-1$ are comparable (see Fig.~\ref{MF_Phase_diagram}), we now take the limit $\epsilon \to 0$ with $\tilde{J}-1 \sim \epsilon$ and $\tilde{K}\sim \epsilon$.
At first order in $\epsilon$, Eqs.~(\ref{MF_polar}--\ref{MF_radial}) become
\begin{subequations}    
\begin{align}
\partial_{t}r & =(\tilde{J} - 1)\,r-\frac{1}{4}\left(1+\frac{1}{3}\cos(4\theta)\right)r^{3}\; , \\
\partial_{t}\theta & =-\tilde{K}+\frac{1}{12}\sin(4\theta)r^{2}.
\end{align}
\end{subequations}
In a static-order state, $\partial_t r = \partial_t \theta = 0$, with $r$ given by
\beq
r=2\sqrt{\frac{\tilde{J}-1}{1+\frac{1}{3}\cos(4\theta)}} \;.
\label{r_SNIC}
\eeq
The equation for $\theta$ then becomes
\beq
\partial_{t}\theta=-\tilde{K}+\frac{\sin(4\theta)}{3+\cos(4\theta)}(\tilde{J}-1).
\eeq
Since $\sin(4\theta)/(3+\cos(4\theta))$ is bounded between $-1/(2\sqrt{2})$ and $1/(2\sqrt{2})$, a stationary state (with $\partial_t \theta=0$) can only exist if ${\tilde{K}<(\tilde{J}-1)/(2\sqrt{2})}$. Therefore, in the $\tilde{J}-1 \ll 1$ regime, the SNIC bifurcation line corresponds to
\beq
\tilde{K}=\frac{\tilde{J} - 1}{2\sqrt{2}}.
\label{SNIC_bifurcation}
\eeq
Note that, in Fig.~\ref{MF_Phase_diagram}, the thick blue line corresponding to the SNIC bifurcation was obtained numerically for the full mean-field dynamics Eq.~(\ref{mean_field_pure}).
Equation~(\ref{SNIC_bifurcation}) only predicts the tangent of this thick blue line close to the pitchfork bifurcation point ($\tilde{ J}=1$, $\tilde{K}=0$).

Finally, the fixed points emerging at the SNIC bifurcation are obtained from the $\theta$ values that maximize $\sin(4\theta)/(3+\cos(4\theta))$, which are
\begin{equation}
    \theta_n=\left[-\tan^{-1}\left(2\sqrt{2}\right)+\pi\left(2n+1\right)\right]/4    
\end{equation}
with $n=0,1,2,3$. Together with their corresponding $r$ values derived from Eq.~(\ref{r_SNIC}) we obtain
\begin{equation}
\left(m_A,m_B\right)=3\sqrt{\frac{\tilde{J}-1}{2}} \left(\cos(\theta_n),\sin(\theta_n)\right).
\end{equation}
Approaching the SNIC bifurcation from the side of the limit cycle (swap phase), $\partial_t \theta$ approaches zero when the system passes at $\theta=\theta_n$. As a result, the oscillation period diverges at the bifurcation (see Fig.~\ref{FigMF2}d-f, where the period increases as we approach the bifurcation). This stands in contrast with the Hopf bifurcation in which the oscillation period remains finite.

\section{Monte-Carlo simulations}
\label{MonteCarlo_App}

To go beyond the mean-field description, we perform large-scale Monte-Carlo simulations of the nonreciprocal Ising model in both two and three dimensions (results in one dimension are summarized in Appendix~\ref{1D_app}).

In our 2D discrete-time Monte-Carlo (MC) simulations, we use the standard Glauber algorithm: at each MC step a spin $\sigma_i^\alpha$ is chosen at random and then flipped with probability
\begin{equation} \label{prob}
p_{i}^{\alpha}=\tau w(F^{\alpha}_{i}\vec{\sigma}|\vec{\sigma})    
\end{equation}
(with $w$ defined in Eq.~(\ref{rates})).

Since three-dimensional simulations are much more costly, in 3D we update spins using a checkerboard algorithm to implement Eq.~(\ref{prob}) instead of random picking. In the checkerboard algorithm non-interacting spins are flipped in parallel~\cite{yang2019high}. 
To generalize the checkerboard approach to the two-species case, we defined the quantity $\mathcal{P}=x_1+x_2+x_3+\alpha$ where $x_n$ is the integer position in the $n$-th dimension of the cubic lattice, and $\alpha=0,1$ labels the two species.
Spins with the same parity of $\mathcal{P}$ (even or odd) are then updated simultaneously. This allows us both to parallelize the simulation and avoid random number generation in choosing the spins. 
We did not observe qualitative changes when changing the update rules from random picking to paralleled checkerboard, as well as when using other update rules, see Appendix~\ref{different_update}.

Unless mentioned otherwise, we initialize the system in an ordered state where all spins of the same species are either up or down, and let the system evolve until a steady state is reached. Averages are then calculated in the steady state, excluding the initial transient. Time is measured in simulation sweeps, where a single sweep is defined as $2\times L^d$ (the number of spins) MC steps (individual attempted flips). To best visualize phenomena of interest, simulation snapshots are presented either with the spin variables $\sigma^A$ and $\sigma^B$, or with the angle variable $\theta$.

Transitions from disorder to time-dependent order and from time-dependent order to stationary order can be quantified by two order parameters.
First, the synchronization order parameter
\begin{equation} \label{R_definition}
R\equiv \left\langle s\right\rangle_{t,\Omega}
\,\,\,\,\text{with}\,\,\,
s \equiv  \sqrt{\frac{M_A ^2 + M_B ^2}{2}}
\end{equation}
quantifies how much spins are aligned with each other within each species.
Here, $M_\alpha=\sum_i \sigma_i^\alpha/L^d$ are the total magnetizations and the average $\langle ...\rangle_{t,\Omega}$ is over time ($t$) and realizations ($\Omega$).
Note that $s$ can also be written as 
\begin{equation}
    s=\frac{1}{L^d}\bigg|\sum_{j}{\rm e}^{i\theta_j}\bigg|,
\end{equation}
similarly to the order parameter quantifying the synchronization of oscillations in, for example, the Kuramoto model~\cite{acebron2005kuramoto}. Second, the phase-space angular momentum
\begin{align} \label{Angular_momentum}
\mathcal{L}\equiv\langle 
 \ell\rangle_{t,\Omega}\,\,\,{\rm with}\,\,\,\ell & \equiv M_{B}\partial_{t}M_{A}-M_{A}\partial_{t}M_{B} \nonumber\\
 & =M_{B,t}M_{A,t+1}-M_{A,t}M_{B,t+1}
\end{align}
quantifies the macroscopic time-reversal symmetry breaking. It is related to the entropy production~\cite{seara2021irreversibility}, see Appendix~\ref{EntProd} for details.

\subsection*{Mean-field behavior of the order parameters}

\begin{figure}[ht]
\centering
{\includegraphics[width=0.5\textwidth,draft=false]{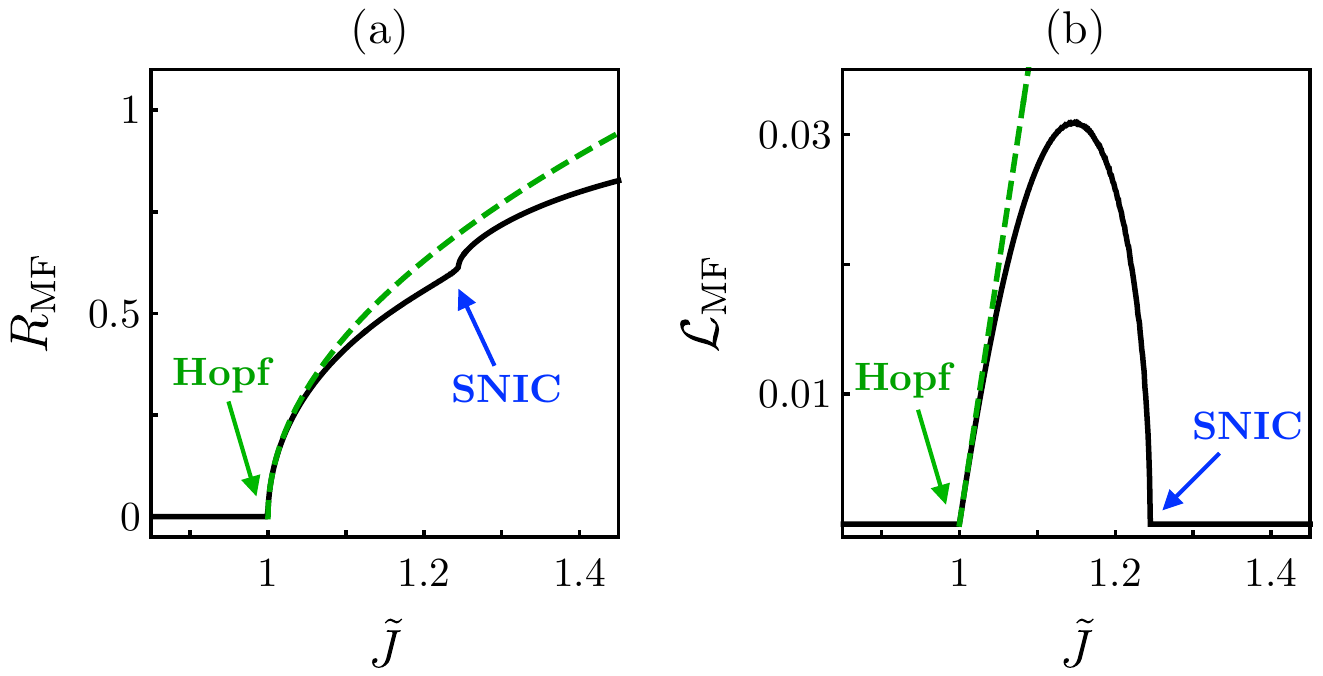}}
\caption{
\textbf{Mean-field order parameters}.
(a) The synchronization order parameter $R_{\rm MF}$ (Eq.~(\ref{MF_order_par1})) and (b) the phase-space angular momentum $\mathcal{L}_{\rm MF}$ (Eq.~(\ref{MF_order_par2})), shown as a function of $\tilde{J}$ and for fixed $\tilde{K}=0.1$. Black lines are obtained from numerical solutions of the spatially homogeneous Eq.~(\ref{mean_field_pure}) in steady state. Dashed green lines are approximate solutions that are valid near the Hopf bifurcation, given by Eq.~(\ref{RMF_LMF}).}
\label{Order_parameters_MF}
\end{figure}

In mean-field and for a spatially uniform state, the synchronization order parameter defined in Eq.~\eqref{R_definition} is
\begin{equation} \label{MF_order_par1}
    R_{\rm MF}=\frac{\langle r\rangle_t}{\sqrt{2}}
\end{equation}
and the phase-space angular momentum defined in Eq.~\eqref{Angular_momentum} is
\begin{equation}
\label{MF_order_par2}
    \mathcal{L}_{\rm MF}=-\langle r^2\dot{\theta}\rangle_t.
\end{equation}
The sign is chosen so that $\mathcal{L}$ is positive when $K$ is positive. In Fig.~\ref{Order_parameters_MF}, $R_{\rm MF}$ and $\mathcal{L}_{\rm MF}$, calculated from a numerical steady-state solution of the spatially homogeneous Eq.~(\ref{mean_field_pure}), are plotted as a function of $\tilde{J}$ for a fixed value of $\tilde{K}$. It is shown that $R_{\rm MF}$ goes from zero to non-zero values when crossing the Hopf bifurcation at $\tilde{J} = 1$. When crossing the SNIC bifurcation, $R_{\rm MF}$ remains finite, but there is a sharp change in the slope: the slope increases when approaching the bifurcation and decreases in the static order regime. Conversely, $\mathcal{L}_{\rm MF}$ is identically zero in both the disordered and static-order phases, but deviates from zero in the swap phase. Both $R_{\rm MF}$ and $\mathcal{L}_{\rm MF}$ are continuous at the bifurcation points, as in a second-order phase transition. In the swap phase and near the Hopf bifurcation, we can use Eq.~(\ref{thetat_Hopf}) and Eq.~(\ref{r_of_theta2}) to obtain
\begin{equation} \label{RMF_LMF}
R_{\rm MF}=\sqrt{\frac{2(\tilde{J}-1)}{1+\tilde{K}^{2}}}
\quad
\text{and}
\quad
\mathcal{L}_{\rm MF}=\frac{4\tilde{K}(\tilde{J}-1)}{1+\tilde{K}^{2}}.
\end{equation}
In the swap phase and near the SNIC bifurcation, $\mathcal{L}_{\rm MF}$ scales with $\mu^{1/2}$ where $\mu$ is the distance from the bifurcation line (because the period scales as $T \sim \mu^{-1/2}$ while $r$ remains finite at the bifurcation~\cite{Strogatz2018}).
We now analyze how these mean-field results hold in finite-dimensional systems using the Monte-Carlo simulations.

\begin{figure}[ht]
\centering
{\includegraphics[width=0.45\textwidth,draft=false]{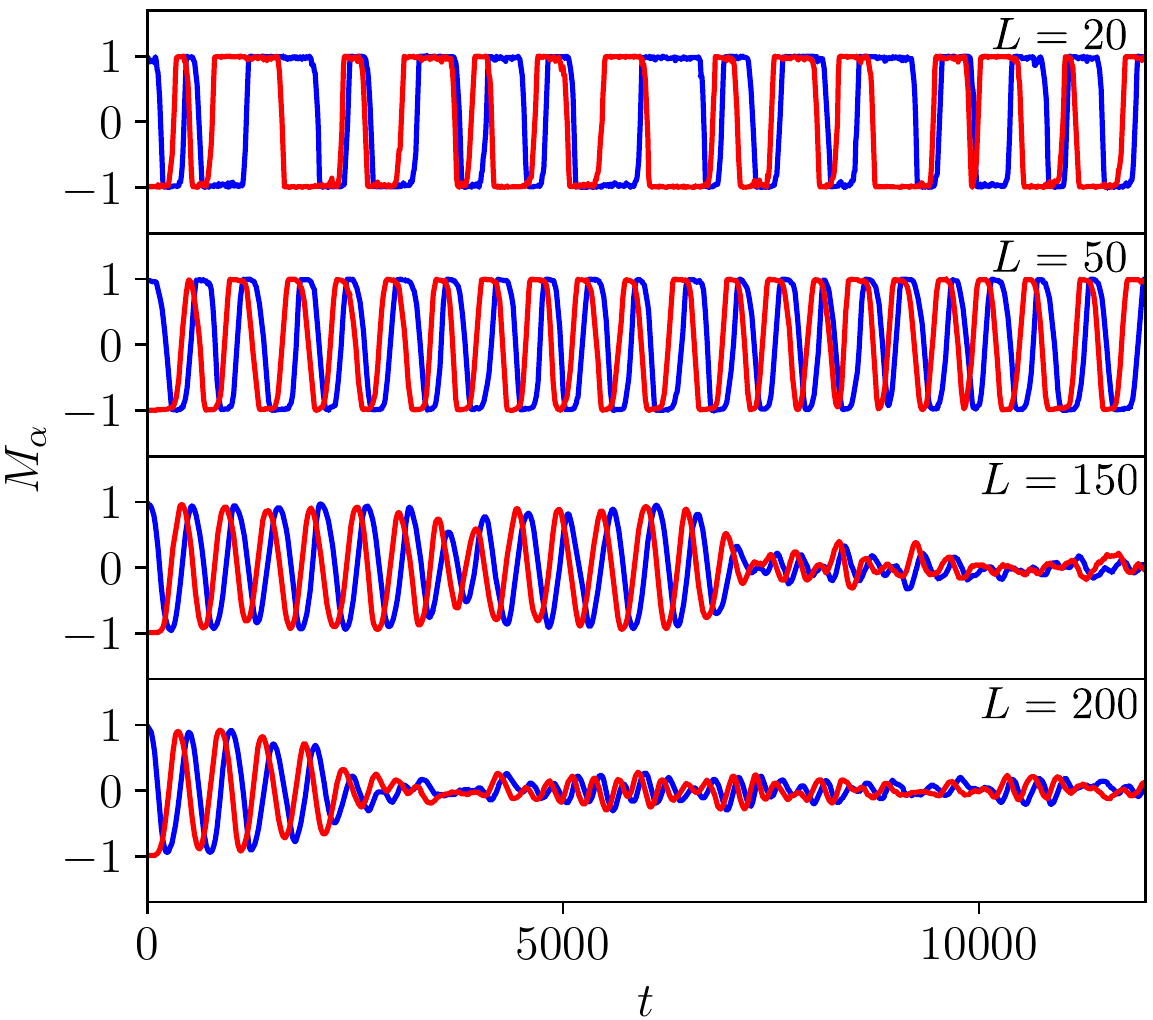}}
\caption{
\textbf{Swap phase in finite 2D systems.}
The magnetizations $M_A$ (blue) and $M_B$ (red) as a function of time, for increasing system sizes (from top to bottom: $L=20,50,150,200$). Obtained from MC simulations with the parameters: $\tilde{J}=2.8$, $\tilde{K}=0.3$.
}
\label{2D_L_increase}
\end{figure}

\section{Destruction of the swap phase in 2D by spiral defects} \label{Sec_2D}
\begin{figure*}[htbp!]
\centering
{\includegraphics[width=0.95\textwidth,draft=false]{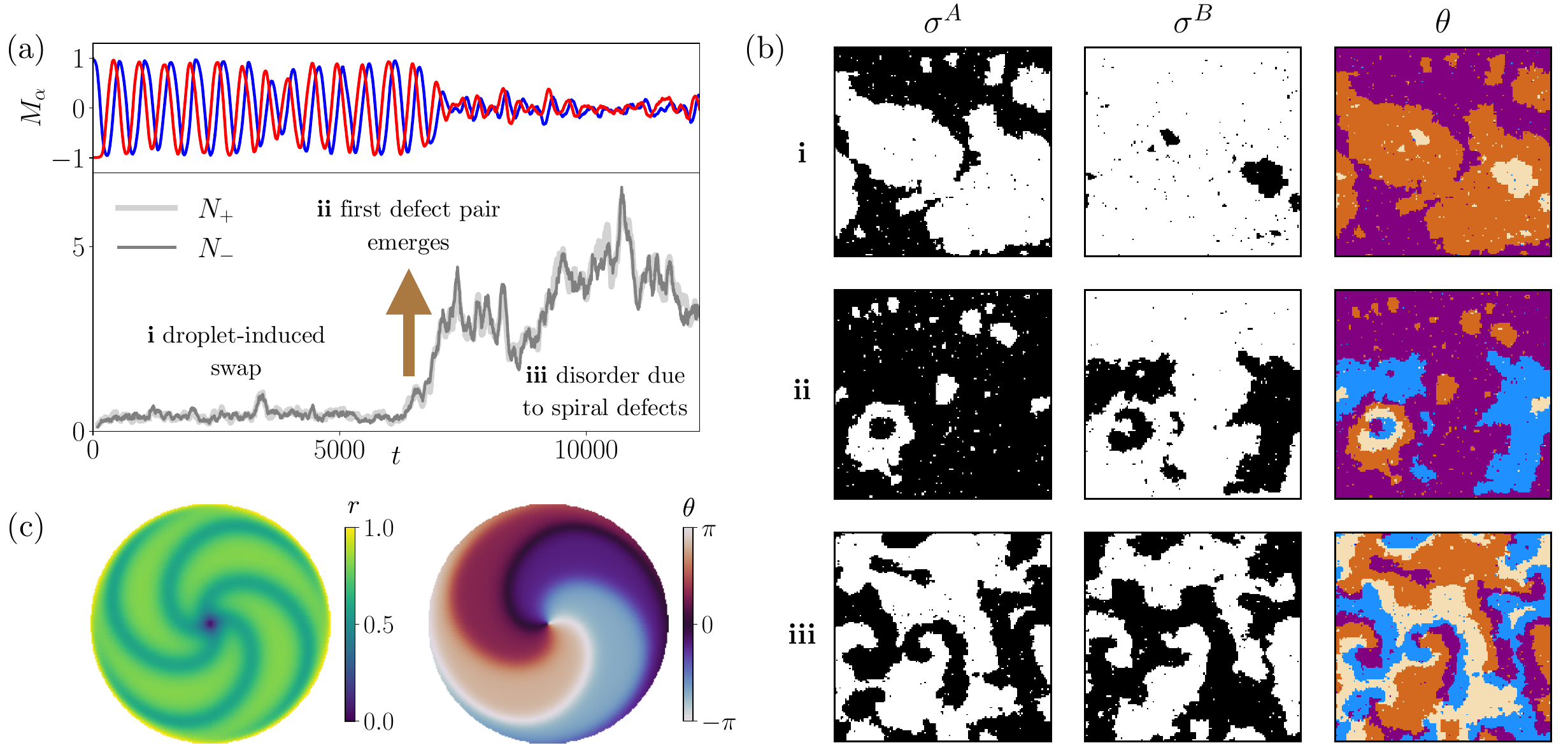}}
\caption{
\textbf{Destruction of the swap phase by spiral defects in 2D.}
(a) Top: the magnetizations $M_A$ (blue) and $M_B$ (red) as a function of time (the same as in Fig.~\ref{2D_L_increase}, third row). Bottom: the number of + and - defects $N_+$ and $N_-$ averaged over 100 time steps, as a function of time. We identify defects as adjacent 2 by 2 sites with a cycle of $\uparrow \uparrow\to\uparrow \downarrow\to\downarrow \downarrow\to\downarrow\uparrow$, either clockwise ($+$ defects) or anti-clockwise ($-$ defects). (b) Snapshots of $\sigma^A$, $\sigma^B$, and $\theta$ (i) before, (ii) during, and (iii) after the proliferation of spiral defects. Black and white are up and down spins, respectively, and the $\theta$ color code corresponds to Fig.~\ref{theta}. See Movie 1 for the full evolution. (c) Snapshot of a spiral in the mean-field equation, showing $r$ (left) and $\theta$ (right). Note the 4-fold rotational symmetry.
Panels a-b are produced from the same MC simulation, with the parameters: $\tilde{J}=2.8$, $\tilde{K}=0.3$, and $L=150$.
Panel c is obtained by solving numerically Eq.~(\ref{MeanFieldExpanded}) on a disk of radius $R_{\text{sim}}$ with boundary conditions $\psi(r=R_{\text{sim}},\theta) = \psi_0 e^{i \theta}$ ($r$ and $\theta$ are polar coordinates) that enforce the existence of a topological defect.
We have set $\tilde{J}=2$, $\tilde{K}=0.6$, $D=0.5$, $R_{\text{sim}}=10$.
Equation~(\ref{MeanFieldExpanded}) is integrated using the pseudospectral solver Dedalus \cite{Burns2020} on a disk domain discretized with $N_\phi=201$ azimuthal modes and $N_r=200$ radial modes using a 1st-order 1-stage diagonally implicit-explicit Runge-Kutta scheme (RK111 in Dedalus) \cite{Ascher1997} with a fixed time step $dt = \num{1e-2}$ for \num{20} simulation time units.
}
\label{Fig2}
\end{figure*}
In small 2D systems, simulations show states in which $M_A$ and $M_B$ oscillate in time (like in the mean-field swap phase). These oscillations are irregular, and typically occur through nucleation of droplets of opposite magnetization by the \enquote{unsatisfied} species, in alternating order (see Movie 1).

However, as $L$ increases these oscillations become a transient that quickly destabilizes (see Fig.~\ref{2D_L_increase}) due to the formation and proliferation of spiral defects. Figure~\ref{Fig2} and Movie 1 show the emergence of defects after an oscillatory transient. 
The defects are manifest in the $\theta$ variable and appear to have four distinct arms, ordered according to the cycle $\uparrow \uparrow\to\uparrow \downarrow\to\downarrow \downarrow\to\downarrow\uparrow$ that can be followed either clockwise ($+$ defects) or anti-clockwise ($-$ defects). The 4-fold rotational symmetry in the spiral shape stems from the $C_4$ symmetry of the model in the fully anti-symmetric case, and is even more visible in the mean-field equation, see Fig.~\ref{Fig2}c displaying a snapshot of a spiral obtained from integrating numerically Eq.~(\ref{MeanFieldExpanded}) with a boundary condition that imposes a winding number. In terms of the complex order parameter $\psi = m_A + i m_B$, the $C_4$ symmetry of the mean-field equation translates into defects satisfying $e^{-i \theta} \, \psi(R_\theta \vec{r},t) = \psi(\vec{r},t)$ for $\theta=0,\pi/2,\pi,3\pi/2$, where $R_\theta$ is the rotation matrix with angle $\theta$.

Despite having 4-fold rather than full rotational symmetric shape, the spiral waves observed in the nonreciprocal Ising model appear to behave similarly to those emerging in the CGL equation~\cite{aranson2002world,aranson1998spiral,Aranson1993,grinstein1993temporally,altman2015two,wachtel2016electrodynamic,chate1996phase,tan2020topological,liu2021topological,Michaud2022,Brauns2021,zhang2023pulsating,manacorda2023pulsating}. Defects are created and annihilated in pairs and opposite topological charges ($+$ and $-$ defects) lead to opposite angular velocities  (see Movie 1).
Finally we note that in mean-field, a homogeneously oscillating system is stable, but given sufficiently strong added noise spirals emerge, as in our Monte-Carlo simulations (see Appendix~\ref{2D_MF_noise}).

\begin{figure}[htbp!]
\centering
{\includegraphics[width=0.5\textwidth,draft=false]{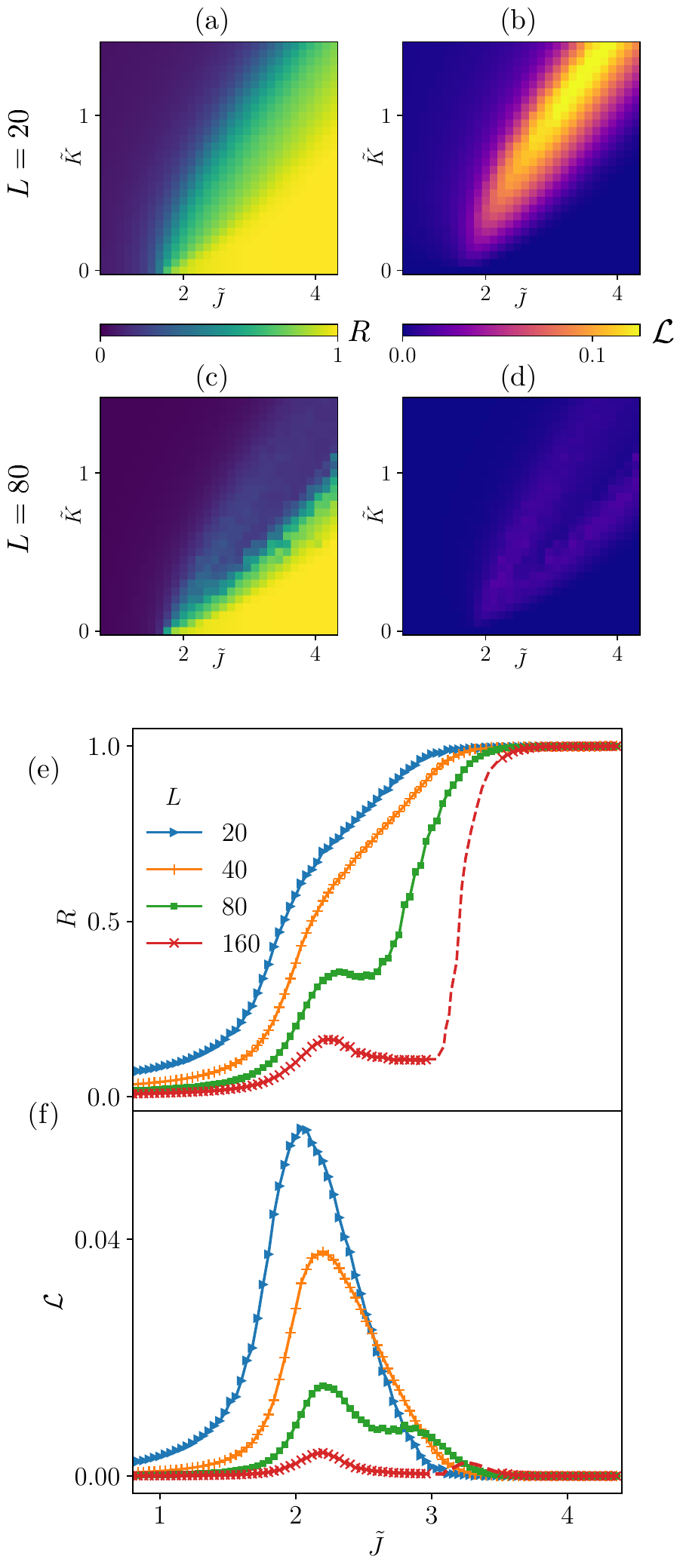}}
\caption{\textbf{Instability of the swap phase in 2D as seen by the order parameters.} Color map of the (a) synchronization order parameter $R$ and (b) phase-space angular momentum $\mathcal{L}$, as a function of $\tilde{J}$ and $\tilde{K}$, for a 2D system with $L=20$. (c-d) The same with $L=80$. (e) $R$ and (f) $\mathcal{L}$, shown as a function of $\tilde{J}$ for $\tilde{K}=0.3$, for different linear system size $L$. Dashed red line shows upper bound for $R$ and $\mathcal{L}$ for points that did not converge during simulation running time.
Obtained from MC simulations.}
\label{R_and_L_2D}
\end{figure}

Quantitatively, the absence of a swap phase in the thermodynamic limit can be seen from the behavior of the order parameters $R$ and $\mathcal{L}$ (Eq.~(\ref{R_definition}) and~(\ref{Angular_momentum}), respectively) as system size increases. In Fig.~\ref{R_and_L_2D}a-d color maps of $R$ and $\mathcal{L}$ are shown as a function of $\tilde{J}$ and $\tilde{K}$ for small ($L=20$) and large ($L=80$) systems. For the $L=20$ system, there is a region in which both $R$ and $\mathcal{L}$ are non-zero, compatible with the swap phase. However, for the $L=80$ system, both $R$ and $\mathcal{L}$ diminish in this region, indicating instead a disorder phase. 
Figures~\ref{R_and_L_2D}e and~\ref{R_and_L_2D}f show the order parameters as function of $\tilde{J}$ for fixed $\tilde{K}$, at different system sizes.
The intermediate region ($1.5\lesssim\tilde{J}\lesssim 3$) corresponds to the finite-system swap state, but both $R$ and $\mathcal{L}$ approach zero as $L$ increases. In addition, the putative critical $\tilde{J}$ marking the transition from disorder to swap depends on the system size, signifying the absence of a well-defined phase transition in the thermodynamic limit.

To conclude, our numerical simulations indicate that the mean-field swap phase is destroyed by spiral defects in 2D where it is replaced by a disordered state.

\section{Existence of a stable swap phase in 3D} \label{Sec_3D}
\begin{figure}[ht]
\centering
{\includegraphics[width=0.45\textwidth,draft=false]{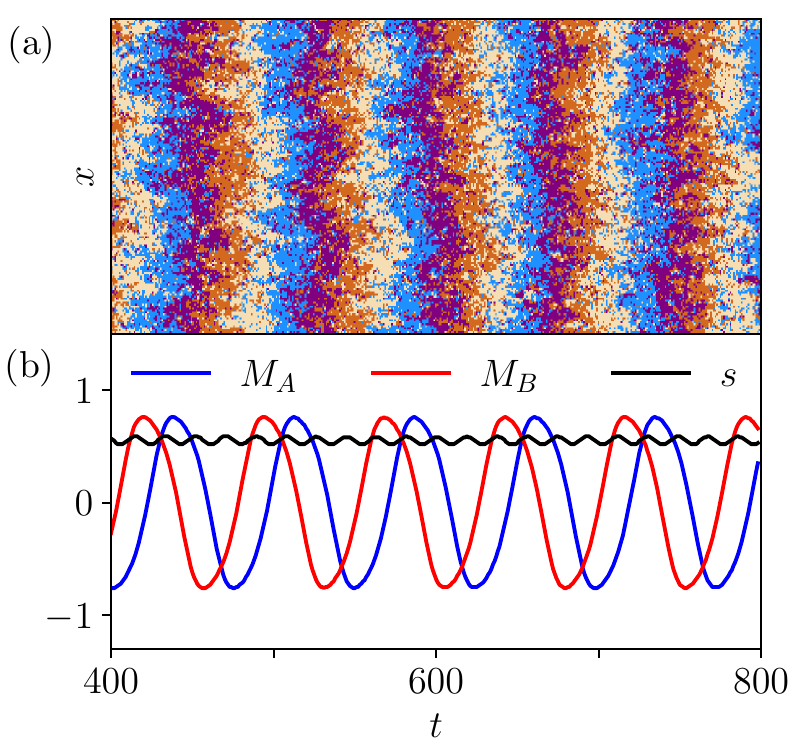}}
\caption{
\textbf{Time evolution of a stable swap state in 3D.} (a) Kymograph of $\theta$ for a single 1D row in the 3D system. (b) $M_A$, $M_B$ and $s=\sqrt{\left(M_A^2+M_B^2\right)/2}$ as a function of time.  Obtained from MC simulation with parameters: $\tilde{J} = 1.5$, $\tilde{K} = 0.1$ and $L=160$.}
\label{kymograph}
\end{figure}

In 3D, numerical simulations  suggest that a stable swap phase does exist. 
Its behavior is demonstrated in Fig.~\ref{kymograph} where we show a kymograph of the $\theta$-field of a single 1D row in the 3D lattice as well as a plot of the total magnetizations and $s$ (Eq.~(\ref{R_definition})) as a function of time. A full visualization of this phase is shown in Movie 2. 
Unlike the 2D droplet-induced swap persisting at finite systems (Movie 1), here the oscillations are noisy and spatially homogeneous, without any apparent structures.
Moreover, $M_A(t)$ and $M_B(t)$ display coherent oscillations with a fixed period and phase shift between the species. 
 
\begin{figure}[ht]
\centering
{\includegraphics[width=0.4\textwidth,draft=false]{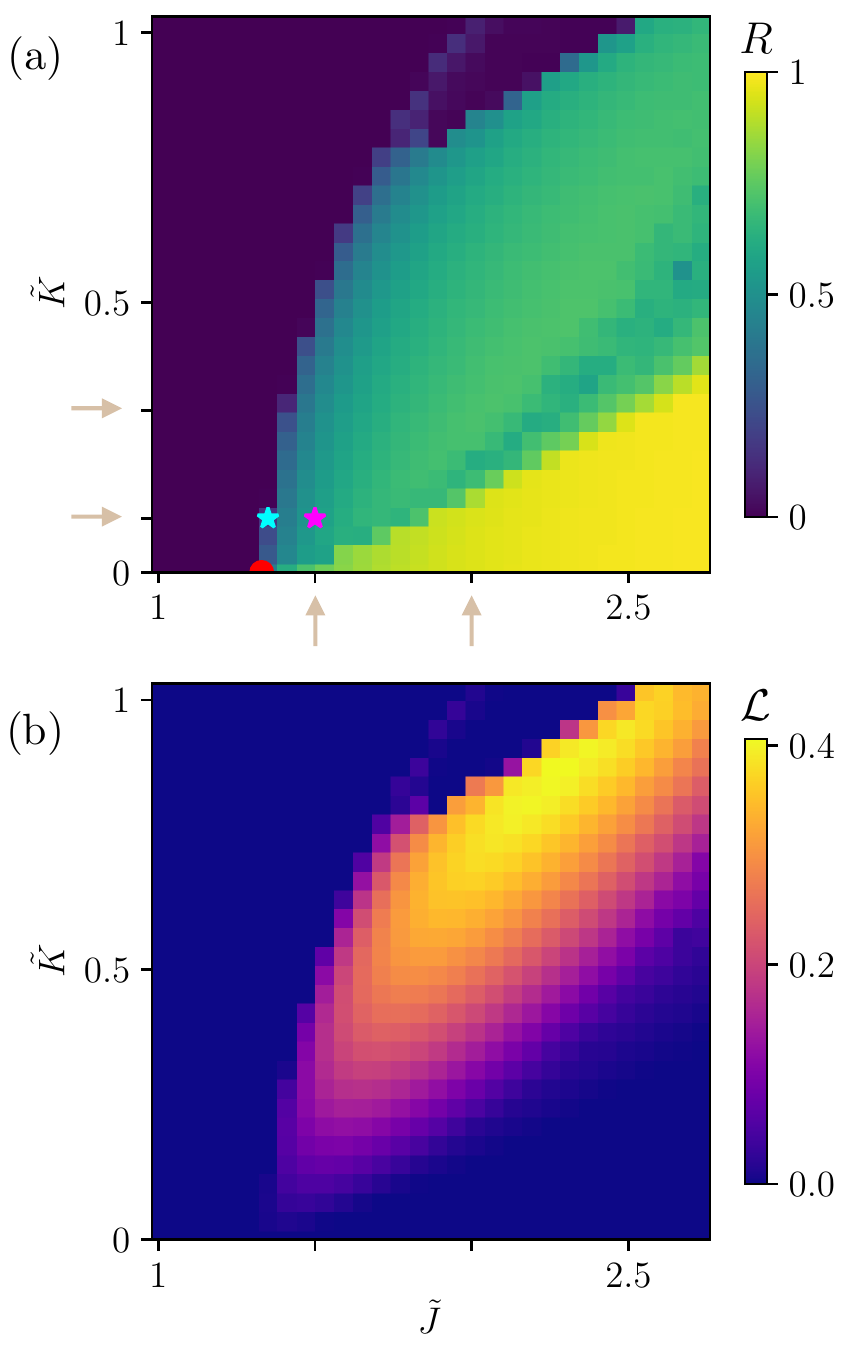}}
\caption{\textbf{Order parameters in a large 3D system.} Color map of (a) $R$ and (b) $\mathcal{L}$ as a function of $\tilde{J}$ and $\tilde{K}$ for a 3D system with $L=320$. In (a), red dot marks the 3D Ising phase-transition point ($\tilde{J}_c\approx 1.33$ and $\tilde{K}=0$), arrows corresponds to fixed $\tilde{K}$ and fixed $\tilde{J}$ cuts analyzed in Figs.~\ref{R_vs_J_3D},~\ref{FSS_fig} and~\ref{R_vs_K_3D_2}  and stars depict points analyzed in Fig.~\ref{trajectory}. Obtained from MC simulations.}
\label{Phase_diagram}
\end{figure}

\begin{figure}[htbp!]
\centering
{\includegraphics[width=0.46\textwidth,draft=false]{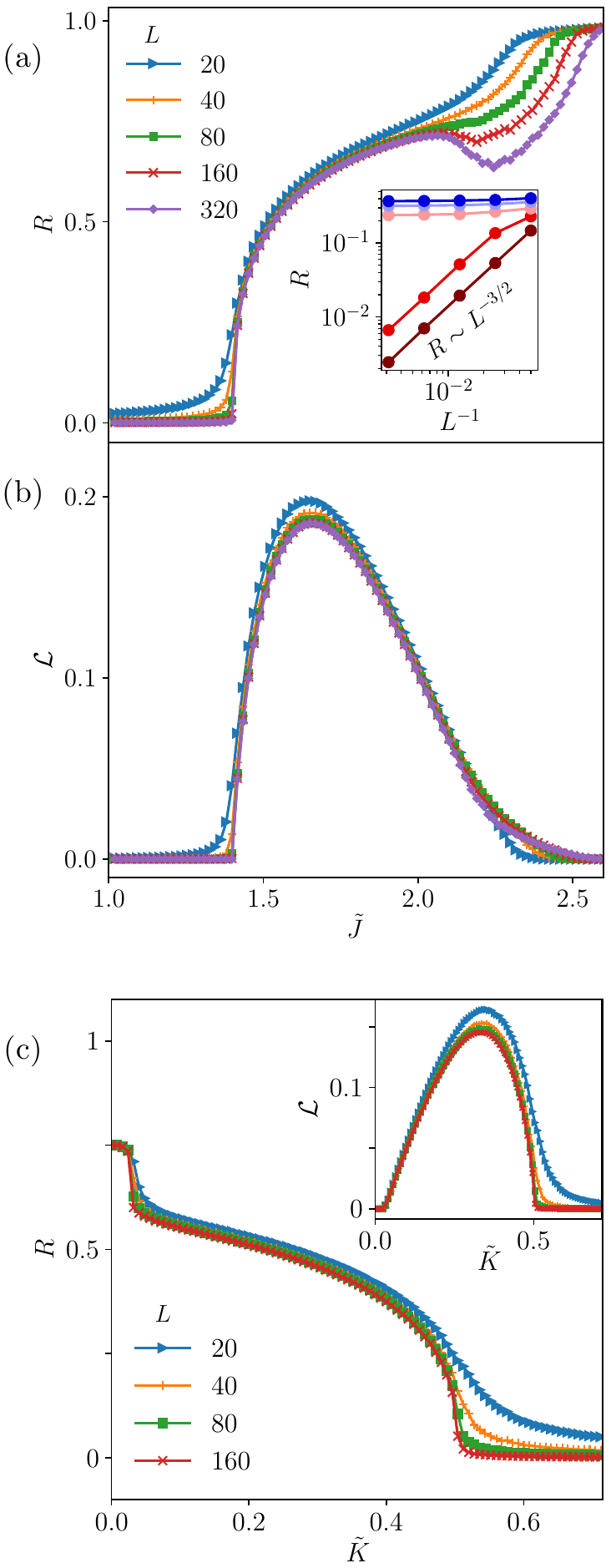}}
\caption{\textbf{Phase transition from disorder to swap in 3D.}
(a) $R$ as a function of $\tilde{J}$ for $\tilde{K} = 0.3$ shown for different linear system size $L$. 
Inset: $R$ vs. $L^{-1}$ for $\tilde{K}=0.3$ and $\tilde{J}=1.38, 1.40, 1.42,  1.43, 1.45$ from bottom to top on a log-log scale. (b) $\mathcal{L}$ as a function of $\tilde{J}$ with the same parameters as panel a. (c) $R$ vs. $\tilde{K}$ for $\tilde{J}=1.5$ for different $L$. Phase-space angular momentum is shown in the inset. Obtained from MC simulations.}
\label{R_vs_J_3D}
\end{figure}

Simulations with varying system sizes indicate there is a well-defined phase transition between a disordered phase with $R=\mathcal{L}=0$ to a swap phase with non-zero $R$ and $\mathcal{L}$. 
Figure~\ref{Phase_diagram} shows color maps of $R$ and $\mathcal{L}$ in the $(\tilde{K},\tilde{J})$ space, for linear system size $L=320$, while Fig.~\ref{R_vs_J_3D}a-b shows a cut of the color maps for $\tilde{K}=0.3$ with different $L$ values. The inset of Fig.~\ref{R_vs_J_3D}a shows two different behaviors depending on whether $\tilde{J}$ is above or below a critical value $\tilde{J}_c$, unlike in 2D (Fig.~\ref{R_and_L_2D}).
Below $\tilde{J}_c$, $R$ approaches zero with $R\sim L^{-d/2}$ scaling as $L$ increases: this corresponds to the disordered phase.
Above $\tilde{J}_c$, $R$ converges to a finite non-zero value as $L$ increases: this corresponds to an ordered phase, here the swap phase.
Moreover, the phase transition appears to be second-order in the sense that $R$ and $\mathcal{L}$ change continuously.

The critical $\tilde{J}_c$ separating disorder from swap is $\tilde{K}$ dependent (see Fig.~\ref{Phase_diagram}), in contrast with the mean-field prediction (Fig.~\ref{MF_Phase_diagram}). In the $\tilde{K}\to \infty$ limit, and for any $\tilde{J}$, the swap phase is unstable and both $R$ and $\mathcal{L}$ are zero. This stems from the fact that when $\tilde{K}\gg \tilde{J}$, each spin responds to the local force imposed by the other species and disregards alignment with its neighbors. As a result, each site performs rapid cycles ($\uparrow \uparrow \to \uparrow \downarrow \to \downarrow \downarrow \to \downarrow \uparrow \to \uparrow \uparrow$), but neighboring sites are not synchronized.
A phase transition similar to the one in Fig.~\ref{R_vs_J_3D}a-b (continuous in $R$ and $\mathcal{L}$) is observed for a fixed $\tilde{J}$ cut (see Fig.~\ref{R_vs_J_3D}c in which $\tilde{J}=1.5$). Both $R$ and $\mathcal{L}$ are zero at large $\tilde{K}$ and become finite below some critical $\tilde{K}_c$ ($\tilde{K}\approx0.5$ in Fig.~\ref{R_vs_J_3D}c).

Note that the second-order phase transition between disorder and swap is limited to small $\tilde{J}$ and $\tilde{K}$ values. At high values of $\tilde{J}$ and $\tilde{K}$, the transition appears to be discontinuous.
This is discussed further in Sec.~\ref{Sec_high_K}.


\subsection{Critical exponents} \label{critical_exponents_section}
Having numerically confirmed the existence of a continuous phase transition from disorder to swap in 3D, we move on to determine the related critical exponents, focusing on the order parameter $R=\langle s \rangle$ (averages here and in the following discussion are with respect to time and realizations but for brevity, we remove the $t$ and $\Omega$ subscripts). Note that $\mathcal{L}$ is not expected to yield novel critical exponents because at the Hopf bifurcation the angular velocity in phase space is finite (see Eq.~(\ref{thetat_Hopf}) for mean-field and Fig.~\ref{T_vs_J_3D} for simulation evidence), and so the phase-space angular momentum scales with $\langle s^2\rangle$.

\begin{figure}
\centering
{\includegraphics[width=0.4\textwidth,draft=false]{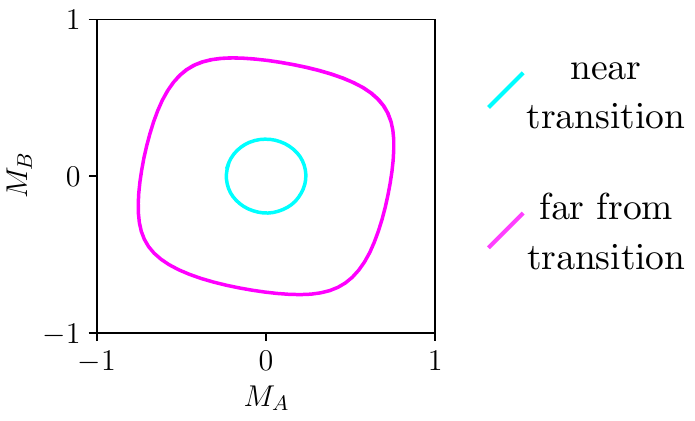}}
\caption{\textbf{Phase-space trajectories close to and far from the transition from disorder to swap in 3D}. $M_B$ as a function of $M_A$ in steady state for $\tilde{J}=1.35$ (cyan) and $\tilde{J}=1.5$ (magenta), with $\tilde{K}=0.1$ and $L=320$ (see points marked by stars in Fig.~\ref{Phase_diagram}). Obtained from MC simulations.}
\label{trajectory}
\end{figure}

 As predicted by the mean-field analysis, the phase-space trajectories become circular when approaching the transition line (Fig.~\ref{trajectory}), supporting a CGL-like behavior as explained in Sec.~\ref{sec_Hopf}. Moreover, renormalization group studies based on the $\epsilon$-expansion suggest that the synchronization transition of the CGL equation with additive noise belongs to the same universality class as the phase transition from disorder to long-range order in the XY model~\cite{risler2004universal,risler2005universal}. Thus, in what follows we will compare our results to the 3D XY critical exponents and to the 3D Ising critical exponents corresponding to the $\tilde{K}=0$ case.

According to the theory of finite-size scaling~\cite{newman1999monte,binder1992monte}, in a finite system the order parameter $R$, the susceptibility $\chi$, defined as
\beq
\chi=L^{d}\left(\langle s^{2}\rangle-\langle s\rangle^{2}\right),
\eeq
and the Binder cumulant~\cite{binder1981finite} defined as
\beq
U = 1-\frac{\langle s^4\rangle }{3\langle s^2\rangle^2},
\eeq
satisfy the following relations in the vicinity of the critical point $\tilde{J}_c$ (assuming fixed $\tilde{K}$) and for large system size:
\beq
\begin{split}
\label{FSS_relations1}
R & = L^{-\beta/\nu}\hat{R}(L^{1/\nu}(\tilde{J}-\tilde{J}_c))\\
\chi & =L^{\gamma/\nu}\hat{\chi}(L^{1/\nu}(\tilde{J}-\tilde{J}_{c}))\\
U & = \hat{U}(L^{1/\nu}(\tilde{J}-\tilde{J}_{c})).
\end{split}
\eeq
Here, $\hat{\chi}(x)$, $\hat{R}(x)$, and $\hat{U}(x)$ are called the scaling functions, and $\nu$, $\gamma$ and $\beta$ are the critical exponents $\xi\sim |\tilde{J}-\tilde{J}_{c}|^{-\nu}$ where $\xi$ is the correlation length, $\chi\sim|\tilde{J}-\tilde{J}_{c}|^{-\gamma}$, and $R\sim|\tilde{J}-\tilde{J}_{c}|^{\beta}$ (for $\tilde{J}>\tilde{J}_{c}$). The scaling functions are independent of $L$ and converge to a finite value at the critical point ({\it i.e.} $\hat{R}(0)$, $\hat{\chi}(0)$ and $\hat{U}(0)$ are constants). In what follows we assume that the finite-size scaling theory holds out of equilibrium, as is often done~\cite{hong2015finite}. We set $\tilde{K}=0.1$ where numerical evidence shows that the transition is continuous (like a second-order transition). 

\begin{figure*}
\centering
{\includegraphics[width=1\textwidth,draft=false]{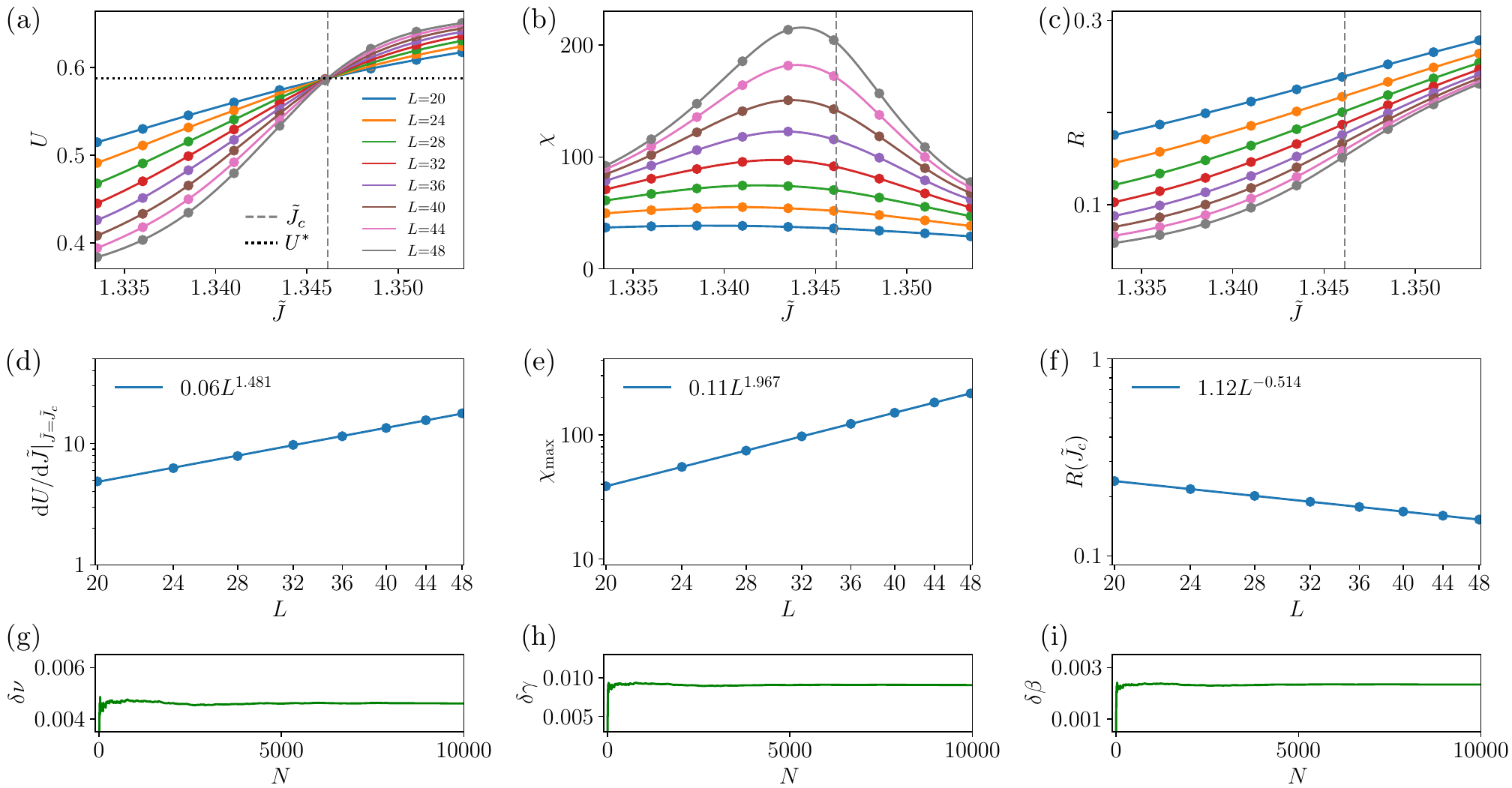}}
\caption{\textbf{Calculation of the critical exponents of the 3D NR Ising model.} We set $\tilde{K}=0.1$. (a-c) The Binder cumulant $U$, the susceptibility $\chi$, and the order parameter $R$ as a function of $\tilde{J}$ for linear system sizes $L=20,24,28,32,36,40,44,48$. The vertical dashed grey line indicates the critical point, $\tilde{J}_c = 1.34614$ while the horizontal dotted black line indicates that value of the Binder cumulant at the critical point, $U^*=0.5877$. Each point is calculated by averaging over $10^6$ samples of $s$. (d-f) ${\rm d}U/{\rm d}{\tilde{J}}|_{\tilde{J}=\tilde{J}_c}$, $\chi_{\rm max}$ and $R(\tilde{J}_c)$ as a function of $L$ on a log-log scale. Dots are extracted from analyzing panels a-c and full lines indicate best linear fits. (g-i) Standard error of $\nu$, $\gamma$, and $\beta$ as a function of the number of bootstrap samples ($10^5$ bootstrap resamplings are performed in total).
}
\label{FSS_fig}
\end{figure*}

The critical point $\tilde{J}_c$ is found using the Binder cumulant. Away from the critical point, the value of $U$ depends on system size, approaching $1/3$ in the disordered and $2/3$ in the ordered regime as $L\to\infty$~\cite{hong2017finite,sarkar2021synchronization}. However, at the critical point $U$ approaches a fixed value $U(\tilde{J}_c)= U^*$ independent of $L$. Hence, the critical point is found from the intersection point of all the $U$ vs. $\tilde{J}$ curves plotted for different system sizes. Figure~\ref{FSS_fig}a shows that indeed all the curves cross at a single point ($\tilde{J}_c = 1.34614\pm 0.00001$ for $\tilde{K} = 0.1$).

\begin{figure}[ht]
\centering
{\includegraphics[width=0.3\textwidth,draft=false]{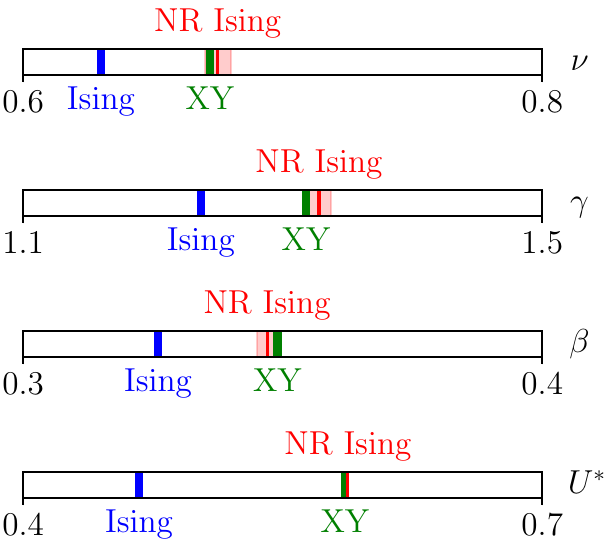}}
\caption{\textbf{Compatibility with XY universality class.}
The critical exponents and the Binder cumulant at the critical point of the 3D nonreciprocal Ising model (red) for $\tilde{K}=0.1$ extracted from finite-size scaling analysis along with their standard deviation represented by a semi-transparent red rectangle, and compared with the 3D Ising model (blue) and the 3D XY model (green).}
\label{comparison}
\end{figure}

In addition to $U$, we plot $\chi$ and $R$ as a function of $\tilde{J}$, see Fig.~\ref{FSS_fig}a-c. The data points are interpolated using a 
cubic spline method with `not-a-knot' boundary conditions as implemented in 
\texttt{scipy.interpolate.CubicSpline}~\cite{2020SciPy-NMeth}. For each linear system size $L$, we find the slope of $U$ at the critical point, ${\rm d}U/{\rm d}\tilde{J}|_{\tilde{J}=\tilde{J}_c}(L)$, the maximum of $\chi$, $\chi_{\rm max}(L)$, and the value of $R$ at the critical point, $R(\tilde{J}=\tilde{J}_c,L)$. We then use the finite-size scaling relations (which can be derived from Eq.~(\ref{FSS_relations1})~\cite{newman1999monte,binder1992monte})
\beq
\begin{split} \label{FSS_relations2}
\frac{{\rm d}U}{{\rm d}\tilde{J}}|_{\tilde{J}=\tilde{J}_c}(L) & =  a L^{1/\nu}\\
\chi_{\rm max}(L) & = b L^{\gamma/\nu}\\
R(\tilde{J}=\tilde{J}_c,L) & = c L^{-\beta/\nu}
\end{split}
\eeq
and perform linear fits by taking the logarithm of both sides of the equations (Fig.~\ref{FSS_fig}(d-f)). The procedure outlined here yields the critical exponents. To ensure good accuracy, all averages are taken using $10^6$ samples.
To make the samples approximately independent, they are separated by two correlation times of $s$ (the correlation times were computed at $\tilde{J}=1.346$). 
Simulation runs are initialized in a fully disordered state. The error is estimated using the bootstrap method~\cite{newman1999monte,moore2007basic}: the same procedure is repeated for multiple samples where each sample is constructed by drawing $10^6$ measurements of $s$ randomly from our data while allowing duplicates. The standard deviation of the critical exponents is estimated by the standard deviation calculated from $10^5$ bootstrap resamplings of the data (Fig.~\ref{FSS_fig}(g-i)). This estimated error is due to finite sampling but does not include other systematic errors ({\it e.g.} finite-size corrections to Eqs.~(\ref{FSS_relations1}--\ref{FSS_relations2})~\cite{binder1992monte}). Our results are
\begin{equation}
\begin{split}
    \nu&=0.675\pm0.005\\
    \gamma&=1.328\pm0.009\\
    \beta&=0.347\pm0.002\,,\\
    \end{split}
\end{equation}
which are all in good agreement with 3D XY critical exponents
 $\nu_{XY}=0.672$, $\gamma_{XY}=1.318$, $\beta_{XY}=0.349$~\cite{pelissetto2002critical,campostrini2001critical}, more so than with 3D Ising critical exponents $\nu_{I}=0.630$, $\gamma_{I}=1.237$, $\beta_{I}=0.326$~\cite{pelissetto2002critical}, see graphical comparison in Fig.~\ref{comparison}. Further evidence that the disorder-to-swap phase transition belongs to the same universality class as the disorder-to-order XY phase transition is given by the value of the Binder cumulant as the critical point, $U({\tilde{J}_c})=U^*$, which is also universal~\cite{binder1981finite}. We find 
 \begin{equation}
 U^*=0.5877\pm0.0002\,,
 \end{equation}
 consistent with studies on the 3D XY model reporting $U^*\approx 0.586$~\cite{hasenbusch1999high}. In contrast, the 3D Ising model yields $U^*\approx 0.467$~\cite{binder2001monte}.

\subsection{Time-crystal behavior}
\label{time_crystal_behavior}
On top of being globally ordered and time-dependent, the 3D swap phase possesses another notable property: its coherence time diverges because of its many-body nature, making it a {\it time crystal}~\cite{Khemani2019,Yao2020,Zaletel2023,winfree1980geometry,Sacha2017,Wilczek2012,Shapere2012,kongkhambut2022observation,wu2024dissipative}.

\begin{figure*}[ht]
\centering
{\includegraphics[width=1\textwidth,draft=false]{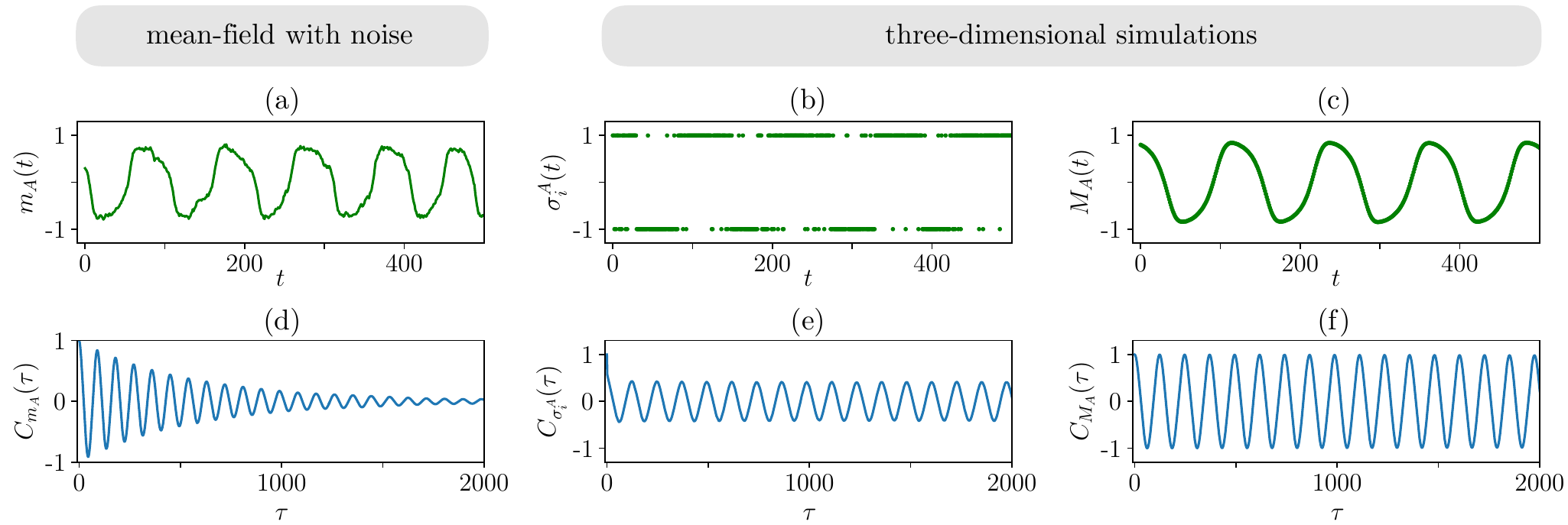}}
\caption{
\textbf{Time correlations in the swap phase of the NR Ising model}.
(a) Time evolution of $m_A$ in the zero-dimensional mean-field equation with added noise. 
(b) Time evolution of a single spin $\sigma^A_i$ in the 3D NR Ising model. (c) Time evolution of the total magnetization of $A$-spins, $M_A$, in the 3D NR Ising model.
(d-f) Auto-correlation function of $m_A$, $\sigma^A_i$, and $M_A$, respectively (defined in Eq.~(\ref{auto_corr})).
Panels a and d are obtained from numerical integration of Eq.~(\ref{0d_with_noise}) with the parameters: $\tilde{J} = 1.2$, $\tilde{K} = 0.12$, $\zeta = 0.0009$, $dt=0.01$. Panels b-c and e-f are obtained from MC simulation with the parameters: $\tilde{J} = 1.6$, $\tilde{K} = 0.1$, $L=80$.
}
\label{time_crystal}
\end{figure*}

To understand this feature, consider a single limit cycle oscillator subject to noise. This can be represented by the zero-dimensional mean-field equation of the nonreciprocal Ising model with added noise, defined as
\begin{equation} \label{0d_with_noise}
    \frac{{\rm d} m_{\alpha}}{{\rm d}t} = -m_\alpha + \tanh(\tilde{J} m_{\alpha} + \tilde{K}\varepsilon_{\alpha \beta}m_\beta ) + \eta_{\alpha}\; ,
\end{equation}
where $\eta_\alpha$ is white noise satisfying $\langle\eta_\alpha(t)\rangle=0$ and $\langle\eta_\alpha\left(t\right)\eta_\beta\left(t'\right)\rangle=\zeta \delta_{\alpha\beta}\delta\left(t-t'\right)$ with $\zeta$ determining the noise amplitude. Figure~\ref{time_crystal} shows a numerical solution of Eq.~(\ref{0d_with_noise}) (panel a) along with the auto-correlation function of $m_A$ (panel d). Such auto-correlation of an observable $x$ is defined here as
\begin{equation} \label{auto_corr}
C_{x}\left(\tau\right)=\frac{\langle x\left(t\right)x\left(t+\tau\right)\rangle_{t}}{\langle x\left(t\right)^{2}\rangle_{t}}.
\end{equation}
Note that $\langle x \rangle_t = 0$. If this was not the case, one should replace $x(t)$ with $x(t) - \langle x \rangle_t$ in Eq.~\eqref{auto_corr}.
As seen in Figure~\ref{time_crystal}, even though the magnetization performs quite regular oscillations, its auto-correlation function decays, thereby indicating that this oscillator eventually forgets its initial phase~\cite{Gaspard2002,Gaspard2002b,Morelli2007,Bagheri2014,delJunco2020,delJunco2020b,Cao2015,Barato2017,Fei2018,Wierenga2018,Nguyen2018,Marsland2019}.

In many-body systems, however, interactions can overcome noise, producing coherent oscillations over arbitrarily long periods~\cite{acebron2005kuramoto,bennett1990stability,grinstein1993temporally,Chate1995,Brunnet1994,Gallas1992,Hemmingsson1993,Grinstein1988,Bohr1987,Binder1992,Lemaitre1996,Chate1991,Chate1992,Gallas1992,Chate1997,Losson1994,Brunnet1994,Grinstein1994,Wendykier2011,Binder1997,wood2006universality,wood2006critical}. Figure~\ref{time_crystal} shows that in the 3D nonreciprocal Ising model, for large system size, temporal correlations of individual spins $\sigma_i^A$ (panel e) and of the total magnetization $M_A$ (panel f) persist and do not decay over large times, making it a robust clock. This robust clock is an emergent feature of our model which does not rely on any periodic drive externally imposing a phase.

The different behaviors of the time correlations in the noisy clock Eq.~(\ref{0d_with_noise}) and in the 3D NR Ising can be respectively compared to spatial density correlations in a liquid (which decay with distance) and a solid (which is maintained over arbitrarily large distances). Such comparison makes Eq.~(\ref{0d_with_noise}) a \enquote{time liquid} and the 3D NR Ising a \enquote{time crystal} which has infinite coherence time. 

\begin{figure}[ht]
\centering
{\includegraphics[width=0.45\textwidth,draft=false]{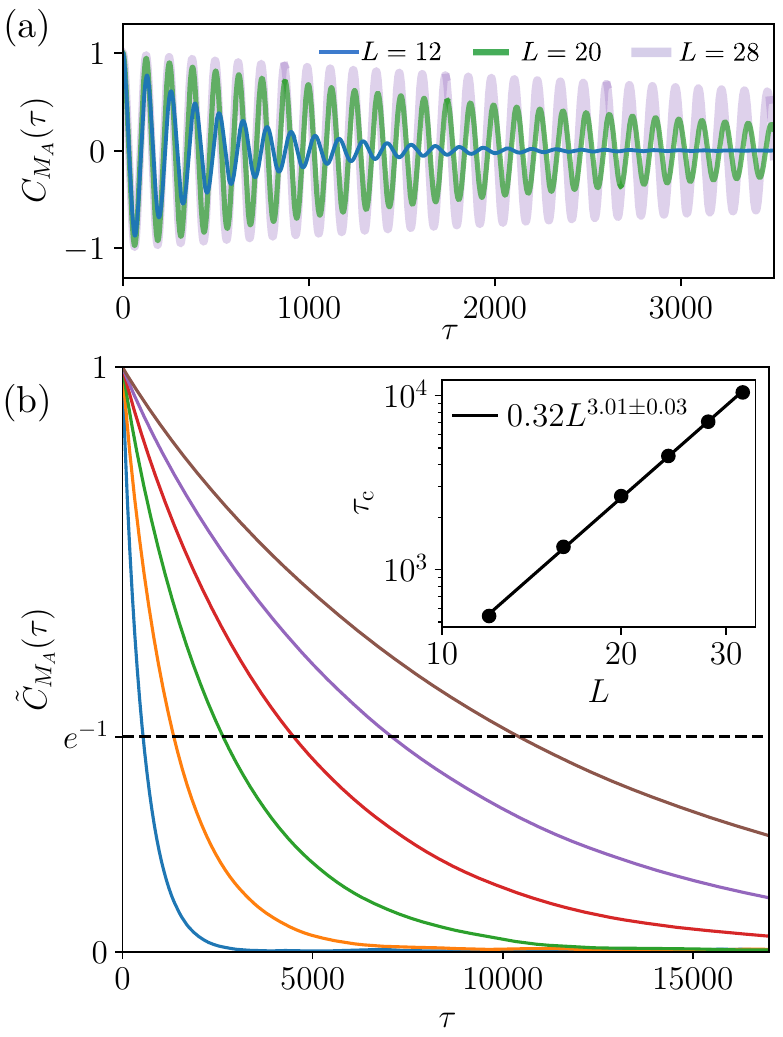}}
\caption{\textbf{Diverging coherence time of the swap phase in the 3D NR Ising model.} (a) The correlation function $C_{M_A}(\tau)$ of the total magnetization of $A$-spins, for different system sizes. Correlations persist over longer times as system size increases. (b) The envelope of the auto-correlation function, $\tilde{C}_{M_A}(\tau)$, shown for linear system sizes $L=12, 16,20,24,28,32$ (from bottom to top). The coherence time $\tau_c$ (Eq.~(\ref{tau_c})) is calculated from the intersection of $\tilde{C}_{M_A}(\tau)$ with $e^{-1}$ (dashed black line). Inset: $\tau_c$ as a function of $L$ on a log-log scale. The dots are calculated from the data in the main panel and the line is the best linear fit, which has a slope of 3.01 with a standard error of 0.03. The coherence time is proportional to $L^d$, supporting a time-crystal behavior. Obtained from MC simulations with the coupling parameters $\tilde{J} = 1.6$ and $\tilde{K} = 0.1$.
}
\label{tau_vs_L}
\end{figure}

To substantiate our claim, we find the coherence time of the swap phase in the 3D NR Ising model by determining the decay rate of the auto-correlation's envelope, $\tilde{C}_{M_A}(\tau)$. We construct $\tilde{C}_{M_A}(\tau)$ by interpolating the peaks of $|C_{M_A}(\tau)|$. We then identify the coherence time $\tau_c$ as the time for which \begin{equation} \label{tau_c}  \tilde{C}_{M_A}\left(\tau_{c}\right)=\exp(-1).
\end{equation} 
Figure~\ref{tau_vs_L} shows that $\tau_c$ increases with system size as 
\begin{equation}
    \label{taucL}
    \tau_c(L) \sim L^3\;,
\end{equation}
{\it i.e.} linearly with the number of spins, thereby proving the perfect time-crystal behavior of the 3D NR Ising in the thermodynamic limit. 
A similar scaling for $\tau_c$ has been observed in a periodically driven 2D system in Ref.~\cite{oberreiter2021stochastic}.

This result suggests the existence of perfectly coherent oscillations in the thermodynamic limit $L \to \infty$, which can be seen as a breaking of ergodicity~\cite{Chen2004,Liggett2012,Martinelli1999}.
Even at arbitrarily long times, the probability distribution function $P(t) \equiv U(t) P_0$ does not necessarily converge to a constant stationary distribution $P^{\text{ss}}$ such that $U(t) P^{\text{ss}} = P^{\text{ss}}$. Here, $U(t)$ is the evolution operator acting on probability distributions, {\it e.g.} defined by the master equation \eqref{MasterEquation}.
Instead, we expect that the system evolves towards a limit-cycle $P_{\text{LC}}$ satisfying $P_{\text{LC}}(t)=P_{\text{LC}}(t+T_{\rm osc})$ where $T_{\rm osc}$ is the oscillation period~\cite{Carollo2022}. 
As a consequence, the distance between the distributions obtained from evolving two initial conditions $P_1$ and $P_2$ does not necessarily go to zero at long times, {\it i.e.} one can have
\begin{equation}
    \lim_{t \to \infty} \lVert U(t) P_1 - U(t) P_2 \rVert \neq 0
    \label{pinfty_nonzero}
\end{equation}
because the many-body oscillator remembers its phase ({\it i.e.} its position on the limit cycle), which is encoded in the probability distribution.

We can combine Eq.~\eqref{taucL} with the conjecture of Ref.~\cite{Oberreiter2022} to obtain a bound on the minimum rate of entropy production required to maintain the oscillations.
This conjecture bounds the average entropy per oscillation $\Delta S$ by the number of coherent oscillations $\mathcal{N} \equiv \tau_c/T_{\rm osc}$ through the inequality $\Delta S \geq 4\pi^2 \mathcal{N}$. 
As a consequence, the rate of entropy production per unit volume
\begin{equation}
\dot{s} \equiv \frac{\Delta S}{L^d T_{\rm osc}}    
\end{equation}
must satisfy
\begin{equation} \label{bound}
    \dot{s} \geq \frac{4 \pi^2}{T_{\rm osc}^2} \, \frac{\tau_c(L)}{L^d}.
\end{equation}
In the stable swap phase $T_{\rm osc}$ does not depend on $L$ (it converges to a constant as $L \to \infty$, see Fig.~\ref{T_vs_J_3D}a in Sec.~\ref{droplet_regime}).
As $\tau_c(L) \sim L^d$, this bound converges to a finite value when $L \to \infty$, meaning that there is a finite minimal cost per oscillator to maintain the time-crystal state.
The fit in the inset of Fig.~\ref{tau_vs_L}b provides $\tau_c/L^d$ for the example considered in the plot. In Appendix~\ref{EntProd} we calculate the entropy production explicitly from simulations under the assumption of local detailed balance, and show that $\dot{s}$ converges to a finite value and that this value indeed exceeds the lower bound provided by Eq.~(\ref{bound}).

Note that ordered oscillations and infinite coherence time are not decoupled properties. It is plausible that a locally coupled system can maintain spatially ordered oscillations over arbitrarily large distances only if all its parts adhere to the same prescribed phase determined by the initial conditions.

\subsection{The phase transition from disorder to swap in the high $(\tilde{J},\tilde{K})$ regime} \label{Sec_high_K}

\begin{figure*}
\centering
{\includegraphics[width=1\textwidth,draft=false]{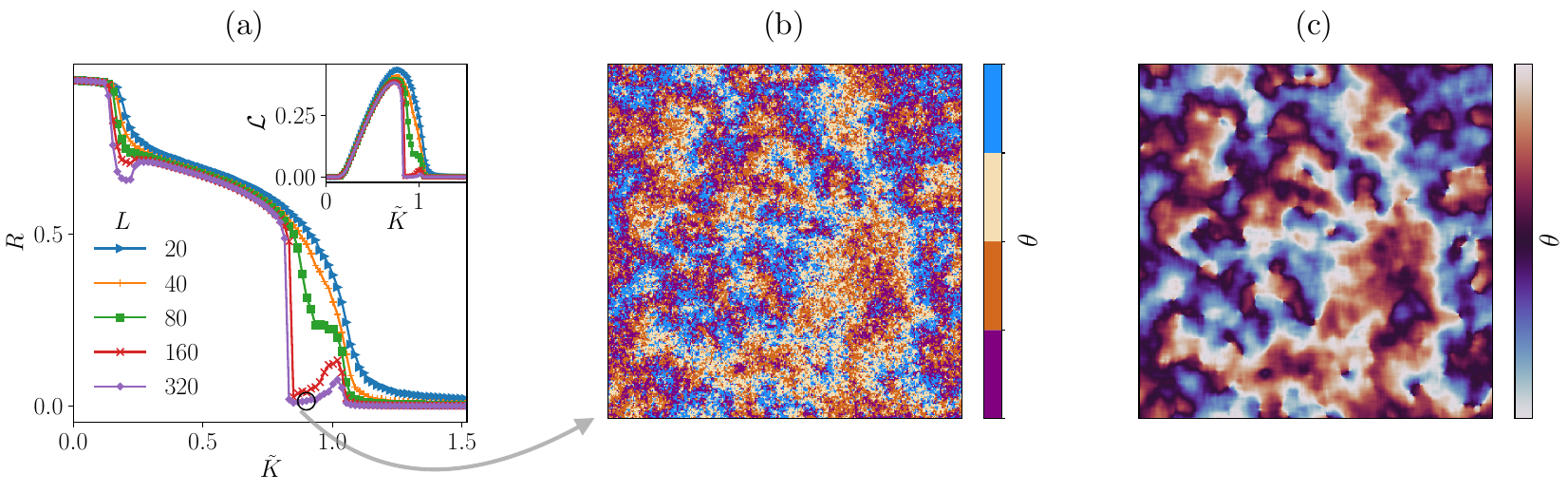}}
\caption{\textbf{Destruction of the second-order phase transition due to scroll waves in the high $\tilde{K}$ and high $\tilde{J}$ regime.} (a) $R$ vs. $\tilde{K}$ for $\tilde{J}=2$. Phase-space angular momentum is shown in the inset. (b) $\theta$ snapshot of a 2D slice from a 3D simulation with the parameters $\tilde{J} = 2$, $\tilde{K} = 0.9$, $L=320$, corresponding to the point marked by a black hollow circle on panel a. (c) The same snapshot coarse-grained over $10\times10$ lattice sites. The snapshots show the destruction of order by scroll waves (appearing as spiral defects in the 2D cut). Obtained from MC simulations.}
\label{R_vs_K_3D_2}
\end{figure*}

At high $\tilde{J}$ and $\tilde{K}$ values, the transition from disorder to swap is different than for small $\tilde{J}$ and $\tilde{K}$ values. This is seen in Fig.~\ref{Phase_diagram} where the change in $R$ and $\mathcal{L}$ at the phase transition appears to become more abrupt in the upper right region, compared to the lower left region.
We further analyze this transition in Fig.~\ref{R_vs_K_3D_2}. Figure~\ref{R_vs_K_3D_2}a shows the order parameters as a function of $\tilde{K}$, for fixed $\tilde{J}=2$, and various system sizes. The plots of $R$ and $\mathcal{L}$ vs. $\tilde{K}$ do not seem to converge to a second-order phase transition in which the order parameters are continuous. Instead, an abrupt change in $R$ and $\mathcal{L}$ emerges at high system sizes, suggesting a discontinuity as in first-order phase transitions. A clear critical point cannot be identified since convergence is not reached at $L=320$, the largest linear system size in our simulations (see $0.8\lesssim\tilde{K}\lesssim1.1$ region in Fig.~\ref{R_vs_K_3D_2}a).

The sharp decrease in synchronization is caused by the emergence of highly noisy scroll waves that destructs the order. This is seen in Fig.~\ref{R_vs_K_3D_2}b showing a 2D-slice snapshot of $\theta$ taken from the 3D simulation. Scroll waves are the 3D analog of spiral waves, having line defects that are typically closed into rings instead of point defects~\cite{winfree1980geometry,siegert1992three,aranson2002world,gabbay1997motion} 
(see also Movie~3 and Fig.~\ref{random_init} in Sec.~\ref{coarsening} in which the scroll waves are more visible).
They also have spiral cross sections, which is seen in the 2D-cut $\theta$ snapshot in Fig.~\ref{R_vs_K_3D_2}b and in its coarse-grained version, Fig.~\ref{R_vs_K_3D_2}c (the latter figure makes the defects more apparent by averaging out the noise). We discuss scroll waves in more detail in Sec.~\ref{coarsening}.

The destruction of the swap phase by scroll waves is slightly reminiscent of the destruction of order in 2D due to spiral defects (compare Fig.~\ref{R_and_L_2D}e-f with Fig.~\ref{R_vs_K_3D_2}a), except that here, it appears that the destruction only occurs for a limited range of high $\tilde{J}$ and $\tilde{K}$ values. Due to the size limitation of our simulations we cannot determine what portion of the 3D swap phase is destructed by the scroll waves or what kind of phase transition is associated with this destruction, although it appears to be first-order-like (in the sense that the order parameters are discontinuous).

\subsection{Why is the swap phase stable in 3D?}
%
\begin{figure*}
\centering
{\includegraphics[width=1\textwidth,draft=false]{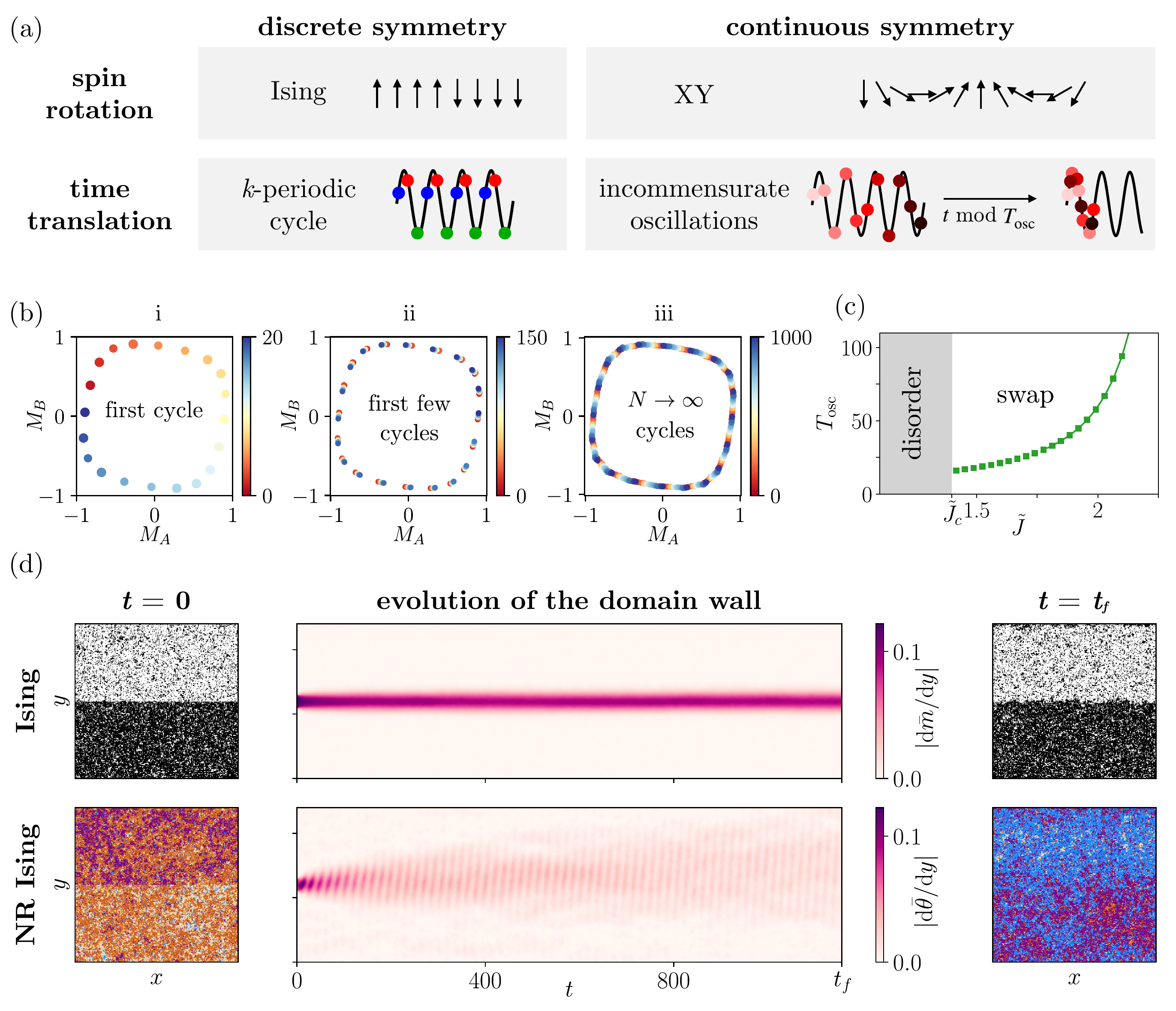}}
\caption{
\textbf{Continuous symmetry breaking of the swap phase.}
(a) Comparison of discrete and continuous symmetry breaking in equilibrium and out-of-equilibrium models. 
(b) $M_B$ vs. $M_A$ for $\tilde{K}=0.5$, $\tilde{J}=2$, and $L=80$. i,ii, and iii show increasing numbers of MC sweeps. Each point is colored according to the sweep it corresponds to. Since the oscillation period is incommensurate, a continuous limit cycle is sampled in the infinite time limit. Hence, a continuous symmetry is broken.
(c) Oscillation period of the magnetizations, $T_{\rm osc}$, in the 3D system as a function of $\tilde{J}$ for $\tilde{K}=0.3$ and $L=80$. $T_{\rm osc}$ is measured as twice the average time between two subsequent sign flips of $M_A$. The continuous dependence of $T_{\rm osc}$ on the coupling parameter indicates incommensurate oscillations. (d) comparison between domain walls in 3D the Ising model (first row) and in the 3D NR Ising model (second row). At $t=0$, the two systems were in steady state and then perturbed by a domain wall. The domain wall is created by flipping the magnetization (Ising) and shifting the oscillation phase (NR Ising) for all spins located at $y>L/2$. System parameters are: $\tilde{J} = 1.4$ (Ising) and $\tilde{J} = 1.5$, $\tilde{K} = 0.1$ (NR Ising), and $L=320$. i and iii are Snapshots of $\sigma$ (Ising) and $\theta$ (NR Ising, see color code in Fig.~\ref{theta}) in a 2D slice taken from a 3D simulation at $t=0$ and $t=1157$, respectively. ii Kymograph of $|{\rm d} {\bar{m}}/ {\rm d}{y}|$ (Ising) and $|{\rm d} {\bar{\theta}}/ {\rm d}{y}|$ (NR Ising), defined by the discrete symmetric derivative ${\rm d}f/{\rm d}y\equiv\left|\left(f_{y+5}-f_{y-5}\right)/10\right|
$, where $\bar{m}$ is the magnetization and $\bar{\theta}$ is the phase, both averaged over the $xz$ plane. The smearing out of the dark pink region indicates the dissipation of the domain wall. Panels b-d are obtained from MC simulations.
} \label{wall}
\end{figure*}
Our numerical evidence points to the existence of a stable swap phase.
This relates to a longstanding question: can time-periodic states of noisy many-body systems with short-range interactions be stable in the thermodynamic limit~\cite{bennett1990stability,grinstein1993temporally,acebron2005kuramoto,Chate1995,Brunnet1994,Gallas1992,Hemmingsson1993,Grinstein1988,Bohr1987,Binder1992,Lemaitre1996,Chate1991,Chate1992,Gallas1992,Chate1997,Losson1994,Brunnet1994,Grinstein1994,Wendykier2011,Binder1997}. 
References~\cite{bennett1990stability,grinstein1993temporally,Grinstein1994} distinguished two cases, depending on whether the spontaneously broken time-translation invariance is discrete or continuous (Fig.~\ref{wall}a). 
In systems that evolve in discrete time as our Monte-Carlo simulation, this depends on the type of oscillations: commensurate vs. incommensurate.
When the period of the oscillations is {\it commensurate}, i.e., a multiple of the discrete time-unite (a case referred to as ``periodic $k$-cycle”), a discrete time-translation symmetry is broken. In that case, it was argued in~\cite{bennett1990stability} that the growth of out-of-phase droplets \footnote{Note that these potential droplets are {\it not} the droplets that destabilize the static-order phase (as in Fig.~\ref{Droplets_change_L}). Instead, these are droplets of a homogeneously oscillating state, immersed in an oscillatory background with a different phase} destabilizes long-range ordered oscillations. That is because flat domain walls are expected to have a non-vanishing velocity, unlike in equilibrium~\cite{coullet1990breaking}, suggesting that a fluctuation producing a large enough droplet, can expand rather than shrink. 
A similar destabilization mechanism has been observed in discrete-symmetry flocks~\cite{benvegnen2023metastability}.

In the {\it incommensurate} case, {\it i.e.} when the oscillation period is not a multiple of the discrete time-unite ({\it e.g.} the time step of the Monte-Carlo simulation), the discrete-time system is equivalent to continuous-time one (the discreteness is irrelevant). The oscillatory state then breaks a continuous time-translation symmetry, which can be seen by the fact that the oscillation’s phase at a given time can take any value between $0$ and $2\pi$. In continuous symmetry breaking, domain walls smear out (as in the XY model) and thus the destabilization-by-droplets argument does not apply~\cite{kardar2007statistical,grinstein1993temporally,Grinstein1994}.

In the nonreciprocal Ising model studied here, the oscillation period is incommensurate with the Monte-Carlo time step. This is seen both in Fig.~\ref{wall}b where a continuous limit cycle is formed in the infinite-time limit, and in Fig.~\ref{wall}c where the period of the oscillation, measured as twice the average time between two subsequent sign flips of $M_A$, is shown to change continuously with the interaction parameters (a continuously changing number is in general incommensurate with $1$). This means that, while stable oscillations are not guaranteed, they are not ruled out by the droplet argument in Ref.~\cite{bennett1990stability}, see also Ref.~\cite{Binder1997}. Figure~\ref{wall}d further exemplifies this: it shows an analysis of 3D simulations of both the Ising model and the nonreciprocal Ising model with a flat domain wall introduced at location $y=0$ at time $t=0$  (the domain wall separates oppositely magnetizaed states for Ising, and two swap states at different phases for nonreciprocal Ising). As time evolves, the domain wall in the nonreciprocal Ising system is shown to widen and dissipate (on top of slightly moving in the $+\hat{y}$ direction), unlike the domain wall in the Ising system which remain steady.
This suggests that there are no stable domain wall between oscillating states in the swap phase of the nonreciprocal Ising model. 

The existence of a broken continuous symmetry (time-translation invariance) suggests a parallel with the Mermin-Wagner theorem~\cite{Mermin1966,Hohenberg1967} in the time domain~\cite{chan2015limit,daviet2024nonequilibrium}: continuous changes in the oscillation phase over space may behave as generalized Goldstone modes and their fluctuations preclude order in 2D yet allow for order in 3D and beyond. Note that this discussion disregards any potential phase transition to quasi-long range order in 2D that would resemble the Berezinskii-Kosterlitz-Thouless (BKT) transition in the XY model. While we have not proven the absence of such a phase transition in the nonreciprocal Ising model, we do not expect it to be present. The similarity with the XY model blurs when we depart from the Hopf bifurcation towards the high $\tilde{J}$ regime (where one might have hoped to observe a phase transition to quasi-long range order), since the discrete symmetry of the spins causes a SNIC bifurcation in mean-field and droplet nucleation in simulations (Sec.~\ref{Sec_static}). Even the CGL equation with additive noise, which is rotationally symmetric and hence more related to the XY model than the NR Ising model, was suggested to not have a BKT transition ~\cite{altman2015two,wachtel2016electrodynamic,aranson1998spiral}.

\begin{figure}
\centering
{\includegraphics[width=0.45\textwidth,draft=false]{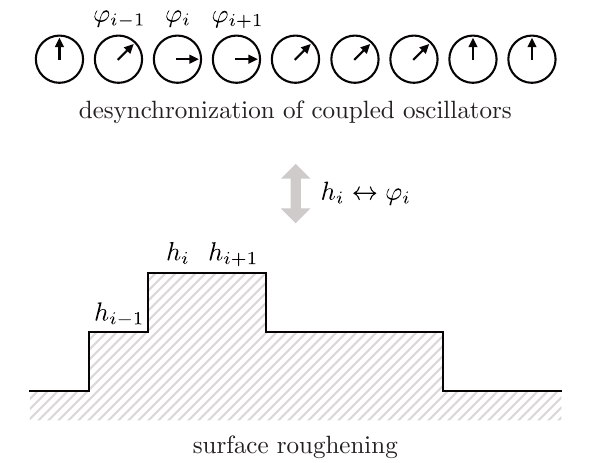}}
\caption{\textbf{Mapping between surface roughening and desynchronisation.}
Schematic drawing of the mapping between coupled limit cycle oscillators (top) and surface roughening described by the KPZ equation (bottom).
The phase $\varphi_i$ of the oscillator at position $i$ is mapped to a height $h_i$.
} \label{KPZ}
\end{figure}

An analogy between (de)synchronization and surface roughening provides further reasons to expect the swap phase to be stable in 3D and not in 2D.
If we add random noise to the mean-field equation~\eqref{MeanFieldExpanded}, it takes the form of a noisy reaction-diffusion system
\beq \label{reac_diff}
\frac{\partial\vec{X}}{\partial t}=F(\vec{X})+D\nabla^{2}\vec{X}+{\rm noise},
\eeq
where $\vec{X}=x_1,..,x_n$ and $D$ can be a matrix. Now assume that the noiseless and diffusion-less version of Eq.~(\ref{reac_diff}) has a limit cycle in steady state. To test whether the addition of diffusion and noise allows for synchronized oscillations, we assume that we are deep within the synchronized regime, where fluctuations from the global limit cycle are small and can be treated perturbatively. 
In this case, we can focus on the phase variable that describes the position of the system on the limit cycle, while neglecting other degrees of freedom. This is known as a phase reduction, see Refs.~\cite{kuramoto1984chemical,Teramae2009,Kuramoto2019,Nakao2015,
Pietras2019}.
Under certain approximations, Eq.~(\ref{reac_diff}) can then be mapped to the Kardar–Parisi–Zhang (KPZ) equation~\cite{Kardar1986}
\beq
\label{kpz}
\frac{\partial\Psi}{\partial t}=\alpha\nabla^{2}\Psi+\beta\left(\nabla\Psi\right)^{2}+{\rm noise}
\eeq
where the deviation $\Psi$ of the local phase with respect to the phase of the deterministic limit cycle plays the role of the surface height in the KPZ equation (Fig.~\ref{KPZ}), see Refs.~\cite{kuramoto1984chemical,grinstein1993temporally,altman2015two,wachtel2016electrodynamic,Gutierrez2023}.
The KPZ equation predicts that interfaces are always rough in 2D, but that there is a phase transition between rough and smooth interfaces in 3D~\cite{kamenev2023field}. 
This suggests, by analogy, that a spatially coherent oscillating phase can exist in 3D, but not in 2D.
Note, however, that the variable $\Psi$ in Eq.~\eqref{kpz} is an angle (defined on a compact space, the circle $S^1$), so the results on the usual KPZ equation cannot be directly applied.
In particular, the fact that the phase $\Psi$ is compact allows for the existence of topological defects, that may change the physics.
We refer to Refs.~\cite{altman2015two,wachtel2016electrodynamic,grinstein1993temporally,Sieberer2016,Zamora2017,Sieberer2018} for further details on the compact KPZ equation.
In 3D, defect rings are expected to be small (at least for certain range of $\tilde{J}$ and $\tilde{K}$) and hence not to completely destroy long-range order~\cite{grinstein1993temporally}.

\section{Destruction of the static-order phase via droplets} \label{Sec_static}

In finite systems, a ferromagnetic-like state with static order is observed both in 2D and 3D. As system size increases, however, our numerical simulations indicate that this static state is destabilized. This is supported by Figs.~\ref{R_and_L_2D}e-f and~\ref{R_vs_J_3D}a-b (right parts) showing $R$ and $\mathcal{L}$ as a function of $\tilde{J}$ (for a fixed $\tilde{K}$). The $\tilde{J}$-transition-point between static order and oscillations in finite systems (indicated by a sharp change in the slope of $R$ and $\mathcal{L}$ becoming zero) increases with system size in a way that does not seem to converge to a finite $\tilde{J}$. Figures~\ref{R_vs_J_3D}c and~\ref{R_vs_K_3D_2}a showing $R$ and $\mathcal{L}$ as a function of $\tilde{K}$ (for a fixed $\tilde{J}$) have a very small region of static order (when $\tilde{K}$ is close to zero) and it is hard to see what is its fate in the thermodynamic limit. However, this region seems to shrink when $L$ increases in Fig~\ref{R_vs_K_3D_2}a, suggesting a destabilization of the static-order phase.

\begin{figure*}
\centering
{\includegraphics[width=0.8\textwidth,draft=false]{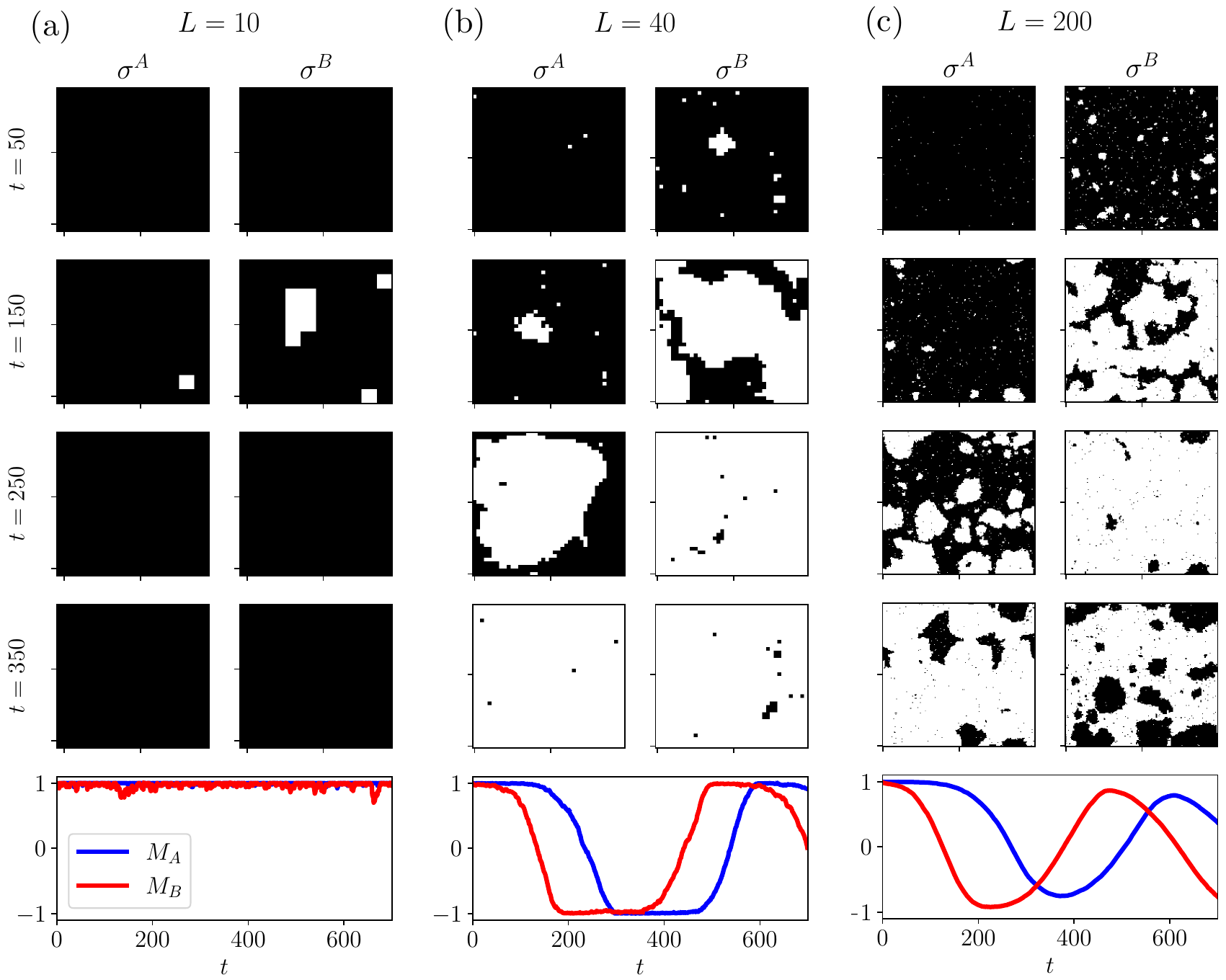}}
\caption{
\textbf{Instability of the static-order phase in the thermodynamic limit due to droplet nucleation}. Top four rows: Snapshots of $\sigma^A$ and $\sigma^B$ in a 2D system with linear system size (a) $L=10$, (b) $L=40$, (c) $L=200$. Time evolves from top to bottom. See Movie 4 for the full evolution. Fifth row: the corresponding total magnetizations, $M_A$ and $M_B$, as a function of time. Obtained from MC simulations with parameters $\tilde{J}=2.8$ and $\tilde{K}=0.3$, and with initialization at $t=0$ with all spins up.
 }
\label{Droplets_change_L}
\end{figure*}

The static-order phase is destabilized by nucleation of droplets that grow and flip the magnetization, in alternating order of A- and B-spins. This mechanism is demonstrated in Fig.~\ref{Droplets_change_L}, where snapshots of the spins at different times are shown for increasing system sizes, in a 2D system. The full evolution is shown in Movie 4. For $L=10$ (Fig.~\ref{Droplets_change_L}a), both species do not flip their magnetization. There are attempts of B-spins to nucleate droplets, but those shrink and disappear. For $L=40$ (Fig.~\ref{Droplets_change_L}b), the system is no longer static: a handful of B-spins droplets (one clearly observed in the figure) surpass a certain critical droplet size, grow and flip the magnetization. Then, the same occurs for the A-spins, and so on, leading to an oscillatory behavior of $M_A(t)$ and $M_B(t)$, but with intermediate periods in which $M_A(t)$ and $M_B(t)$ are constants. When $L=200$ (Fig.~\ref{Droplets_change_L}c), multiple droplets ($\gg1$) drive the magnetization flipping of each species, making the $M_A(t)$ and $M_B(t)$ curves smooth at all times, {\it i.e}, there is no finite period of time in which the magnetization is constant and the system could be called static. We note that the figure shows a 2D system but a similar behavior is observed in 3D.

\begin{figure}
\centering
{\includegraphics[width=0.35\textwidth,draft=false]{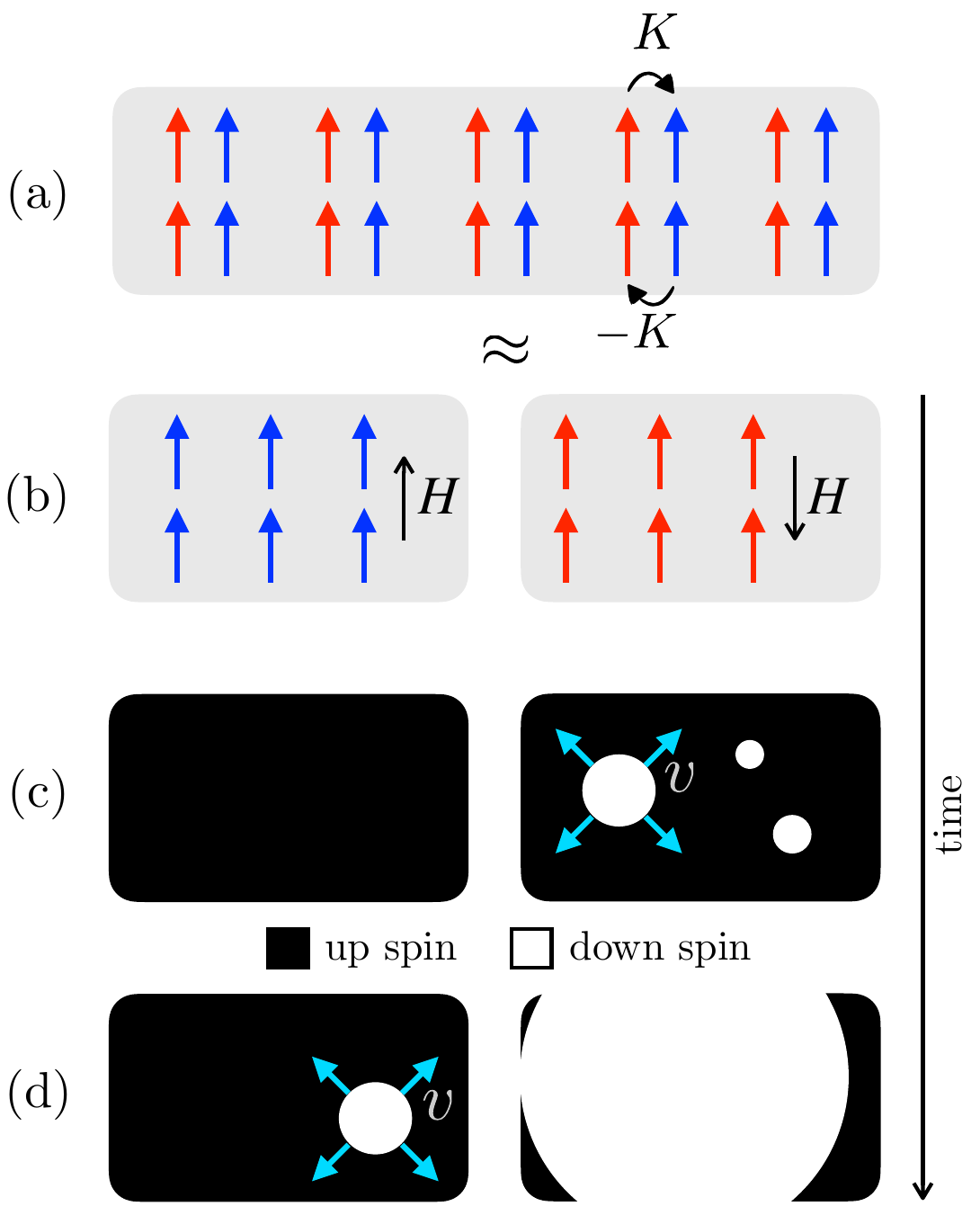}}
\caption{
\textbf{Droplet argument for the instability of the static-order phase}. (a) A system in the static-order phase can be reduced into (b) two equilibrium Ising models subject to magnetic fields, where one system is in a stable equilibrium while the other in a metastable state. (c) The metastable system nucleates droplets that expand and flip the magnetization. (d) The stable system becomes metastable and nucleates droplets that expand with the same velocity, and so on.}
\label{Magnetic_field}
\end{figure}

To see why nonreciprocity allows droplets to grow, in contrast with the equilibrium Ising model where they tend to shrink, assume first that the system is in the static-order phase, so most spins in both lattices are up (Fig.~\ref{Magnetic_field}a)~\cite{assis2011infinite}. We further assume $K$ is very small, which is when static order is most likely to be stable. As long as the system is static, it can be mapped into two equilibrium Ising models with opposite magnetic fields, with energy
\beq
E = -J\sum_{\langle i,j\rangle} \sigma_i \sigma_j - H\sum_{i}\sigma_i
\eeq
where $H\approx K$ for A-spins and $H\approx-K$ for B-spins (see Eq.~(\ref{SelfishEnergy}) and Fig.~\ref{Magnetic_field}b). While sub-system A is in a global free energy minimum in the effective equilibrium system, sub-system B is in a metastable state, as it prefers a state with opposite magnetization. Provided that the system size is large enough, sub-system $B$ then transitions to its global minimum by nucleating droplets larger than a critical size $\rho_c$ above which the effective magnetic field, which acts on the volume of the droplet, overcomes the nearest neighbor attraction, imposing surface tension, which makes the droplet grow~\cite{rikvold1994metastable,assis2011infinite,bennett1990stability,krapivsky2010kinetic} (Fig.~\ref{Magnetic_field}c). After a droplet has expanded beyond $\rho_c$, the stable sub-system becomes metastable and nucleates droplets, and so on (Fig.~\ref{Magnetic_field}d). Crucially, due to the symmetry between $A$ and $B$ spins, both droplet types expand with the same speed. Therefore, a nested droplet remains much smaller than its \enquote{mother-droplet} until the latter reaches the system size or collides with other droplets (see Sec.~\ref{Sec_K_+} in which this symmetry is absent). A simplified view of this mechanism can be found in defining a \enquote{selfish free energy} for the zero-dimensional mean-field equation, see Appendix~\ref{Selfish_free_energy}.

The argument sketched above relies on $\rho_c$ being finite, which is the case in any finite dimension (and finite couplings $\tilde{J}$ and $\tilde{K}$) because the volume term ultimately dominates at sufficiently large droplet sizes. Hence, we conjecture that in the fully anti-symmetric nonreciprocal Ising model, the static-order phase is unstable in the thermodynamic limit in any finite dimension.

\subsection{Between static order and swap: the droplet regime}
\label{droplet_regime}

The droplet mechanism explained above means that as system size increases, a regime of droplet-induced swap develops instead of the static-order phase. In 2D, this regime separates static-order from disorder, while in 3D, it separates static-order from the noisy
homogeneous swap regime.

In the droplet-induced swap regime, the critical droplet size increases as $\tilde{J}$ increases (and as $\tilde{K}$ decreases). As a result, the droplets become more sparse and the distance they travel before colliding with each other increases. Eventually, $\tilde{J}$ becomes large enough (or $\tilde{K}$ small enough) that the critical droplet size becomes larger than the system size.
In that case, the magnetization can flip when fluctuations are of the order of the system size, in the same way as in the usual equilibrium Ising model.
 
\begin{figure*}
\centering
{\includegraphics[width=0.9\textwidth,draft=false]{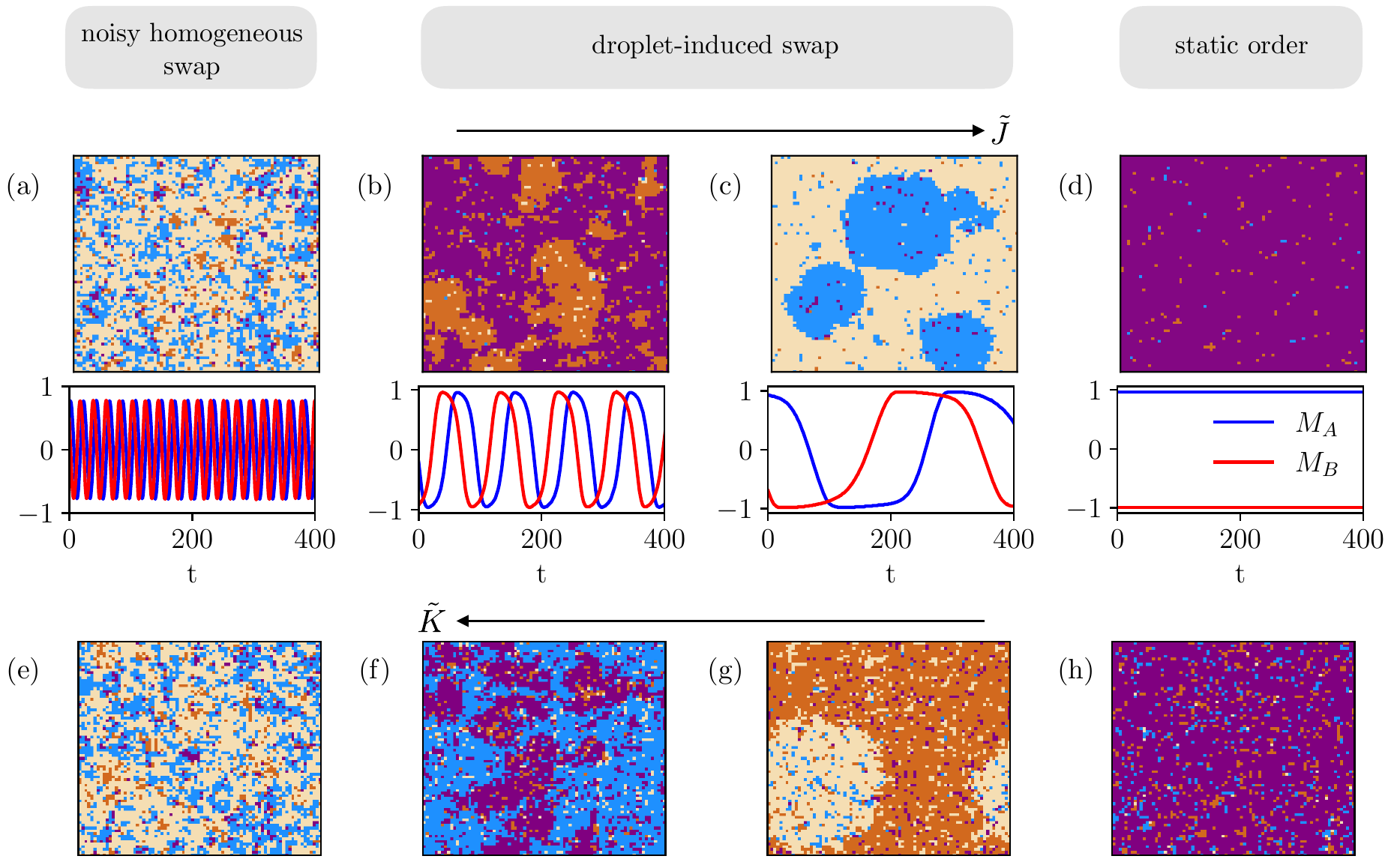}}
\caption{\textbf{Transition from swap to static order in a finite 3D system, as a function of $\tilde{J}$ and $\tilde{K}$}.
(a-d) Top: 2D-slice snapshots of the $\theta$ field, obtained from MC simulations with increasing values of $\tilde{J}$. Bottom: The corresponding total magnetizations $M_A$ (blue) and $M_B$ (red) as a function of time measured in simulation sweeps. (a) $\tilde{J}=1.6$. (b) $\tilde{J}=2.2$. (c) $\tilde{J}=2.3$. (d) $\tilde{J}=2.5$, all with $\tilde{K}=0.3$ and $L = 80$. As $\tilde{J}$ increases, a noisy homogeneous swap phase is replaced by droplet-induced swap and then becomes static and ordered. (e-f) The same as panels a-d (top) but with $\tilde{J}=1.6$ and decreasing values of $\tilde{K}$. (e) $\tilde{K}=0.3$. (f) $\tilde{K}=0.08$. (g) $\tilde{K}=0.05$. (h) $\tilde{K}=0.02$.
} \label{Droplets_change_J}
\end{figure*}

This is demonstrated in Fig.~\ref{Droplets_change_J} where 2D-slice snapshots of the $\theta$ field, as well as the magnetizations of the two species as a function of time, are shown for a 3D system with linear size $L=80$ at increasing values of $\tilde{J}$ (and decreasing values of $\tilde{K}$). For small $\tilde{J}$ (high $\tilde{K}$), the 3D system is in a noisy homogeneous swap phase. There are no apparent droplets but rather noisy patterns without a well-defined shape (Fig.~\ref{Droplets_change_J}a,e). Put differently, independent fluctuations do not have time to expand significantly before colliding. For larger $\tilde{J}$ (smaller $\tilde{K}$) values we observe droplets that drive the magnetization flips (Fig.~\ref{Droplets_change_J}b-c,f-g). For an even larger $\tilde{J}$ (smaller $\tilde{K}$), no droplets are observed and the magnetizations do not change over a long period of time (Fig.~\ref{Droplets_change_J}d,h).

\begin{figure}
\centering
{\includegraphics[width=0.45\textwidth,draft=false]{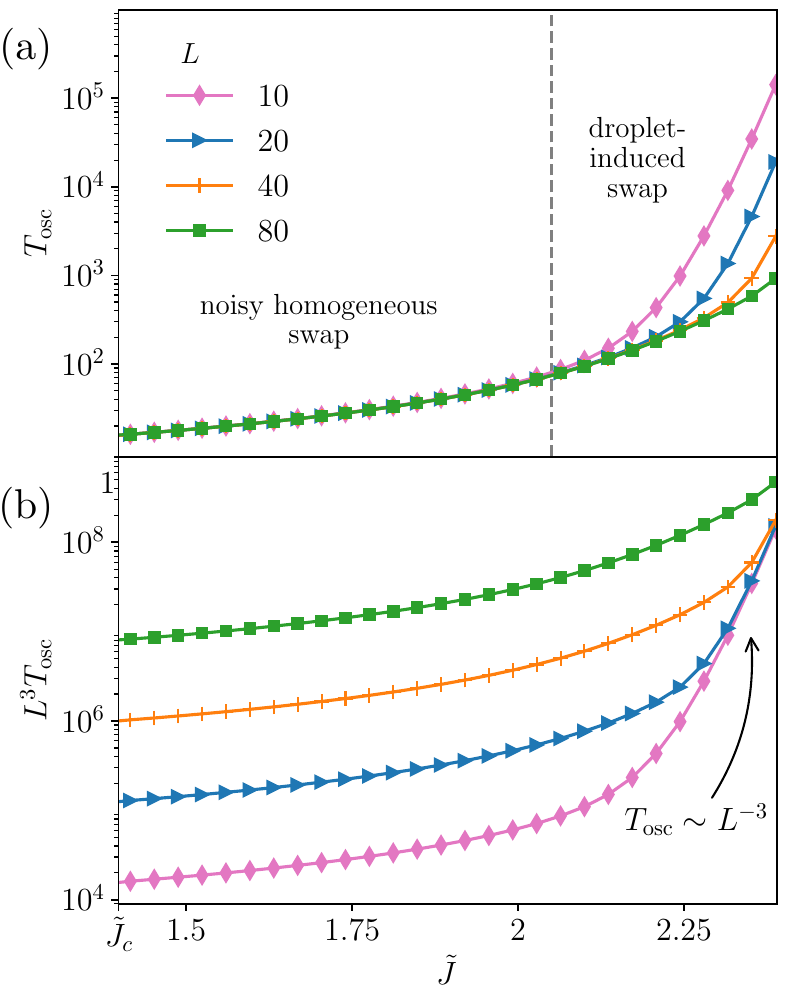}}
\caption{\textbf{Behavior of the oscillation period in the 3D swap phase}. (a) The oscillation period of the total magnetizations $T_{\rm osc}$, on a logarithmic scale, as a function of $\tilde{J}$ for fixed $\tilde{K}=0.3$ and different system sizes. (b) The same with a rescaled oscillation period, $L^3 T_{\rm osc}$. Dashed grey line in panel a separates a regime in which the oscillations are driven by well-defined droplets (right) from 
a regime in which they do not (left). $\tilde{J}_c$ marked on the x-axis is the critical point for the disorder-to-swap phase transition. The period is finite at $\tilde{J}_c$, supporting a Hopf-like behavior. Obtained from MC simulations. 
} \label{T_vs_J_3D}
\end{figure}

The droplet regime is characterized by a distinct behavior of the oscillation period, $T_{\rm osc}$, as a function of system size. In Fig.~\ref{T_vs_J_3D}a, the oscillation period in 3D is shown as a function of $\tilde{J}$ with a fixed $\tilde{K}$, for increasing $L$ values. In the non-droplet regime, {\it i.e.} small $\tilde{J}$, $T_{\rm osc}$ is independent of system size (note that the period is finite at the critical point $\tilde{J}_c$, supporting a Hopf-like behavior), while in the droplet-induced swap regime (large $\tilde{J}$) $T_{\rm osc}$ decreases with system size at intermediate $L$ but converges in the $L\to\infty$ limit to a finite value, independent of $L$ (see overlap of orange and green lines in Fig.~\ref{T_vs_J_3D}a up to $\tilde{J}\approx 2.3$). At larger $\tilde{J}$ values, in a finite system, droplets are so rare that each magnetization flip is driven by a single droplet (here we assume that $\tilde{J}$ is not {\it too} large, such that the critical droplet size is still smaller than the system size). In this regime, the single droplet can emerge in any lattice position at a given time step, thus $T_{\rm osc}$ scales with $L^{-3}$~\cite{rikvold1994metastable,assis2011infinite}, see Fig.~\ref{T_vs_J_3D}b.

The different oscillation regimes share similarities with the relaxation of the magnetization in an Ising model from an initial state magnetized opposite to the applied field $H$.
Reference~\cite{rikvold1994metastable} identifies four regimes, depending on the field strength and system size. The \enquote{strong-field} regime in which \enquote{the droplet picture is inappropriate}, and three other regimes which are classified using droplet theory by the relation between three length scales: the critical droplet radius $\rho_c$, the mean droplet separation $\rho_0$, and the linear system size $L$. These are the multi-droplet regime $L\gg \rho_0 \gg \rho_c$, the single-droplet regime $\rho_0\gg L \gg \rho_c$, and the \enquote{coexistence region} $\rho_0 \gg \rho_c \gg L$
where the system behaves as if there is no applied field.
The average metastable lifetime calculated in Ref.~\cite{rikvold1994metastable} is shown to be independent of system size in strong field and multi-droplet regime yet highly dependent on system size in the single-droplet and coexistence regime, in a way reminiscent of Fig.~\ref{T_vs_J_3D}a. We note that the analogy drawn here is not perfect because nonreciprocal interactions are different from an externally applied field in that they change once the magnetization flips.

\subsection{The fate of the droplet regime} \label{fate_droplet_regime}
We now discuss the fate of the droplet-induced swap regime in the thermodynamic limit. In 2D, this regime is destabilized by spiral defects making it disordered, as was shown in Sec.~\ref{Sec_2D}.

\begin{figure}
\centering
{\includegraphics[width=0.45\textwidth,draft=false]{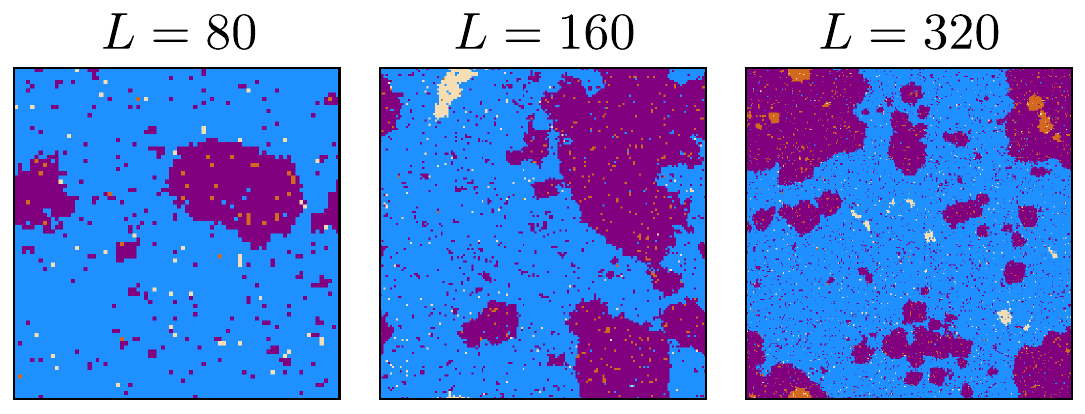}}
\caption{
\textbf{Loss of synchronization in the droplet regime with increased system size}. Snapshots of the $\theta$ field (color code in Fig.~\ref{theta}) in 2D slices taken from a 3D MC simulation with $L=80,160$, and $320$. Other system parameters are $\tilde{K}=0.3$ and $\tilde{J}=2.25$.
We see $n=(2, 3, 4)$ distinct $\theta$-valued regions for $L=(80,160,320)$, respectively.
} \label{loss_sync}
\end{figure}

In 3D, the scenario is more complex. While droplets persist at large system size, distant regions become less synchronized and the system loses its global order. This is illustrated in Fig.~\ref{loss_sync} which shows snapshots of the $\theta$ field, taken from simulations in the droplet-induced swap regime, with different system sizes.
In Fig.~\ref{loss_sync}, we see that at system sizes $L = (80, 160, 320)$ there are regions with $n=(2, 3, 4)$ distinct $\theta$-valued  (i.e. $n$ different colors in the plot), respectively. This decreases the synchronization. Indeed, the synchronization order parameter, $R$, is shown to decrease with $L$, in the right part of Fig.~\ref{R_vs_J_3D}a, corresponding to the droplet-induced swap region. 
While $R$ decreases with system size, it does not converge. Thus, it is unclear to what extent order is lost in the thermodynamic limit, and whether it remains finite or is ultimately deemed disordered ($R=0$). If the latter is correct, it remains an open question what is the type of phase transition that occurs and where lies the critical line between swap and disorder in this droplet regime. 

The decrease in order in the droplet-induced-swap regime causes an intriguing behavior, observed only for large system size, ($L\geq 160$): synchronization is non-monotonic as a function of $\tilde{J}$ (or $\tilde{K}$). See $L=320$ line in Fig.~\ref{R_vs_J_3D}a, where $R$ increases monotonically 
with $\tilde{J}$ beyond the Hopf bifurcation until $\tilde{J}\sim 2.05$, and then starts decreasing. It then reaches a local minimum at $\tilde{J}\sim 2.25$ and increases again until saturation at $\tilde{J}\sim 2.55$. A similar behavior is observed in the left part Fig.~\ref{R_vs_K_3D_2}a, for $L=320$, where $R$ is non-monotonic as a function of $\tilde{K}$. In both cases the anomalous decrease starts when the swap phase transitions from noisy homogeneous swap to droplet-induced swap.

\begin{figure}
\centering
{\includegraphics[width=0.48\textwidth,draft=false]{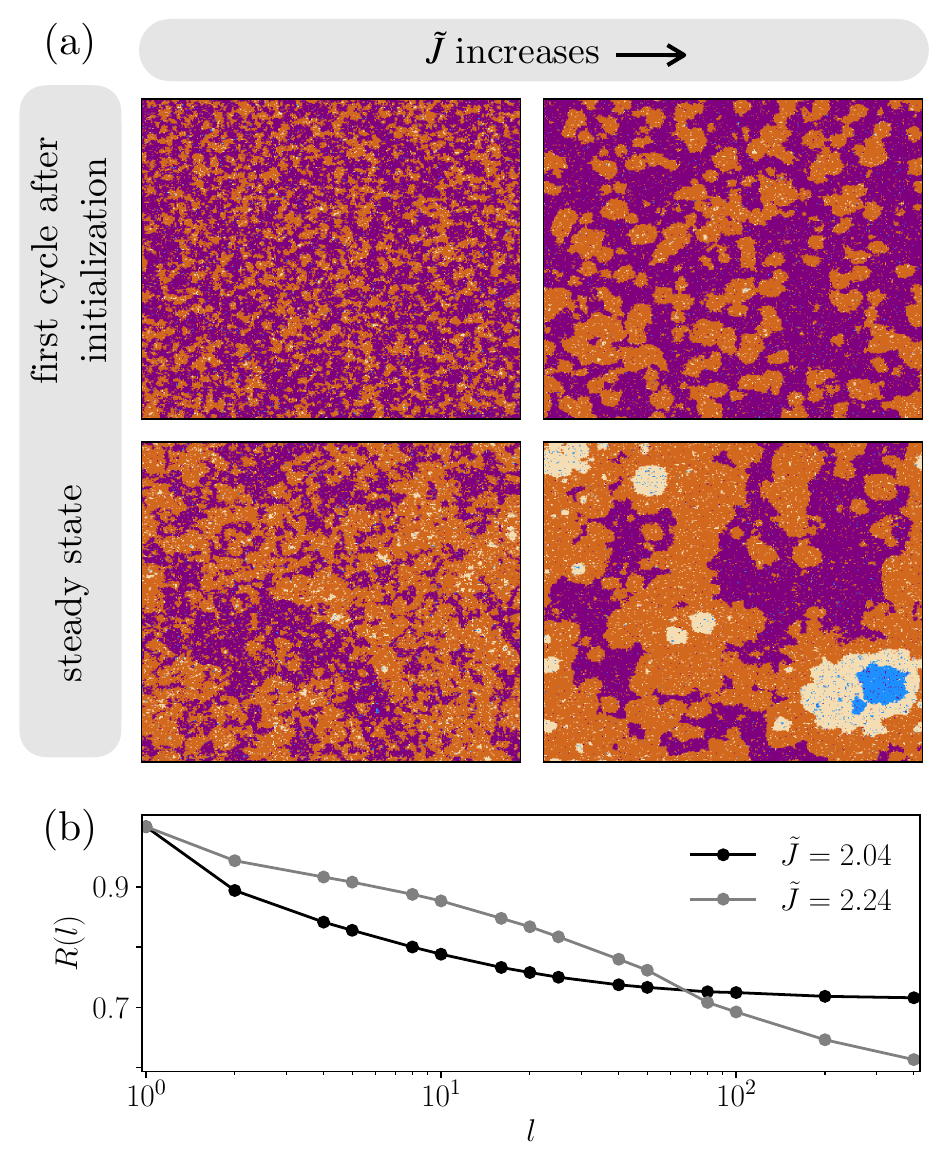}}
\caption{\textbf{Decrease in global order in the droplet regime as a function of $\tilde{J}$.} (a) Snapshots of 2D slices taken from a 3D simulation with $\tilde{K}=0.3$ and $L=400$, comparing two different $\tilde{J}$ values, $\tilde{J}=2.04$ (left) and $\tilde{J}=2.24$ (right). The top snapshots are taken shortly after initialization while the bottom snapshots are taken after 10000 MC sweeps. (b) The synchronization parameter $R(l)$ defined in Eq.~(\ref{R_of_l}), measured in steady state as a function of $l$, for the two $\tilde{J}$ values in panel a. Averages are taken over $(L/l)^2$ squares and 1000 MC sweeps.
} \label{non_mono}
\end{figure}

The mechanism under which global synchronization decreases with increased $\tilde{J}$ (or decreased $\tilde{K}$) in the droplet regime is shown in Fig.~\ref{non_mono}, which compares two simulations with the same $\tilde{K}=0.3$ value: one at the onset of droplet-induced swap ($\tilde{J}=2.04$), and the other well within the droplet-induced swap ($\tilde{J}=2.24$). Simulation snapshots (Figure~\ref{non_mono}a) reveal that while the system is more correlated over short distances at larger $\tilde{J}$ values, it loses {\it global order} due to large regions that are out-of-phase with the rest of the system. This can be quantified by a size-dependent synchronization parameter,
\begin{equation} \label{R_of_l}
    R(l)\equiv \left\langle s(l)\right\rangle
\,\,\,\,\text{with}\,\,\,
s(l) \equiv \frac{1}{l^d}\bigg|\sum_{j}{\rm e}^{i\theta_j}\bigg|
\end{equation}
where the sum is over all sites inside a box of linear size $l \leq L$, and the average is taken with respect to different boxes and time. The size dependent synchronization parameter is shown in Fig.~\ref{non_mono}b, for squares instead of cubes to match with the 2D snapshots. For small $l$, $R(l)$ is larger for the system with the higher $\tilde{J}$, whereas for large $l$, $R(l)$ is larger for the system with the lower $\tilde{J}$. 
Moreover, while $R(l)$ at low $\tilde{J}$ decays exponentially with $l$, in the high $\tilde{J}$ regime it does not and the second derivative of $R(l)$ changes signs. This means that within the droplet-induced swap regime, the system cannot be described by a single correlation length, and instead has multiple length scales related to the droplet parameters such as the critical droplet size, the typical distance between droplets, and the correlation length inside a single droplet.

%

\section{Coarsening dynamics: scroll waves and planar waves} \label{coarsening}

\begin{figure*}
\centering
{\includegraphics[width=1\textwidth,draft=false]{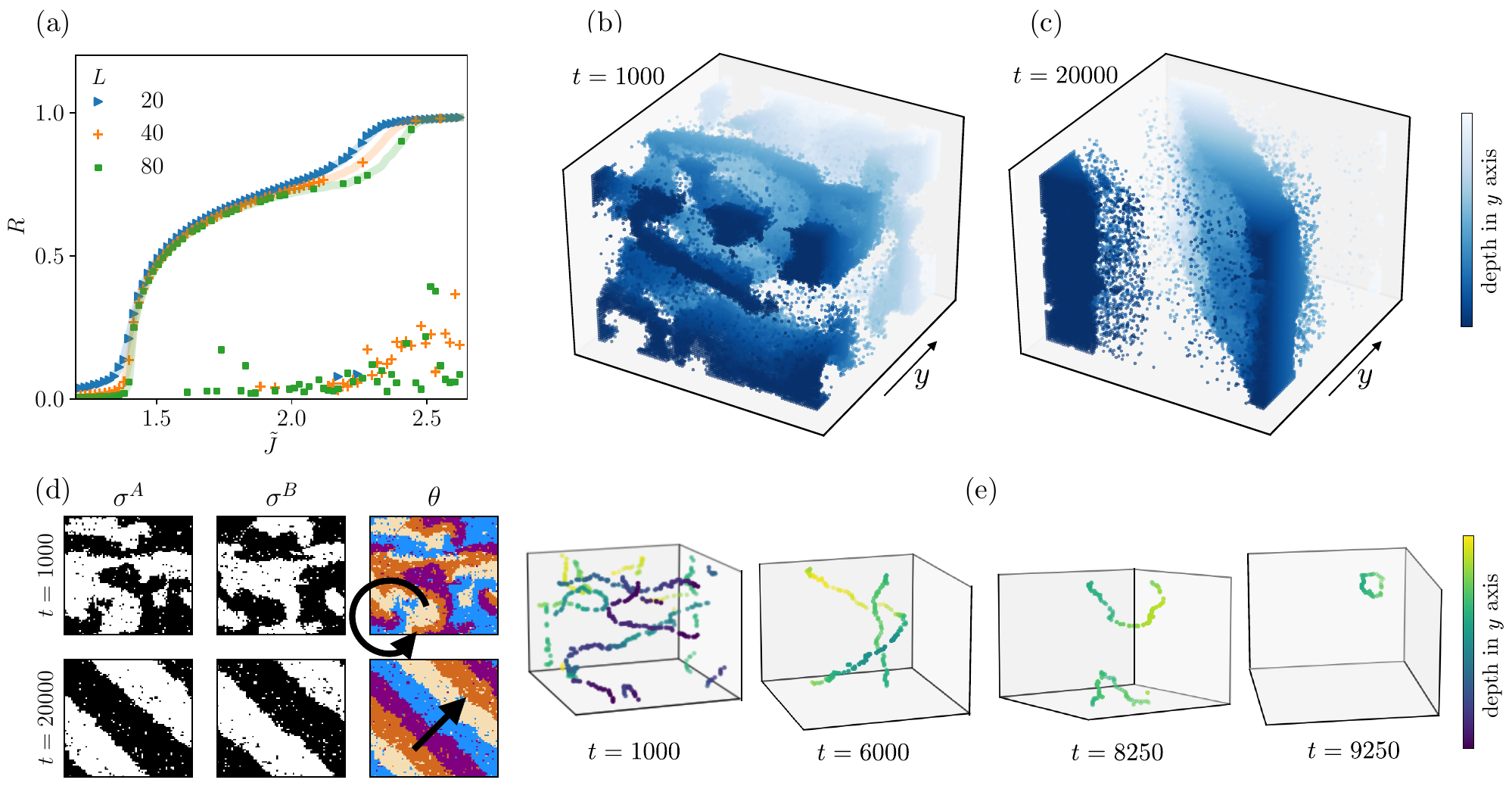}}
\caption{\textbf{
Random initial conditions lead to long-lived scroll waves and planar waves in 3D.} (a) $R$ as a function of $\tilde{J}$ shown for different linear system size $L$, and for $\tilde{K} = 0.3$. Scattered symbols are measured by averaging over the last $10^5$ MC sweeps in a single $10^6$ sweeps long simulation, with random initial conditions for the spins. Semi-transparent lines are the results from Fig.~\ref{R_vs_J_3D}a  in which ordered initial conditions were used and an average was taken over many simulations. 
(b-e) Analysis of a simulation with random initial conditions and system parameters: $\tilde{J} = 2.28$, $\tilde{K} = 0.3$ and $L=80$, with an update scheme in which spins are chosen at random.
(b-c) 3D images at different times in the evolution, showing scroll waves and planar wave, respectively. Sites in $\uparrow \uparrow$ state are marked in blue while the rest of the sites are not shown. (d) 2D-slice snapshots of the magnetization of $A$- and $B$-spins, and of $\theta$, at two different times, showing 2D cuts of the scroll waves (top) and the planar wave (bottom). Black arrows indicate the direction in which the stripes and spiral patterns move over time. (e) Evolution of the defect filaments. As time evolves, the filaments merge, forming rings that shrink and disappear. Note the periodic boundary conditions. Defects are found using the condition $s(l)<s_{\rm max}$ (Eq.~(\ref{R_of_l})) with $l=3$ and $s_{\rm max}=0.2$.
}
\label{random_init}
\end{figure*}

We now qualitatively discuss the way in which the system coarsens into an ordered state in 3D. In the simulations used to calculate the order parameters $R$ and $\mathcal{L}$ (Figs.~\ref{Phase_diagram}, \ref{R_vs_J_3D} and \ref{R_vs_K_3D_2}), the spins were initialized in an ordered initial condition (all spins up), making the transition to steady state in the swap and static-order phases relatively fast.
When fully disordered initial conditions are used (each spin is set to $+1$ or $-1$ randomly), however, the system can coarsen into long-lived metastable states, before reaching steady state. To show this, we present in Fig.~\ref{random_init}a the same plot as Fig.~\ref{R_vs_J_3D}a ($R$ vs. $\tilde{J}$ in 3D) but with random initial conditions, and average taken over a single simulation running for a very long time. In both the swap phase ($1.4\lesssim \tilde{J}\lesssim 2.3$) and static-order phase ($\tilde{J}\gtrsim 2.3$) we see that for some of the realizations the system reaches the same steady state as with ordered initialization (shown in semi-transparent lines). However, sometimes the measured $R$ is much lower than the one obtained with ordered initialization, suggesting that the system got trapped in a metastable state.

The states in which the system gets trapped in are scroll waves and planar waves, shown in Fig.~\ref{random_init}b-d and Movie 3 in both 3D snapshots and 2D-slice snapshots. Here the scroll waves are much less noisy than those discussed in Sec.~\ref{Sec_high_K}. 

The planar waves in the nonreciprocal Ising model are analogous to what happens in the equilibrium Ising model, in which the system often reaches stripes or minimal surfaces in the final stages of the coarsening dynamics~\cite{krapivsky2010kinetic,olejarz2012fate}. Here, however, the stripes move in a particular direction that guarantees, for each site, the $\uparrow \uparrow \to \uparrow \downarrow \to \downarrow \downarrow \to \downarrow \uparrow \to \uparrow \uparrow$ cycle on average. 
A planar wave state breaks both time translation symmetry and space translation symmetry (in one direction).

The coarsening dynamics is most easily understood by following the defect filaments of the scroll waves (Fig.~\ref{random_init}e). The defects are found using the size-dependent synchronization parameter, $s(l)$, defined in Eq.~(\ref{R_of_l}). Note that $s(l)$ is 1 if all the sites in the box are identical, and it is close to zero if the box has a mixture of sites in all four possible states. Thus, we classify defects as sites on the coarse-grained grid in which $s(l)$ is smaller than a threshold $s_{\rm max}$. The box size $l$ and the $s$-threshold are fine-tuned to detect the defect filaments while filtering out noise ($l=3$ and $s_{\rm max}=0.2$ in Fig.~\ref{random_init}e).

In the example shown in Fig.~\ref{random_init}e, the system coarsens into a complex mixture of defect lines shortly after initialization with disordered initial conditions. After a while, only two defect lines survive. The lines are connected onto themselves by virtue of the periodic boundary conditions, having the topology of rings on a 3-dimensional torus with opposite winding numbers ($1$ and $-1$). Thus, they cannot shrink by themselves to a point. Finally, the two filaments collide, forming a new ring with zero winding number that shrinks and disappears. The system then stays in a planar waves state for a very long time. Movie 3 shows another example of the filament dynamics that ends in a defect-free swap phase.

The occurrences in which the system escapes scroll-waves state into swap or static order, without any observation of transitions from the swap or static-order phase into scroll waves, suggest that these are metastable states. This is not the case, however, in the high $\tilde{K}$ limit, where the scroll waves appear to be stable, as explained in Sec.~\ref{Sec_high_K}.

\section{Schematic phase diagram in the fully anti-symmetric case}
\label{summary_purely_antisymmetric}
The results from Secs.~\ref{Sec_2D},~\ref{Sec_3D}, and~\ref{Sec_static} enable us to construct a schematic phase diagram in both 2D and 3D for the fully anti-symmetric case.

In 2D, our numerical results suggest that both the swap phase and the static-order phase are unstable, as long as $K\neq0$. This leads to the conclusion that there is a single stable phase in the thermodynamic limit: disorder (see Fig.~\ref{2D_3D_PD}a). 
The same conclusion applies to the 1D case, as is shown in Appendix~\ref{1D_app}.

In 3D (see Fig.~\ref{2D_3D_PD}b), our numerical results suggest that static order is unstable but that the swap phase is stable, with a second-order-like phase transition from disorder to swap featuring 3D XY critical exponents. Regions with high $\tilde{J}$ or $\tilde{K}$ have not been revealed in full: 
in the large $\tilde{J}$ limit, synchronization decreases, possibly to zero, due to the difficulty of droplet-induced swap to synchronize. In the high $\tilde{K}$ regime, scroll waves completely destabilize the long-range order. Both mechanisms are capable of destroying a large portion of the stable swap phase regime, and there is also a possibility that they would meet in the high $\tilde{J}$ and high $\tilde{K}$ regime, leading to a compact swap phase in $(\tilde{J},\tilde{K})$ space rather than an infinite one. Due to the limitations of our simulations, these interesting possibilities lie beyond the scope of this work, which has focused mostly on the low $\tilde{J}$ and $\tilde{K}$ regime, where the swap phase appears to be stable and critical exponents can be calculated.

\begin{figure}
\centering
{\includegraphics[width=0.48\textwidth,draft=false]{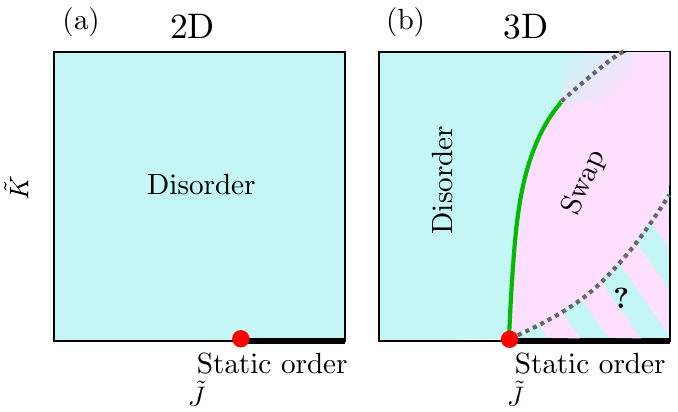}}
\caption{\textbf{Schematic phase diagram of the fully anti-symmetric case}. (a) Schematic 2D phase diagram and (b) schematic 3D phase diagram. To be compared with the mean-field prediction, Fig.~\ref{MF_Phase_diagram}. The green line is the second-order-like phase transition with critical exponents compatible with those of the 3D XY model. Grey dashed lines are phase transitions that have not been revealed in full and are possibly first-order-like.
In the striped area, it is not clear from numerical simulations whether the system is in the swap or disordered phase.
} \label{2D_3D_PD}
\end{figure}
%

\section{General asymmetric couplings} \label{Sec_K_+}
In the previous sections, we have focused on the fully anti-symmetric case where $K_+=0$ in Eq.~\eqref{K_matrix}, for which our model has a $C_4$ symmetry.
In this section, we test whether our results rely on this additional symmetry by considering $K_+ \neq 0$, for which the symmetry is reduced to the diagonal $\mathbb{Z}_2$ described in Eq.~\eqref{symmetrygen1}.

\subsection{Mean-field analysis} \label{MF_KP_section}
\begin{figure*}
\centering
{\includegraphics[width=1\textwidth,draft=false]{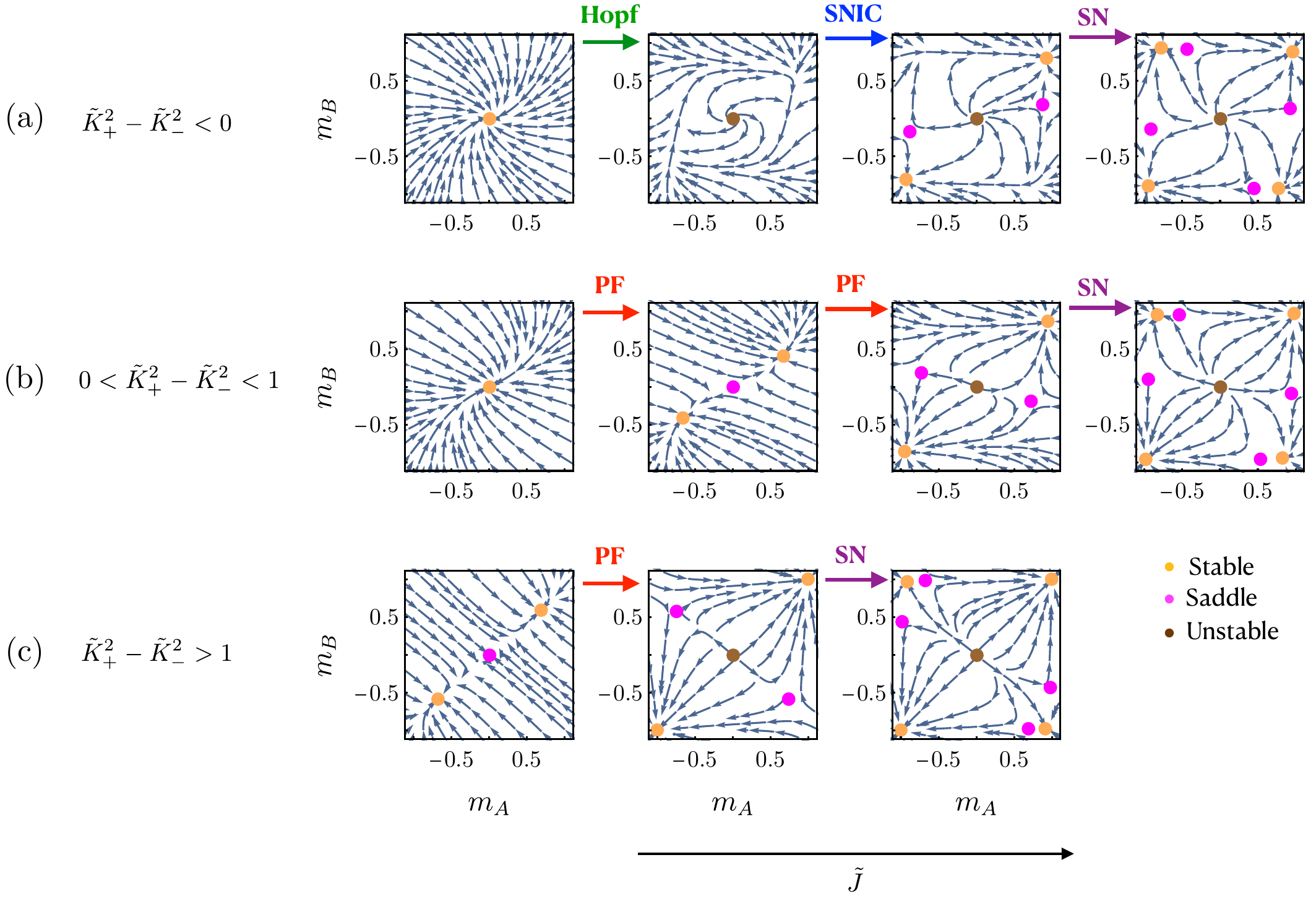}}
\caption{\textbf{Bifurcations in the general asymmetric case}. Flow in $\left(m_A,m_B\right)$ space calculated from the spatially homogeneous full mean-field equation, Eq.~(\ref{mean_field}), for $\tilde{K}_-=0.2$, and (a) $\tilde{K}_+=0.1$, (b) $\tilde{K}_+=0.3$, and (c) $\tilde{K}_+=1.1$. In each panel, $\tilde{J}$ is increased from left to right to capture the different phase portraits. In (a) $\tilde{J}=0.4,1.2, 1.5,1.7$, in (b) $\tilde{J}=0.4,0.9,1.4,2$ and in (c) $\tilde{J}=0.1,2.3,3.1$. The arrows mark the bifurcations that connect every two configurations: Hopf, saddle-node on an invariant circle (SNIC), saddle-node (SN) or pitchfork (PF).} \label{with_reciprocal}
\end{figure*}

Expanding Eq.~(\ref{mean_field}) as was done for the fully anti-symmetric case (Sec.~\ref{Bifurcations}), we obtain,
\begin{align} \label{MeanFieldExpanded2}
\partial_t m_{\alpha}  =&-(1-\tilde{J})m_{\alpha}+ \tilde{K}_{\alpha\beta}m_{\beta}+D \nabla^{2}m_{\alpha}\\
 & -\frac{1}{3}\left(\tilde{J}m_{\alpha}+\tilde{K}_{\alpha\beta} m_{\beta}\right)^{3}  \nonumber 
\end{align}
where $\tilde{K}_{\alpha \beta} \equiv K_{\alpha \beta} / (k_BT)$ and $K_{\alpha\beta}$ is given by Eq.~(\ref{K_matrix}). Assuming a spatially homogeneous state, the eigenvalues $\lambda_1$ and $\lambda_2$ of the Jacobian matrix $\mathcal{J}$ at $m_A=m_B=0$ are
\begin{equation}
\lambda_{1,2}=-1+\tilde{J}\pm\sqrt{\tilde{K}_+^2-\tilde{K}_-^2}.
\end{equation}
Thus, upon changing $\tilde{J}$, the possible phases and bifurcations depend on the value of $\tilde{K}_+^2-\tilde{K}_-^2$. This is demonstrated in Fig.~\ref{with_reciprocal}, which shows phase portraits of the spatially homogeneous full mean-field equation, Eq.~(\ref{mean_field}), for different $\tilde{K}_+^2-\tilde{K}_-^2$ values.

For $\tilde{K}_+^2-\tilde{K}_-^2<0$
(which is equivalent to $\tilde{K}_+<\tilde{K}_-$ when both $\tilde{K}_+$ and $\tilde{K}_-$ are assumed to be positive), Fig.~\ref{with_reciprocal}a, the system is disordered when $\tilde{J}<1$, and undergoes a Hopf bifurcation at $\tilde{J}=1$, transitioning from disorder to a limit cycle. Unlike the $K_+=0$ case studied throughout most of the paper, here the limit cycle is not rotationally-symmetric even in the $\tilde{J}-1\ll 1$ limit. This is seen when writing the spatially homogeneous mean-field equation in polar coordinates,
\begin{equation}
\begin{split}
\label{r_and_theta}
\partial_{t}r & =-\left(1-\tilde{J}-\tilde{K}_{+}\sin(2\theta)\right)r+\mathcal{O}\left(r^{3}\right)\\
\partial_{t}\theta & =-\tilde{K}_{-}+\tilde{K}_{+}\cos(2\theta)+\mathcal{O}\left(r^{2}\right).
\end{split}
\end{equation}
The leading order term in $r$ is $\theta$-dependent, and does not simply scale with $\tilde{J}-1$. Hence, a separation between the time scales of $r$ and $\theta$ cannot be invoked as in Sec.~\ref{Bifurcations}. Instead, the limit cycle is elliptical near the Hopf bifurcation~\cite{kuramoto1984chemical}.

 Increasing $\tilde{J}$ leads to additional bifurcations. First, a ($\mathbb{Z}_2$-symmetric) SNIC bifurcation where the limit cycle is destroyed by the emergence of two (rather than four for the $K_+=0$ case) pairs of stable and saddle points. The stable points that emerge lie in the first and third quadrants of the $(m_A,m_B)$ plane, corresponding to states where the two species are aligned. Then, for even larger values of $\tilde{J}$, a saddle-node bifurcation occurs where two additional pairs of stable and saddle points emerge, yet static order is maintained. The new stable points lie in the second and fourth quadrants, representing anti-aligned states. This behavior is observed for $\tilde{K}_+$ and $\tilde{K}_-$ much smaller than $1$ ($0.1$ and $0.2$ respectively in the Fig.~\ref{with_reciprocal}). For larger values, a saddle-node bifurcation can occur such that the ferromagnetic phase coexists with the limit cycle (this scenario is explored in Ref.~\cite{guislain2023discontinuous}).

For $0\leq\tilde{K}_+^2-\tilde{K}_-^2<1$ (Fig.~\ref{with_reciprocal}b), the system is disordered when $\tilde{J}<1-\sqrt{\tilde{K}_+^2-\tilde{K}_-^2}$ and undergoes a pitchfork bifurcation at $\tilde{J}=1-\sqrt{\tilde{K}_+^2-\tilde{K}_-^2}$ transitioning from disorder to static order with two stable fixed points of aligned states. In that case, there is no oscillatory phase. When $\tilde{J}=1+\sqrt{\tilde{K}_+^2-\tilde{K}_-^2}$, an additional pitchfork bifurcation occurs where two saddle points emerge from the origin $(0,0)$, which transitions from a saddle to an unstable point. As $\tilde{J}$ increases further, a saddle-node bifurcation occurs where two pairs of anti-aligned stable and saddle points appear. These two additional bifurcations maintain the system in a static-order phase. In the $\tilde{K}_+^2-\tilde{K}_-^2>1$ case (Fig.~\ref{with_reciprocal}c), the bifurcations are the same as in the $0\leq\tilde{K}_+^2-\tilde{K}_-^2<1$ case, except that there are already two stable fixed points at $\tilde{J}=0$. As a result, the system remains in the static-order phase for any value of $\tilde{J}$ (although $\tilde{J}$ must greater than zero for coarse-graining to take place). The absence of oscillations for $\tilde{K}_+>\tilde{K}_-$ arises because, despite having nonreciprocal interactions, the two species \enquote{agree} on whether they prefer to align or anti-align. This makes a magnetization flip unfavorable for either species.

\begin{figure}
\centering
{\includegraphics[width=0.34\textwidth,draft=false]{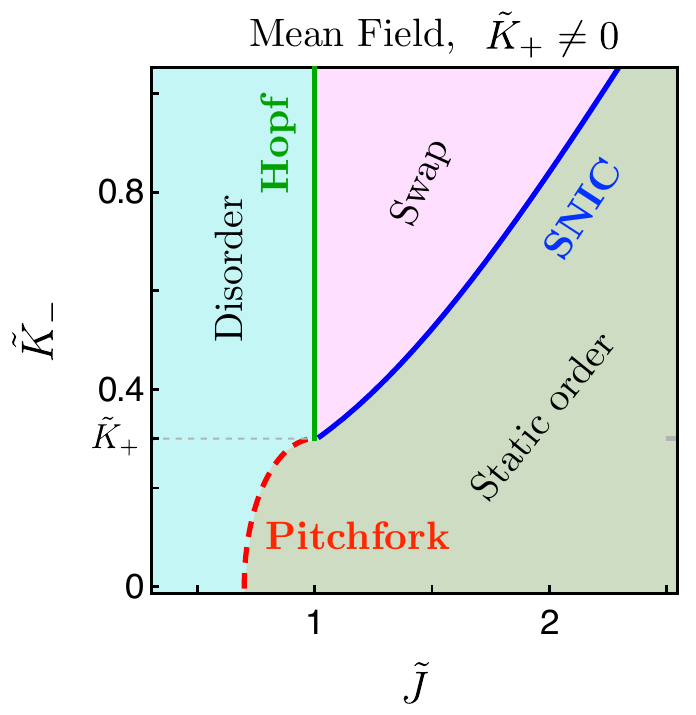}}
\caption{\textbf{Mean-field phase diagram of the general asymmetric case.} The diagram is shown as a function of $\tilde{J}$ and $\tilde{K}_-$, with the reciprocal inter-species interaction coupling set to $\tilde{K}_+=0.3$. The saddle-node bifurcations illustrated in Fig.~\ref{with_reciprocal} are not drawn in this plot. Obtained from numerical analysis of the spatially homogeneous Eq.~(\ref{mean_field}).} \label{MF_with_Kp}
\end{figure}

Figure~\ref{MF_with_Kp} shows a slice through the three-dimensional phase diagram (which depends on $\tilde{J},\tilde{K}_+$, and $,\tilde{K}_-$), with $\tilde{K}_+$ fixed at $0.3$. The main difference between this phase diagram and the one in Fig.~\ref{MF_Phase_diagram} (which shows the $\tilde{K}_+=0$ case) is the pitchfork bifurcation line separating the paramagnetic and ferromagnetic phases in the $\tilde{K}_-\leq\tilde{K}_+$ region, which is just a point in Fig.~\ref{MF_Phase_diagram}.

\subsection{Swap phase} \label{swap_Kp}
\begin{figure}[ht]
\centering
{\includegraphics[width=0.4\textwidth,draft=false]{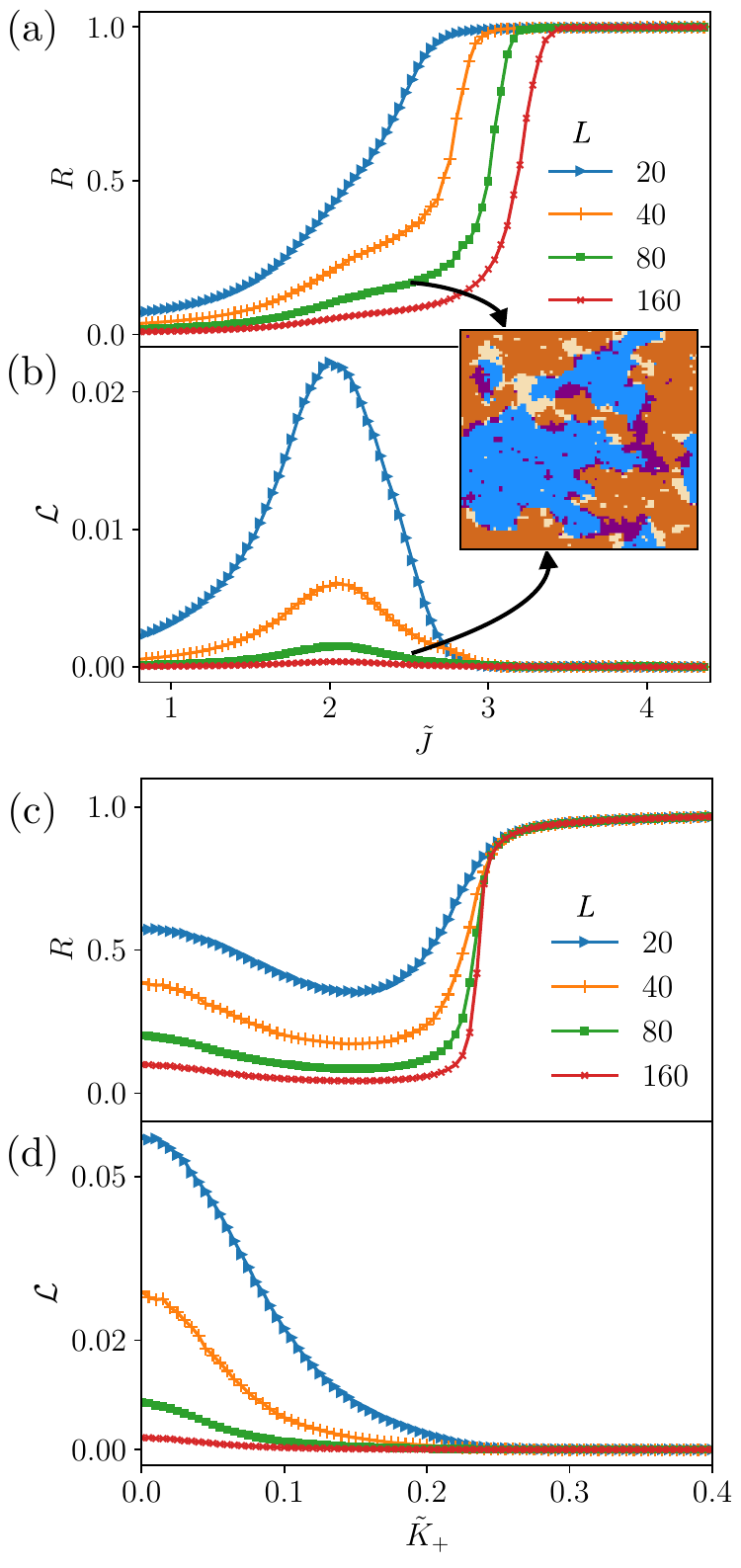}}
\caption{\textbf{No stable swap phase in 2D in the general asymmetric case}.
(a) $R$ and (b) $\mathcal{L}$  as a function of $\tilde{J}$ for $\tilde{K}_+ = 0.1$ and $\tilde{K}_- = 0.3$ shown for different linear system size $L$. Inset: Snapshot of the $\theta$ field (taken after steady state is reached) in a simulation with $\tilde{J}=2.5$ and $L=80$, and the same $\tilde{K}_+$ and $\tilde{K}_-$ as the main plots. The snapshot reveals defects with blue and brown regions ($\uparrow \uparrow$ and $\downarrow \downarrow$, respectively), being more abundant than yellow and purple regions ($\downarrow \uparrow$ and $\uparrow \downarrow$, respectively). (c) $R$ and (d) $\mathcal{L}$  as a function of $\tilde{K}_+$ for $\tilde{J} = 2$ and $\tilde{K}_- = 0.3$ and for different system sizes. Obtained from MC simulations. 
}
\label{R_and_L_2D_with_K_p}
\end{figure}

We perform Monte-Carlo simulations in both 2D and 3D for the general asymmetric case, {\it i.e.} with the inclusion of reciprocal inter-species interactions. In 2D, we find that the swap phase is unstable, as in the case of fully anti-symmetric interactions (Sec.~\ref{Sec_2D}). This is shown in Fig.~\ref{R_and_L_2D_with_K_p} in which $R$ and $\mathcal{L}$ are plotted as a function of $\tilde{J}$ with fixed $\tilde{K}_+$ and $\tilde{K}_-$ (a-b) and as a function of $\tilde{K}_+$ with fixed $\tilde{J}$ and $\tilde{K}_-$ (c-d). In the region in which $\mathcal{L}\neq 0$ (indicating the swap phase), both $R$ and $\mathcal{L}$ decrease with system size. Furthermore, in appendix~\ref{2D_colormaps_appendix} we show color maps of $R$ and $\mathcal{L}$ as a function of $\tilde{J}$ and $\tilde{K}_-$ with a fixed non-zero value of $\tilde{K}_+$. These color maps show that, similarly to the $\tilde{K}_+=0$ case shown in Fig.~\ref{R_and_L_2D}a-d, $R$ and $\mathcal{L}$ diminish as system size increases in the swap regime.

The destabilized phase is shown to be contaminated with defects (inset in Fig~\ref{R_and_L_2D_with_K_p}), but it does not seem to support spiral waves. This seems to result from the lack of symmetry between the four two-spin states, which is required for the stability of a spiral waves: since $A$-spins want to align with $B$-spins more than $B$-spins wish to anti-align with $A$ spins, $\uparrow\uparrow$ and $\downarrow\downarrow$ states are more favorable than $\uparrow\downarrow$ and $\downarrow\uparrow$, and thus the four \enquote{arms} surrounding a defect move at different velocities, destroying the potential spiral configuration. Instead of spirals we observe a complex behavior in which large homogeneous regions of either $\uparrow\uparrow$ or $\downarrow\downarrow$ state continuously move and deform, while defects are rapidly created and annihilated at the boundaries between them.

\begin{figure}[ht]
\centering
{\includegraphics[width=0.5\textwidth,draft=false]{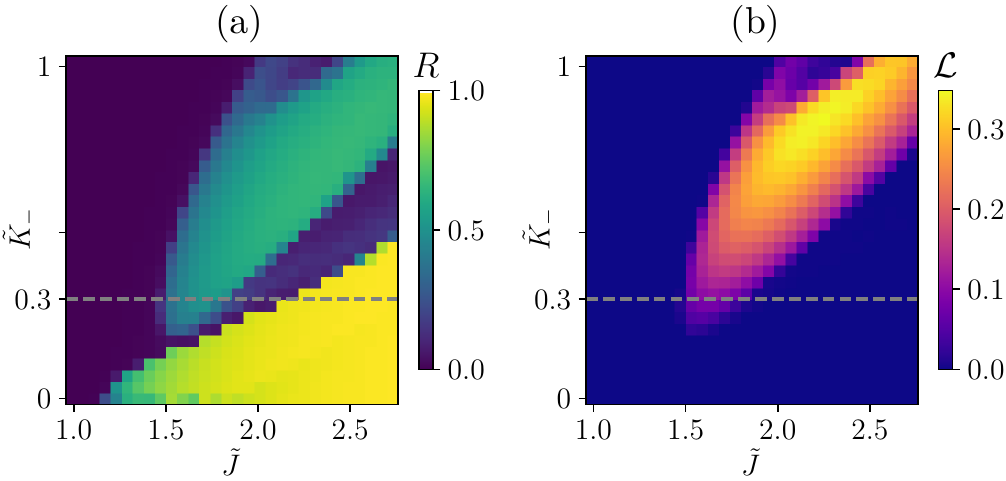}}
\caption{\textbf{Order parameters in 3D in the general asymmetric case}. (a) $R$ and (b) $\mathcal{L}$ color map as a function of $\tilde{J}$ and $\tilde{K}_-$ with $\tilde{K}_+ = 0.1$ and $L=80$. Dashed grey lines show the fixed $\tilde{K}_-$ cut analyzed in Fig.~\ref{3D_R_vs_R_and_L_with_Kp}. Obtained from MC simulations.}
\label{3D_PD_with_Kp}
\end{figure}

\begin{figure}[ht]
\centering
{\includegraphics[width=0.4\textwidth,draft=false]{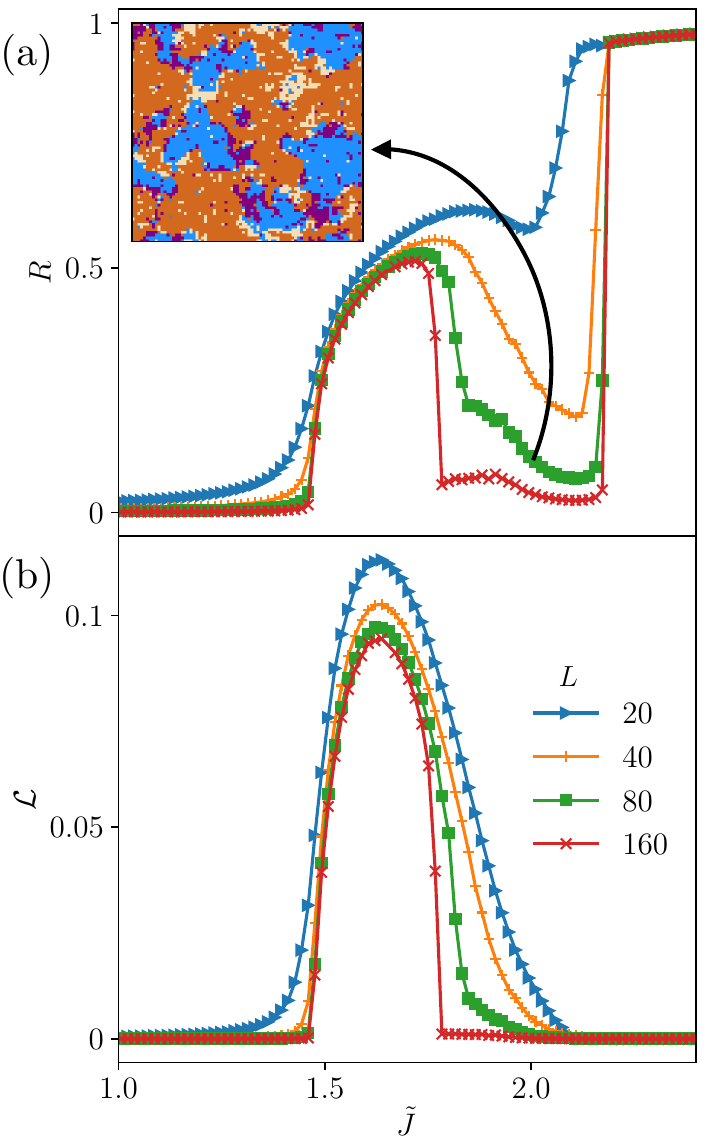}}
\caption{\textbf{Phase transition from disorder to swap in 3D in the general asymmetric case}. (a) $R$ and (b) $\mathcal{L}$ as a function $\tilde{J}$ with $\tilde{K}_+=0.1$ and $\tilde{K}_-=0.3$, for different system sizes.
The inset in panel a shows a snapshot of $\theta$ in a 2D slice of the 3D system (taken after steady state is reached), with $\tilde{J}=2$, $L=80$, and the same $\tilde{K}_+$ and $\tilde{K}_-$ values as in the main panel. Obtained from MC simulations.}
\label{3D_R_vs_R_and_L_with_Kp}
\end{figure}

In 3D, Monte-Carlo simulations suggest that a stable swap phase still exists, even when reciprocal interactions are added. Figure~\ref{3D_PD_with_Kp} shows color maps of $R$ and $\mathcal{L}$ for a fixed $\tilde{K}_+$ and $L=80$. A region with non-zero $R$ and $\mathcal{L}$, corresponding to a swap phase, is observed. Figure~\ref{3D_R_vs_R_and_L_with_Kp} then analyzes a line cut through the color maps for a fixed value of $\tilde{K}_-$. As $\tilde{J}$ is varied, a continuous phase transition from disorder to swap is observed (at $\tilde{J}\approx1.45$) and becomes more sharp as $L$ increases. The swap phase is stable for $\tilde{J}\lesssim1.7$ but destabilizes at higher values, as is discussed next.

To conclude, the instability of the swap phase in 2D and its stability in 3D are not limited to the case of fully anti-symmetric interactions in which $\tilde{K}_+=0$, but extends to the general asymmetric case.

\subsection{Stability of the static-order phase: a droplet-capture mechanism}

While the fate of the swap phase is robust to the addition of reciprocal inter-species interactions, our Monte-Carlo simulations suggest this is not the case for the static-order phase.

First, let us examine the right part of each panel of Fig.~\ref{R_and_L_2D_with_K_p}, showing the transition to static order ($R\neq 0$ and $\mathcal{L}\approx 0$) for a finite-size system in 2D. Figure~\ref{R_and_L_2D_with_K_p}a-b shows that the $\tilde{J}$-transition-point between the unstable swap regime and static order increases with system size and it is hard to conclude whether static order is stable in the thermodynamic limit. In Fig.~\ref{R_and_L_2D_with_K_p}c-d, where $\tilde{K}_+$ is varied, $R$ seems to converge to finite non-zero values (while $\mathcal{L}$ approaches zero) for $\tilde{K}_+\gtrsim 0.24$, which might indicate a stable phase transition from disorder to static order. Surprisingly, the supposed phase transition occurs even though $\tilde{K}_+$ is lower than $\tilde{K}_-$ ($\tilde{K}_-=0.3$), which is the case where one species favors alignment while the other favors anti-alignment. In 3D, Fig.~\ref{3D_R_vs_R_and_L_with_Kp} gives an even stronger indication that there is a well-defined phase transition from disorder to static order: for linear system sizes $L=40,80,160$, there is an abrupt transition from a region in which $R$ and $\mathcal{L}$ approach zero (see $1.8\lesssim\tilde{J}\lesssim2.2$ in Fig.~\ref{3D_R_vs_R_and_L_with_Kp}) to a regime in which $R$ is finite and $\mathcal{L}\to 0$ ($\tilde{J}\gtrsim 2.2$ in Fig.~\ref{3D_R_vs_R_and_L_with_Kp}), and the transition seems to converge at $\tilde{J}\approx 2.2$. Again, the transition occurs when $\tilde{K}_+<\tilde{K}_-$ ($\tilde{K}_+=0.1$ while $\tilde{K}_-=0.3$). These indications raise the question: does the droplet destabilization argument sketched in Sec.~\ref{Sec_static} for the fully anti-symmetric case apply here?

Let us assume again a static-order scenario in which both spin species are up on average. The species favoring anti-alignment ($B$) prefers a state in which all of its spins are flipped. As a result, it will nucleate droplets with opposite magnetization. As in the fully anti-symmetric case, we expect that once these droplets will surpass a critical size, they will expand instead of shrinking. After certain amount of time, the same event will repeat itself for the $A$ species: a critical droplet will nucleate and expand.

\begin{figure}[ht]
\centering
{\includegraphics[width=0.4\textwidth,draft=false]{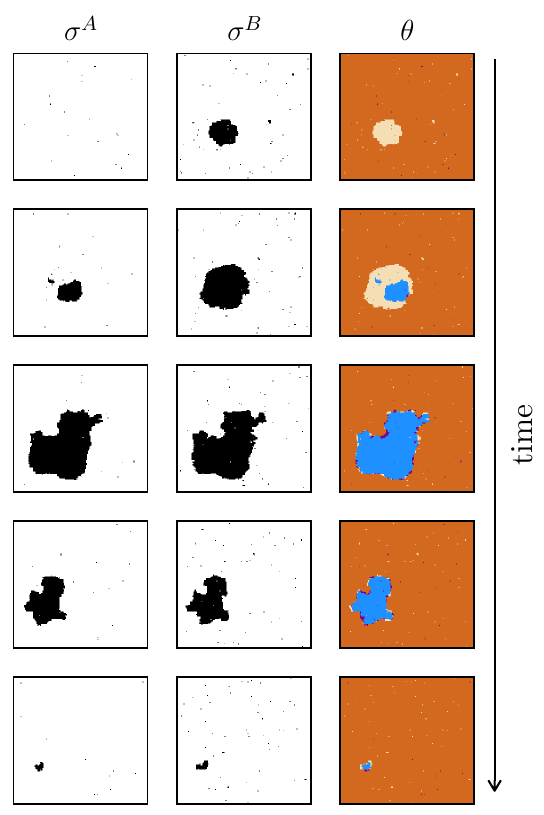}}
\caption{\textbf{Droplet capture mechanism}. MC simulation snapshots of a 2D system with $\tilde{J}=3.3$, $\tilde{K}_+=0.1$, $\tilde{K}_-=0.3$, and $L=150$. See Movie 5 for the full evolution. Each row displays a distinct time step, with time ascending from top to bottom. Species $B$ nucleates a droplet, but before it reaches system size, a nested droplet of $A$-spins nucleates and expands at faster speed, eventually catching up with the boundaries of the $B$-droplet, making it shrink and disappear. This mechanism can stabilize the static-order phase when $\tilde{K}_+<\tilde{K}_-$.
}
\label{Droplet_eating_droplet}
\end{figure}

However, because of the asymmetry between the two species imposed by $K_+$, the critical droplet size $\rho_c$ and more importantly the velocity in which droplets expand $v(\rho)$, is different for $A$-droplets and $B$-droplets. Here, $A$-spins interact more strongly with $B$-spins than the other way around, leading to smaller $\rho_c$ and larger $v(\rho)$ for $A$-droplets~\cite{rikvold1994metastable}. As a result, a nested $A$-droplet inside a $B$-droplet will expand more quickly and may eventually reach the boundaries of the $B$-droplet. Once this happens, our simulations show that the $B$-droplet can disintegrate, and the magnetization flip it prevented. Movie 5 shows a 2D simulation in which this mechanism takes place. Snapshots from the simulation are shown in Fig.~\ref{Droplet_eating_droplet}.
Note that the simulation parameters were chosen to enhance the visibility of the mechanism, allowing the $B$-droplet to persist for an extended period, but the mechanism is expected to be more efficient in suppressing droplet growth as the asymmetry between the species increases (larger $K _+$).

To summarize, when inter-species interactions have opposite signs (the $\tilde{K}_+<\tilde{K}_-$ case) but are not fully anti-symmetric, there is a competing mechanism that prevents droplets from growing. This suggests a re-stabilization of the static-order phase in the general asymmetric case, for $d\geq 2$.
 
 \subsection{Absence of infinite-period phase transition} \label{NO_SNIC}

In the fully anti-symmetric case ($K_+ = 0$), there is an oscillating (swap) phase in 3D, but there is no static-order phase.
As a consequence, there is no infinite-period phase transition that would correspond to the SNIC bifurcation observed in the mean-field dynamics. 
A similar situation has been described in Ref.~\cite{assis2011infinite}.
In the general asymmetric case, however, both the swap phase and static-order phase exist in 3D. 
Nevertheless, we did not observe any direct phase transition between the two. 
Figures~\ref{3D_PD_with_Kp} and~\ref{3D_R_vs_R_and_L_with_Kp} show that the transition between them, when $\tilde{K}_+$ is fixed and $\tilde{J}$ and $\tilde{K}$ are varied, is mediated by an intermediate disordered region. Unlike the 3D droplet regime with fully anti-symmetric interactions whose fate is left undetermined (Sec.~\ref{fate_droplet_regime}), here the intermediate regime shows convergence to a state with $R=0$, {\it i.e.}, disorder.
A snapshot of $\theta$ of a 2D slice in the intermediate disordered regime is shown in the inset in Fig.~\ref{3D_R_vs_R_and_L_with_Kp}a. Large regions of aligned spins, either in $ \uparrow\uparrow $ or $\downarrow\downarrow $ state, coexist with some defects on the boundaries between them, similar to the 2D disorder discussed in Sec.~\ref{swap_Kp}.
In Appendix~\ref{3D_PD_cuts}, the order parameters on additional cuts through the phase diagram are shown (one with a fixed value of $\tilde{J}$ and another with a fixed value of $\tilde{K}_-$). These cuts support the same conclusion: the transition between swap to static order we observed is mediated by disorder.

\begin{figure}[ht]
\centering
{\includegraphics[width=0.5\textwidth,draft=false]{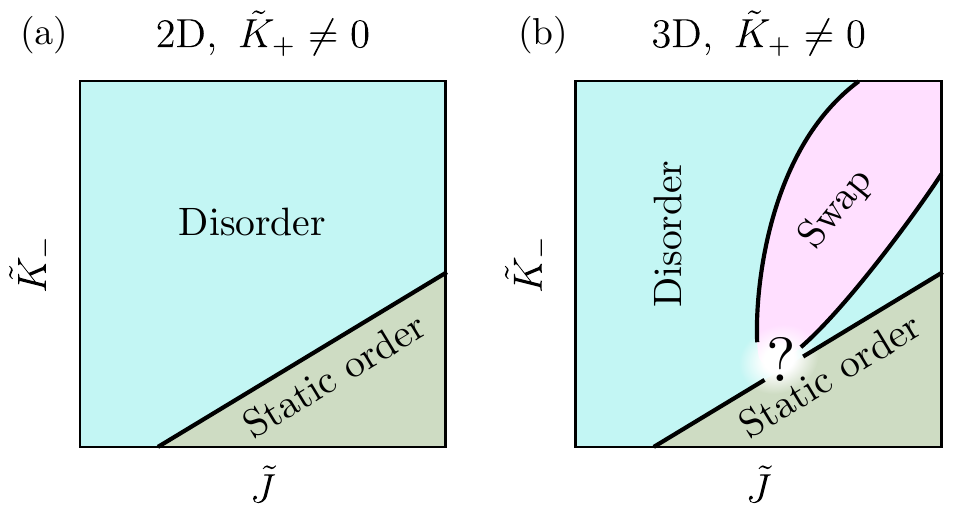}}
\caption{\textbf{Schematic phase diagram of the general asymmetric case} in (a) 2D and (b) 3D, shown as a function of $\tilde{J}$ and $\tilde{K}_-$, with fixed (non-zero) $\tilde{K}_+$. The nature of the phase transition lines are left unspecified, see discussion in Sec.~\ref{Schematic_general}. The question mark represents the uncertainty regarding whether the swap and static-order phases meet at a point or are strictly separated by disorder.}
\label{PD_sketch_with_KP}
\end{figure}
%
\subsection{Schematic phase diagram in the general asymmetric case} \label{Schematic_general}
A schematic drawing of the 2D and 3D phase diagram with a fixed non-zero $\tilde{K}_+$ is shown in Fig.~\ref{PD_sketch_with_KP}. 
In 2D, our numerical results suggest that there are stable disorder and static-order phases (but no swap), while in 3D, our numerical results suggest that all the phases predicted by mean-field: disorder, swap, and static order, are stable. We expect the disorder-to-swap phase transition to be characterized by the same critical exponents as in Sec.~\ref{critical_exponents_section} (compatible with 3D XY), at least throughout a portion of the phase transition line. Other portions may be described by first-order-like phase transitions.
Similarly, we hypothesize that the phase transition from disorder to static order in both 2D and 3D has Ising critical exponents in the regime where $K_+>K_-$, but might be first-order-like in the opposite $K_+<K_-$ regime (see Fig.~\ref{3D_PD_with_Kp}a in which the $K_+<K_-$ phase transition seems more abrupt than the $K_+>K_-$ phase transition).
Finally, we cannot exclude that the three phases in 3D meet at a tricritical point (or line in the three-dimensional phase diagram).

\section{Conclusions} \label{Sec_conclusions}
We introduced a nonreciprocal Ising model as a minimal model for nonreciprocal many-body dynamics, and studied the stability of its static and time-dependent phases along with the transitions between them. Our main result is the stability of a time-dependent (swap) phase in 3D, which behaves as a time-crystal with infinite coherence time in the thermodynamic limit. The phase transition from disorder to swap is characterized by critical exponents compatible with the 3D XY model, which can be rationalized by the breaking of a continuous symmetry (time translation) in the swap phase. The existence of a swap phase in 3D is robust and holds for fully anti-symmetric interactions between the species as well as the more general asymmetric interactions case. 

Based on simulations and theoretical arguments, we predict that the static-order (ferromagnetic) phase is unstable in any finite dimension for fully anti-symmetric interactions. The instability is caused by the expansion of droplets of opposite magnetization occurring in the metastable species. When the inter-species interaction is not fully anti-symmetric, a competing mechanism of droplet-capture by the other species can re-stabilize the static-order phase. A direct phase transition between a time-dependent phase and a static-ordered phase, however, was not observed.

\medskip
\noindent \emph{Code availability} --- The code used for performing the Monte-Carlo simulations, computing the critical exponents, and plotting the mean-field phase portrait is available under the 2-clause BSD license at \url{https://doi.org/10.5281/zenodo.14816551}

\appendix

\bigskip\bigskip

\begin{acknowledgments}
We thank G. Biroli, S. Diehl, O. Granek, M. Han, T. Khain, P. Littlewood, R. Mandal, D. Mukamel, S. Sethi, S. Sondhi, G. A. Weiderpass, C. Weiss, and T. Witten for helpful discussions.
Y.A., D.S. and M.F. acknowledge support from a MRSEC-funded Kadanoff–Rice fellowship and the University of Chicago Materials Research Science and Engineering Center, which is funded by the National Science Foundation under award no. DMR-2011854. Y.A. acknowledges support from the Zuckerman STEM Leadership Program. 
D.M., M.F., and V.V acknowledge support from the France Chicago center through a FACCTS grant. M.F. acknowledges support from the National Science
Foundation under grant no. DMR-2118415 and the Simons Foundation.
V.V. acknowledges partial support from the Army Research Office under grant nos. W911NF-22-2-0109 and W911NF-23-1-0212 and the Theory in Biology program of the Chan Zuckerberg Initiative. This research was partly supported by the National Science Foundation through the Center for Living Systems (grant no. 2317138) and the National Institute for   Theory and Mathematics in Biology (NITMB).
All the authors acknowledge the support of the UChicago Research Computing Center which provided the computing resources for this work.
\end{acknowledgments}

\section{Measurements of time-reversal symmetry breaking} \label{EntProd}

As shown in Sec.~\ref{sec:BDB}, nonreciprocal coupling breaks detailed balance, and therefore time-reversal symmetry.
This is a statement about the dynamics of our system.
The dynamics can be related to thermodynamic observables -- such as heat, work, and entropy -- under the \textit{local detailed balance} condition~\cite{Seifert2012,VandenBroeck2015,Horowitz2019,Maes2021,Sekimoto2010,Pachter2024}.
As we don't explicitly model the environment of our system, the local detailed balance condition is a hypothesis.
Specifically, local detailed balance relates the entropy produced by a state transition (here, a spin flip) to the transition rates $w(F_i^\alpha \vec{\sigma} | \vec{\sigma})$ via
\begin{equation}
    \label{eq:entropyChange}
    \Delta S_\text{single flip} = \ln \frac{w\left(F_{i}^{\alpha}\vec{\sigma}|\vec{\sigma}\right)}{w\left(\vec{\sigma}|F_{i}^{\alpha}\vec{\sigma}\right)}.
\end{equation}
Here entropy is given in units of $k_B$.
By summing over a series of transitions in a specified trajectory, the above expression gives the total entropy produced by that trajectory.
For example, Eq.~\eqref{eq:BDBCycle} gives the entropy produced by a specific cycle of spin states.

To characterize the steady states of the system, it is more convenient to measure the average entropy production rate.
This is given by (see \textsection6.2.2 of \cite{Seifert2012} for derivation)
\begin{equation}
\dot{S}=\sum_{i,\alpha} \left\langle w\left(F_{i}^{\alpha}\vec{\sigma}|\vec{\sigma}\right)
\ln \frac{w\left(F_{i}^{\alpha}\vec{\sigma}|\vec{\sigma}\right)}{w\left(\vec{\sigma}|F_{i}^{\alpha}\vec{\sigma}\right)} \right\rangle_{t,\Omega},
\end{equation}
where $\langle \ldots \rangle_{t, \Omega}$ is an average over time and realizations.
Using the Glauber transition rate, Eq.~(\ref{rates}), we obtain that the entropy production rate per spin is
\begin{equation}
    \label{eq:EntProd} \dot{s}=-\frac{1}{L^{d}} \sum_{i,\alpha}
    \left\langle \left[1-\tanh\left(\frac{\Delta E^{\alpha}_i}{2k_B T}\right)\right]
\frac{\Delta E_{i}^{\alpha}}{2k_{B}T}\right\rangle_{t,\Omega}.
\end{equation}
where $\tau$ was taken as 1 in simulation time.

The entropy production rate in Eq.~\eqref{eq:EntProd} is a microscopic measurement of time-reversal symmetry breaking.
By contrast, the phase space angular momentum $\mathcal{L}=\langle M_{B}\partial_{t}M_{A}-M_{A}\partial_{t}M_{B}\rangle_{t,\Omega}$ introduced in Eq.~\eqref{Angular_momentum} of the main text is a \textit{macroscopic} measurement of time-reversal symmetry breaking.
Coarse-graining from micro- to macroscopic variables is known to give a lower bound to the true entropy production rate~\cite{Seifert2012}, below we will show that $\mathcal{L}$ is proportional to the coarse-grained (mesoscopic) entropy production rate, denoted by $\dot{\mathcal{S}}$, only for cases where the fluctuations in magnetization are negligible compared with the average magnetization and if one considers the dynamics of the mean-field equation with added noise in the $m_\alpha \ll 1$ limit.

To see this, we work with the mean-field equation of the fully anti-symmetric case expanded around $m_\alpha \ll 1$, Eq.~\eqref{MeanFieldExpanded}, as a stochastic PDE with white noise,
\begin{equation}
\begin{aligned}
\label{eq:noisyMF}
    \partial_t m_{\alpha}  =&-(1-\tilde{J})m_{\alpha}+\tilde{K} \varepsilon_{\alpha\beta}m_{\beta}+D \nabla^{2}m_{\alpha}\\
 & -\frac{1}{3}\left(\tilde{J}m_{\alpha}+\tilde{K} \varepsilon_{\alpha\beta} m_{\beta}\right)^{3} + \eta_\alpha.
\end{aligned}
\end{equation}
where the white noise field satisfies $\langle\eta_{\alpha}(\vec{r},t)\rangle=0$ and $\langle \eta_\alpha(\vec{r}, t) \eta_\beta(\vec{r}\,', t') \rangle = \zeta \delta_{\alpha \beta} \delta(\vec{r} - \vec{r}\,') \delta(t - t')$.
This can be rewritten as
\begin{equation}
\partial_{t}m_{\alpha}=-\frac{\delta \mathcal{F}}{\delta m_{\alpha}}+A_{\alpha} + \eta_{\alpha},
\end{equation}
where $-\delta \mathcal{F} / \delta m_a$ are conservative forces derived by the variation of a free energy
\begin{equation}
\begin{aligned}
\mathcal{F} =\int{\rm d}\vec{r}\bigg[ & \frac{(1-\tilde{J})}{2}\left(m_{A}^{2}+m_{B}^{2}\right) + \frac{\tilde{J}^{3}}{12}\left(m_{A}^{4}+m_{B}^{4}\right)\nonumber\\
 & + \frac{1}{2}\tilde{J}\tilde{K}^{2}m_{A}^{2}m_{B}^{2} +\frac{D}{2}\left(\left|\nabla m_{A}\right|^{2}+\left|\nabla m_{B}\right|^{2}\right) \bigg],
\end{aligned}
\end{equation}
and $A_a$ are the non-conservative forces on the right-hand side of Eq.~\eqref{eq:noisyMF},
\begin{align}
A_{\alpha} & = \sum_{\beta} \varepsilon_{\alpha\beta} a_{\alpha \beta} \\
a_{\alpha \beta} & = 
\tilde{K}m_{\beta} -
\tilde{J}^{2}\tilde{K}m_{\alpha}^{2} m_{\beta} -
\frac{1}{3} \tilde{K}^{3} m_\beta^3\,\,.
\end{align}
Finally, we introduce the path probability functional of observing a trajectory $\mathbf{m}(\vec{r}, t)$ as
\begin{equation}
\mathcal{P}[\mathbf{m}(\vec{r}, t)] = \frac{\exp \left( -\mathcal{A}[\mathbf{m}(\vec{r}, t)] \right)}{\mathcal{Z}},    
\end{equation}
where $\mathbf{m}\equiv(m_A,m_B)$, $\mathcal{Z}$ is a normalization constant, and $\mathcal{A}[\mathbf{m}(\vec{r}, t)]$ is the {action}, which can be found using path-integral techniques~\cite{kamenev2023field,Onsager1953,Martin1973,DeDominicis1975,Janssen1976} to be
\begin{equation}
\label{eq:action}
    \mathcal{A}[\mathbf{m}(\vec{r}, t)] = \dfrac{1}{\zeta} \int {\rm d}\vec{r} {\rm d}t \sum_{\alpha} \left( \partial_t m_{\alpha} + \dfrac{\delta \mathcal{F}}{\delta m_{\alpha}} - A_{\alpha}\right)^2 .
\end{equation}

The average entropy production rate for a trajectory is written in terms of $\mathcal{P}$ and of the probability $\widetilde{\mathcal{P}}$ of observing the trajectory under time-reversal as~\cite{Kawai2007}
\begin{equation}
\label{eq:pathEPR}
    \dot{\mathcal{S}} = \lim_{T \to \infty} \dfrac{1}{T} \left\langle \ln \dfrac{\mathcal{P}}{\widetilde{\mathcal{P}}} \right\rangle_{\Omega} = \lim_{T \to \infty} \dfrac{\langle \widetilde{\mathcal{A}} - \mathcal{A} \rangle_{\Omega}}{T}
\end{equation}
where $T$ here is the trajectory duration and the average is over noise realizations~\cite{nardini2017entropy}.
As we do not have magnetic forces, finding $\widetilde{\mathcal{A}}$ amounts to taking $t \to -t$, which only affects the time derivative terms in Eq.~\eqref{eq:action}, giving
\begin{equation}
\label{eq:actionTR}
    \widetilde{\mathcal{A}}[\mathbf{m}(\vec{r}, t)] = \dfrac{1}{\zeta} \int {\rm d}\vec{r} {\rm d}t \sum_{\alpha} \left( -\partial_t m_{\alpha} + \dfrac{\delta \mathcal{F}}{\delta m_{\alpha}} - A_{\alpha}\right)^2 .
\end{equation}

Plugging Eqs.~\eqref{eq:action} and~\eqref{eq:actionTR} into Eq.~\eqref{eq:pathEPR}, we find
\begin{equation}
\begin{aligned}[b]
    \dot{\mathcal{S}} &= \lim_{T\to\infty}\frac{4}{\zeta T}\left\langle\int{\rm d}\vec{r}{\rm d}t 
    \sum_\alpha \left(A_\alpha - \dfrac{\delta \mathcal{F}}{\delta m_\alpha} \right) \partial_t m_\alpha \right\rangle_{\Omega}\\
    &= \lim_{T\to\infty}\frac{4}{\zeta T}\left\langle\int{\rm d}\vec{r}{\rm d}t \sum_\alpha \left( A_\alpha \partial_t m_\alpha \right) - \Delta\mathcal{F}\right\rangle_{\Omega}\\
    &= \lim_{T\to\infty}\frac{4}{\zeta T}\left\langle\int{\rm d}\vec{r}{\rm d}t \sum_\alpha A_\alpha \partial_t m_\alpha\right\rangle_{\Omega}\\
    &= \lim_{t\to\infty}\frac{4}{\zeta T}\left\langle\int{\rm d}\vec{r}{\rm d}t \left(a_{AB} \partial_{t} m_A  -  a_{BA} \partial_{t} m_B\right)\right\rangle_{\Omega}\\
    &=\frac{4}{\zeta }\int{\rm d}\vec{r}\left\langle a_{AB} \partial_{t} m_A  -  a_{BA} \partial_{t} m_B\right\rangle_{t,\Omega}.
\end{aligned}
\end{equation}
To go from the first to second line, we identified the total time derivative of the free energy, ${\rm d}\mathcal{F}/{\rm d}t = \int {\rm d}\vec{r}\sum_\alpha \partial_t m_\alpha \ (\delta \mathcal{F}/ \delta m_\alpha)$.
To go from the second to the third line, we used the fact that the $\Delta\mathcal{F}$ is a constant, which will contribute zero to $\dot{\mathcal{S}}$ when the limit $T\to\infty$ is taken. In the last equality we replaced $\frac{1}{T}\int {\rm d}t\,...$ with a time average $\langle...\rangle_t$.
For small $m_{\alpha}$, non-linear terms can be dropped and $a_{\alpha \beta} \simeq \tilde{K} m_\beta$, leading to
\begin{equation}
    \dot{\mathcal{S}}=\frac{4\tilde{K}}{\zeta}\int{\rm d}\vec{r}\langle m_{B}\partial_{t}m_{A}-m_{A}\partial_{t}m_{B}\rangle_{t,\Omega}.
\end{equation}
Finally, we assume small fluctuations around the mean magnetization, {\it i.e.} $m_\alpha(\vec{r}) = M_\alpha + \delta m_\alpha(\vec{r})$, $\delta m_\alpha(\vec{r})\ll M_\alpha$, and obtain to leading order,
\begin{equation}
\!\!\!
    \dot{\mathcal{S}} = \dfrac{4\tilde{K}L^d}{\zeta} \left\langle M_B \partial_t M_A - M_A \partial_t M_B \right\rangle_{t,\Omega} = \dfrac{4\tilde{K}L^d}{\zeta} \mathcal{L}.
\end{equation}
In conclusion, $\mathcal{L} \propto \dot{\mathcal{S}}$ for a mesoscale system obeying Eq.~\eqref{eq:noisyMF} if (i) $|m_\alpha| \ll 1$ and (ii) $\delta m_\alpha(\vec{r})\ll M_{\alpha}$.
Under these two approximations, and in the vicinity of the Hopf bifurcation, Eq.~\eqref{eq:noisyMF} can be reduced into the noisy Stuart-Landau equation~\cite{kuramoto1984chemical}
\begin{equation}
    \frac{{\rm d}\mathcal{M}}{{\rm d}t} = (a - b|\mathcal{M}|^2) \mathcal{M} + H
\end{equation}
where $\mathcal{M} = M_A + i M_B$, $H=\eta_A + i \eta_B$, and $a$ and $b$ are complex coefficients.
Writing the Stuart-Landau equation in polar coordinates and eliminating the radial equation yields a drift-diffusion equation that saturates the bound on the entropy production per oscillation $\Delta S \geq 4 \pi^2 \mathcal{N}$ of Ref.~\cite{Oberreiter2022} mentioned in Sec.~\ref{time_crystal_behavior} (here, $\mathcal{N} \equiv \tau_c/T_{\rm osc}$ where $\tau_{\text{c}}$ is the coherence time and $T_{\text{osc}}$ is the period of oscillation).
Indeed, it is shown in Ref.~\cite[SM \S~III]{Chen2024} that in this case, the rate of entropy production is $\dot{S} = f^2 \tau_{\text{c}}$ in which $f=2\pi/T_{\text{osc}}$, and $\Delta S = T_{\text{osc}} \dot{S}$ saturates the bound. We refer to Ref.~\cite{Chen2024} for more details.

While $\mathcal{L}$ can be measured in any state, in general one will have $\mathcal{L} \neq \dot{\mathcal{S}}$, and no conclusion about entropy production in terms of phase space angular momentum can be made.

\begin{figure}[ht]
\centering
{\includegraphics[width=0.5\textwidth,draft=false]{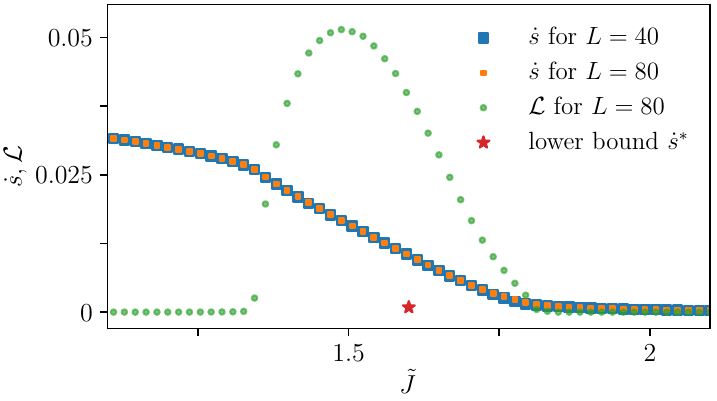}}
\caption{\textbf{Measurements of time-reversal symmetry breaking in the 3D NR Ising model.} {Entropy production rate per spin, $\dot{s}$ (blue and orange squares, calculated from Eq.~(\ref{eq:EntProd})), and phase space angular momentum, $\mathcal{L}$ (green dots), as a function of $\tilde{J}$ for fixed $\tilde{K}=0.1$. Theoretical lower bound for $\dot{s}$, $\dot{s}^*= \frac{4 \pi^2}{T_{\rm osc}^2} \, \frac{\tau_c(L)}{L^d}$ (Eq.~(\ref{bound})), at $\tilde{J}=1.6$, is shown as a red star, where $T_{\rm osc}$ was measured to be $T_{\rm osc} = 120$ and $\tau_c/L^3$ was taken as $0.32$ in accordance with the fit in Fig.~\ref{tau_vs_L}. $\dot{s}^*=9\times10^{-4}$ while $\dot{s}=0.01$ at $\tilde{J}=1.6$.}}
\label{EntProdFig}
\end{figure}

We calculate the entropy production rate per spin in the 3D NR Ising model using MC simulations.
Figure~\ref{EntProdFig} shows $\dot{s}$ for $L=40$ and $L=80$, as a function of the coupling $\tilde{J}$.
The curves overlap, which shows that $\dot{s}$ is independent of system size for large systems. 
$\dot{s}$ decreases monotonically with $\tilde{J}$, as stronger nearest-neighbor coupling slows down the cycle rate of individual spin-pairs.
By contrast, $\mathcal{L}$ is non-zero only in the swap phase.
For $\tilde{J}=1.6$ and $\tilde{K}=0.1$, we compare $\dot{s}$ to the lower-bound given by the temporal autocorrelation function in Eq.~\eqref{bound}, where $T_{\rm osc}$ is found by measuring the average oscillation time and $\tau_c$ is given by the fit in Fig.~\ref{tau_vs_L}. Indeed, $\dot{s}$ exceeds the lower bound by a factor of about 10.

\section{1D NR Ising model} \label{1D_app}
%
\begin{figure}[ht]
\centering
{\includegraphics[width=0.45\textwidth,draft=false]{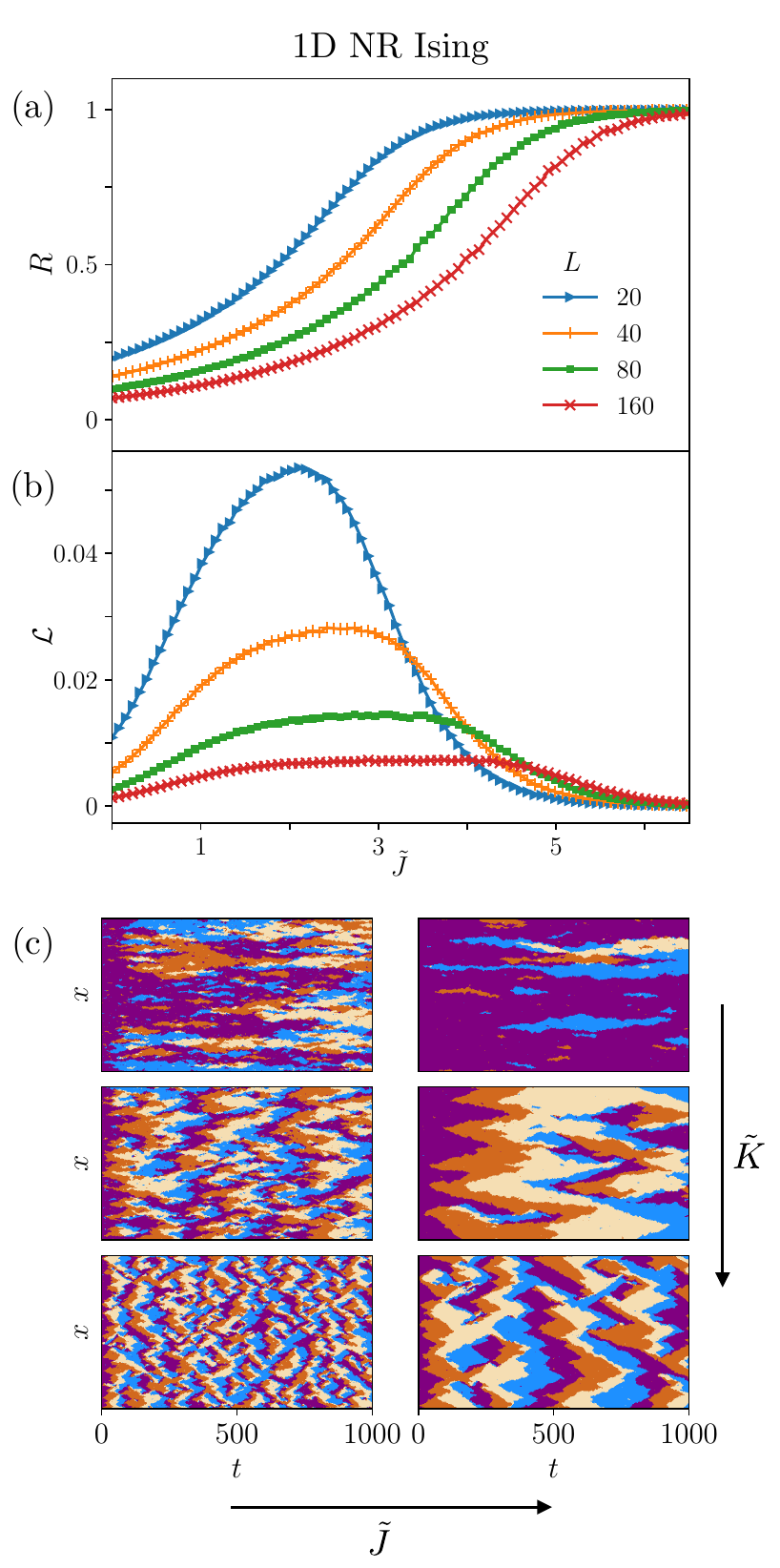}}
\caption{\textbf{1D NR Ising model.} (a) $R$ and (b) $\mathcal{L}$ as a function of $\tilde{J}$ for $\tilde{K}=0.3$ and different linear system size $L$. (c) Kymographs of $\theta$ for $\tilde{J} = 3$ (left) and $\tilde{J} = 4$ (right), comparing the Ising model ($\tilde{K}=0$, top) with the NR Ising model ($\tilde{K}=0.2$, middle and $\tilde{K}=0.5$, bottom). System size is $L=500$. Obtained from MC simulations.}
\label{1D_Fig}
\end{figure}
We simulate the one-dimensional NR Ising model, with fully anti-symmetric interactions, using random picking Glauber dynamics as in the 2D simulations (see Sec.~\ref{MonteCarlo_App}). The order parameters $R$ and $\mathcal{L}$ as a function of $\tilde{J}$ and for fixed $\tilde{K}$ are shown in Fig.~\ref{1D_Fig}a and Fig.~\ref{1D_Fig}b, respectively, while kymographs of $\theta$ with different $\tilde{J}$ and $\tilde{K}$ values are shown in Fig.~\ref{1D_Fig}c.

The 1D NR Ising model does not have a phase transition to an ordered state (either static or time dependent), as is evident from the decrease of $R$ and $\mathcal{L}$ as system size increases. Similarly to the 1D Ising model, as $L$ increases, domains of opposite magnetization emerge and the system becomes disordered. However, unlike the Ising model, the boundaries between the domains do not diffuse but move with non-zero velocity (see Fig.~\ref{1D_Fig}c where the kymograph slope indicates the velocity). Domains expand in a way that can be viewed as the 1D version of droplet growth discussed in Sec.~\ref{Sec_static}.

\section{The effect of different update rules} \label{different_update}
Out of equilibrium, different update rules may change the steady state of the system. To study this effect in the NR Ising model, we performed simulations with random sequential updates, checkerboard updates~\cite{yang2019high}, and synchronous updates (in which all the spins are updated at the same time)~\cite{schonfisch1999synchronous}. Additionally, we used the Metropolis algorithm~\cite{newman1999monte} and compared it to the Glauber which was used throughout the paper. 

\begin{figure}[ht]
\centering
{\includegraphics[width=0.4\textwidth,draft=false]{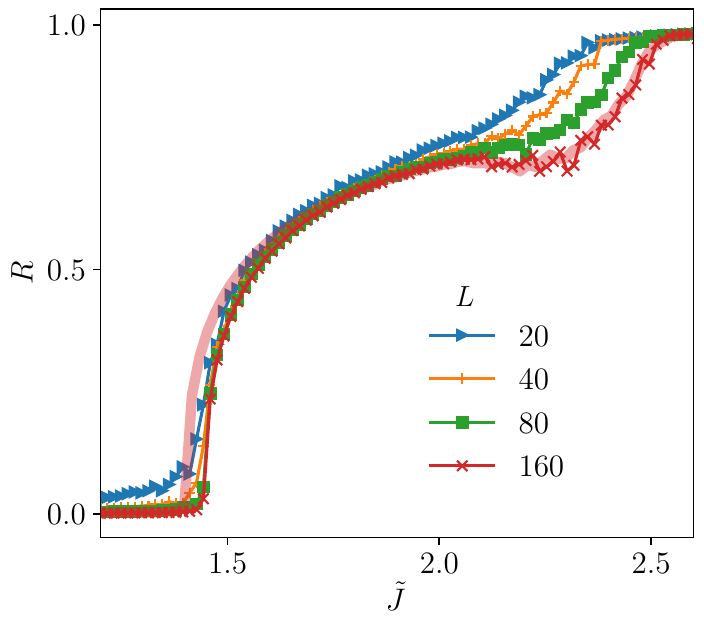}}
\caption{\textbf{Qualitative agreement between two Monte-Carlo update rules.}
$R$ as a function of $\tilde{J}$ for $\tilde{K}=0.3$ and different linear system size $L$, calculated from MC simulations with random spin update rule instead of the checkerboard update which is used in Fig.~\ref{R_vs_J_3D}a. Red semi-transparent line shows the checkerboard update results for $L=160$. Results are qualitatively the same as in Fig.~\ref{R_vs_J_3D}a although exact $\tilde{J}_c$ value varies. Averages are taken over a smaller number of realizations than in Fig.~\ref{R_vs_J_3D}a, making the high $\tilde{J}$ data noisy.}
\label{random_update}
\end{figure}

For fixed coupling parameters ($\tilde{J}$ and $\tilde{K}$), the different update rules changed the dynamics in steady state. However, they did not lead to qualitative changes in the phase diagram. That way, the critical point $\tilde{J}_c$, for some fixed $\tilde{K}$, is altered, but not the existence / destabilization of a phase transition. As an example, we show in Fig.~\ref{random_update} that the phase transition from disorder to swap in 3D, as characterized by the variation of the order parameter $R$ as a function of $\tilde{J}$, is qualitatively the same when using random sequential updates (standard Glauber algorithm), as it is with the checkerboard algorithm, which was used to produced Fig.~\ref{R_vs_J_3D}a. 

We did not find any qualitative difference related to phase transitions.
The only notable difference is that with synchronous updates, fluctuations were considerably larger due to \enquote{checkerboard patterns} (small regions in which each spin is anti-aligned with its same-species neighbors) that frequently appeared and then disappeared.

%

\section{Simulations of the mean-field equation in 2D} \label{2D_MF_noise}
%
\begin{figure*}[ht]
\centering
{\includegraphics[width=0.8\textwidth,draft=false]{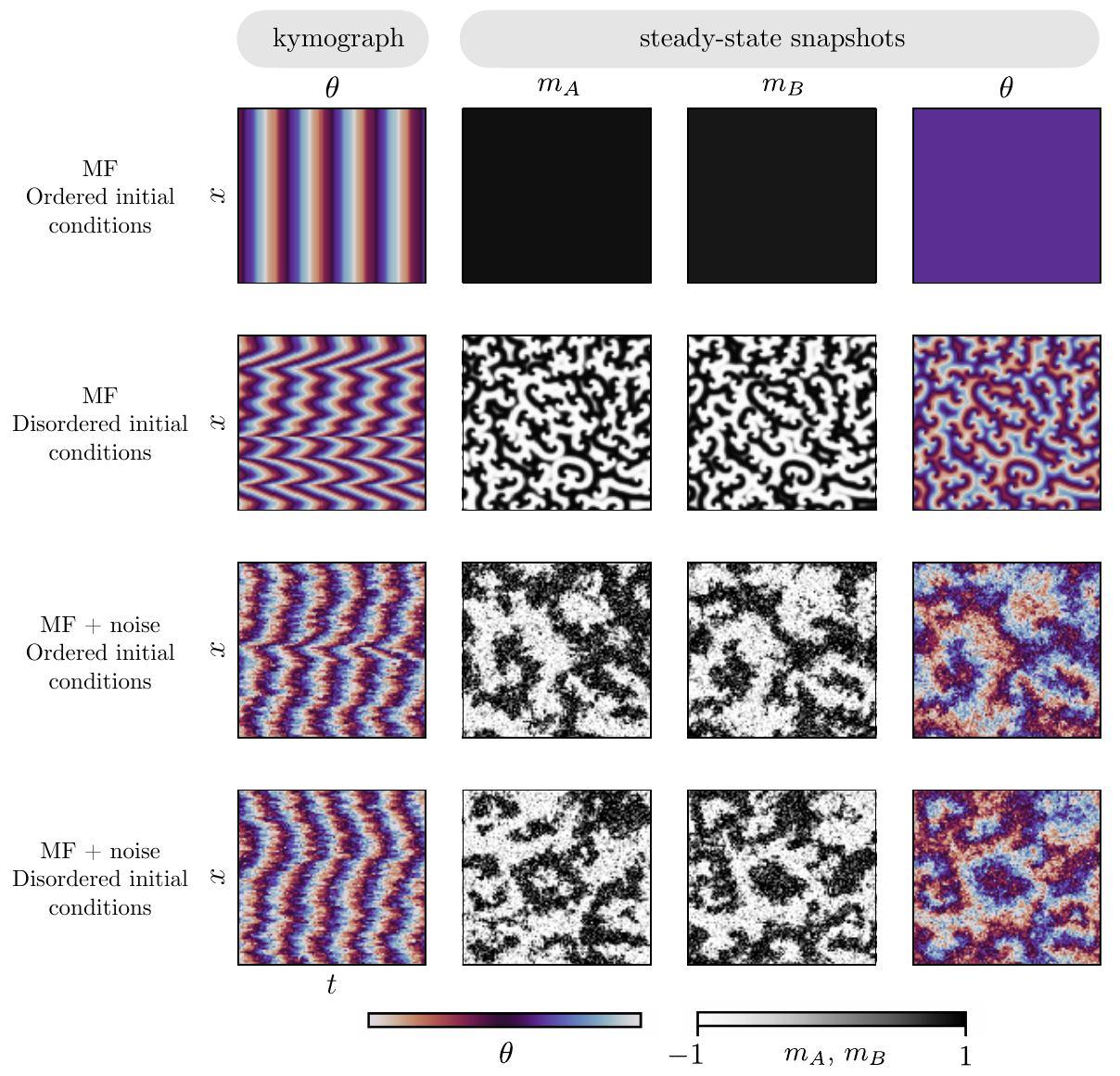}}
\caption{
\textbf{The effect of added noise on the mean-field equation in 2D}.
Simulations of Eq.~(\ref{MF_with_noise}) in 2D comparing the cases of no added noise with added noise and ordered vs. disordered initial conditions. First column is kymographs of the $\theta$ field for a single row on the 2D system. Second, third, and forth columns are snapshots of the $m_A$, $m_B$, and $\theta$ fields, respectively, taken after the system has reached steady state. Without added noise, the steady state is a homogeneous swap phase for ordered initial conditions and an inhomogeneous spiral phase for disordered initial conditions. 
With enough added noise, the swap phase is unstable even with ordered initial conditions and spiral defects are formed for both initialization types. Simulations were performed using the Euler–Maruyama method
with a time step $\Delta t = 0.05$. Noise amplitude for the case of added noise is $\zeta=0.36$, and other system parameters are: $\tilde{J}=2$, $\tilde{K}=0.8$, $L=200$, and $\tau=1$. At $t=0$, we set $m_A,m_B=1 \pm 0.1$ for ordered initial conditions and $m_A,m_B=\pm 0.1$ for disordered initial conditions, where the plus/minus sign is chosen at random for each spin. Kymographs show evolution between $t=9900$ and $t=10000$.}
\label{MF_simulation1}
\end{figure*}
We simulate the full mean-field equation on a grid (Eq.~(\ref{MF_eq0})) in 2D in the fully anti-symmetric case and in the regime corresponding to a stable swap phase (see Fig.~\ref{MF_Phase_diagram}). Results are shown in Fig.~\ref{MF_simulation1}. Depending on initial conditions, the system can end up in different states. An initially ordered state leads to a stable swap phase where the $\theta$ field changes periodically and homogeneously (Fig.~\ref{MF_simulation1}, first row), while an initially fully disordered state ($m_A$ and $m_B$ are chosen randomly at each point to be either $+0.1$ or $-0.1$) reaches a stable state with spiral defects, whose cores barely move over time~(Fig.~\ref{MF_simulation1}, second row).

We test the effect of adding noise to the mean-field equation by writing,
\beq \label{MF_with_noise}
\frac{{\rm d} m_i^{\alpha}}{{\rm d}t}=-\frac{m_i^{\alpha}}{\tau}+\frac{1}{\tau}\tanh\left(\frac{J\sum_{j\,{\rm nn\,of}\,i}m_j^{\alpha}+K\varepsilon_{\alpha\beta}m_i^{\beta}}{k_{B}T}\right)+ \eta_i^\alpha.
\eeq
where $\eta_i^\alpha$ is white noise satisfying $\langle\eta_i^\alpha\rangle=0$ and $\langle\eta_i^\alpha\left(t\right)\eta_j^\beta\left(t'\right)\rangle=\zeta \delta_{\alpha\beta}\delta_{ij}\delta\left(t-t'\right)$.
Above some critical $\zeta$, and for fixed system size, we observed that even when the system is initialized in an ordered state, spiral defects eventually form over time. (see Fig.~\ref{MF_simulation1}, third row). Moreover, they are qualitatively similar to those observed when random initial conditions are implemented (Fig.~\ref{MF_simulation1}, fourth row). In that sense, the mean-field equation with noise exhibits the same qualitative behavior as the Monte-Carlo simulations, where the swap phase is unstable in 2D.

\begin{figure*}[ht]
\centering
{\includegraphics[width=0.9\textwidth,draft=false]{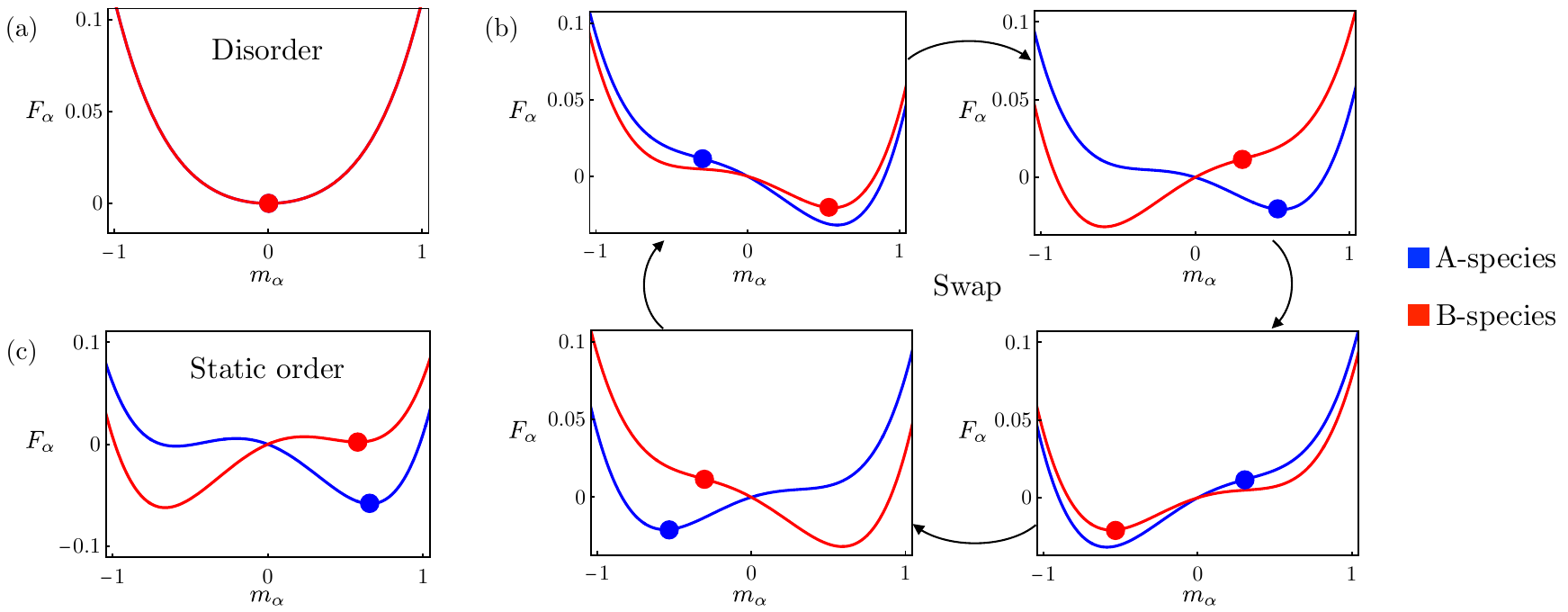}}
\caption{\textbf{Selfish free-energy landscape}. $F_A$ as a function of $m_A$ (blue) and $F_B$ as a function of $m_B$ (red) in zero-dimension NR Ising model. Obtained by solving numerically the spatially homogeneous Eq.~(\ref{MeanFieldExpanded}) in steady state, and substituting $m_A(t)$ and $m_B(t)$ into $F_A$ and $F_B$ defined in Eq.~(\ref{selfish_Free}),
respectively. Actual $m_A$ and $m_B$ values are shown as disks. (a) Disorder, $\tilde{J}=0.9$, (b) swap, $\tilde{J}=1.11$, and (c) static order, $\tilde{J}=1.3$; in all plots $\tilde{K}=0.1$. In panel a, $A$ and $B$ lines as well as disks are identical and therefore only B is shown.}
\label{SelfishFreeEnergy}
\end{figure*}
%

\section{Selfish free energy} \label{Selfish_free_energy}
The dynamics of non-conserved fields $m_{\alpha}(\vec{r},t)$ with $\alpha=1,...,N$ (model A) in equilibrium can be written as~\cite{kardar2007statistical,tauber2014critical,Cates2019}
\begin{equation}
\partial_{t}m_{\alpha}(\vec{r},t)=-\mu \frac{\delta \mathcal{F}[m_1,...,m_N]}{\delta m_{\alpha}(\vec{r},t)} + \eta_{\alpha}(\vec{r},t) 
\end{equation}
where $\mathcal{F}$ is the effective free energy functional, $\mu$ is a generalized mobility, and $\eta_\alpha$ is a white noise field satisfying $\langle\eta_{\alpha}(\vec{r},t)\rangle=0$ and
$\langle \eta_\alpha(\vec{r}, t) \eta_\beta(\vec{r}\,', t') \rangle = \zeta \delta_{\alpha \beta} \delta(\vec{r} - \vec{r}\,') \delta(t - t')$ where $\zeta = 2\mu k_BT$.

For nonreciprocal systems, there is no global $\mathcal{F}$ generating the dynamics. However, we extend the notion of a free energy by writing the dynamics of the two magnetization fields $m_A$ and $m_B$ as 
\begin{equation} \label{SelfishModelA}
\partial_{t}m_{\alpha}(\vec{r},t)=-\mu\frac{\delta \mathcal{F}_{\alpha}[m_{A},m_{B}]}{\delta m_{\alpha}(\vec{r},t)}+\eta_{\alpha}(\vec{r},t)
\end{equation}
where $\mathcal{F}_{\alpha}$ is the {\it selfish free energy} of the $\alpha$-species. We set $\mu=1$ and use Eq.~(\ref{MeanFieldExpanded}) (interpreted as the mean-field version of Eq.~(\ref{SelfishModelA})) to obtain the selfish free energies of the nonreciprocal Ising model in the fully anti-symmetric case,
\begin{align} \label{selfish_Free0}
\mathcal{F}_{A} & =\int {\rm d}\vec{r}\bigg[\frac{1}{2}(1-\tilde{J})\,m_{A}^{2}-\tilde{K}m_{A}m_{B}+\frac{D}{2}\left|\nabla m_{A}\right|^{2}\nonumber\\
 & +\frac{1}{12\tilde{J}}\left(\tilde{J}m_{A}+\tilde{K}m_{B}\right)^{4}\bigg]\\
\mathcal{F}_{B} & =\int {\rm d}\vec{r}\bigg[\frac{1}{2}(1-\tilde{J})\,m_{B}^{2}+\tilde{K}m_{A}m_{B}+\frac{D}{2}\left|\nabla m_{B}\right|^{2}\nonumber\\
 & +\frac{1}{12\tilde{J}}\left(\tilde{J}m_{B}-\tilde{K}m_{A}\right)^{4}\bigg].\nonumber
\end{align}
Equation~(\ref{selfish_Free0}) can be viewed as the free-energy landscape that each species \enquote{sees}, where the free-energy landscape of $A$ changes with $m_B$ and vice versa.

We now restrict ourselves to the zero-dimensional case, where the selfish free energies become the functions
\begin{align} \label{selfish_Free}
F_{A} & =\frac{1}{2}(1-\tilde{J})\,m_{A}^{2}-\tilde{K}m_{A}m_{B}+\frac{1}{12\tilde{J}}\left(\tilde{J}m_{A}+\tilde{K}m_{B}\right)^{4}\\
F_{B} & =\frac{1}{2}(1-\tilde{J})m_{B}^{2}+\tilde{K}m_{A}m_{B}+\frac{1}{12\tilde{J}}\left(\tilde{J}m_{B}-\tilde{K}m_{A}\right)^{4}.\nonumber
\end{align}
In order to visualize the free-energy landscape of each species at a given state, we solve the spatially homogeneous mean-field equation (Eq.~(\ref{MeanFieldExpanded})) and substitute the $m_B(t)$ solution into $F_A$ and the $m_A(t)$ solution into $F_B$. Snapshots of the steady-state free-energy landscape are shown in Fig.~\ref{SelfishFreeEnergy}, for the disordered, swap, and static-order phases. The actual values of $m_A$ and $m_B$ in steady state are shown as disks moving on the free-energy landscape.

In the disordered phase (Fig.~\ref{SelfishFreeEnergy}a), both $m_A$ and $m_B$ are in the global minimum of their respective free energies. In the swap phase (Fig.~\ref{SelfishFreeEnergy}b), the free-energy landscape of each species changes continuously. While one species is momentarily in a global minimum, the other species is descending along the free-energy landscape which changes the free-energy landscape of the first species and so on. In the static-order phase (Fig.~\ref{SelfishFreeEnergy}c), one of the species is at its global minimum while the second is in a local minimum, that is not the global one. This configuration is stable in mean-field, but once noise is introduced, it becomes metastable: the second species will \enquote{jump} over the barrier and reach global minimum, making the first species metastable and the cycle repeats itself. In a spatially extended system, this transition happens via droplet nucleation and growth as depicted in Sec.~\ref{Sec_static}.

\section{Color maps of the order parameters in 2D in the general asymmetric case} \label{2D_colormaps_appendix}
\begin{figure}[ht]
\centering
{\includegraphics[width=0.5\textwidth,draft=false]{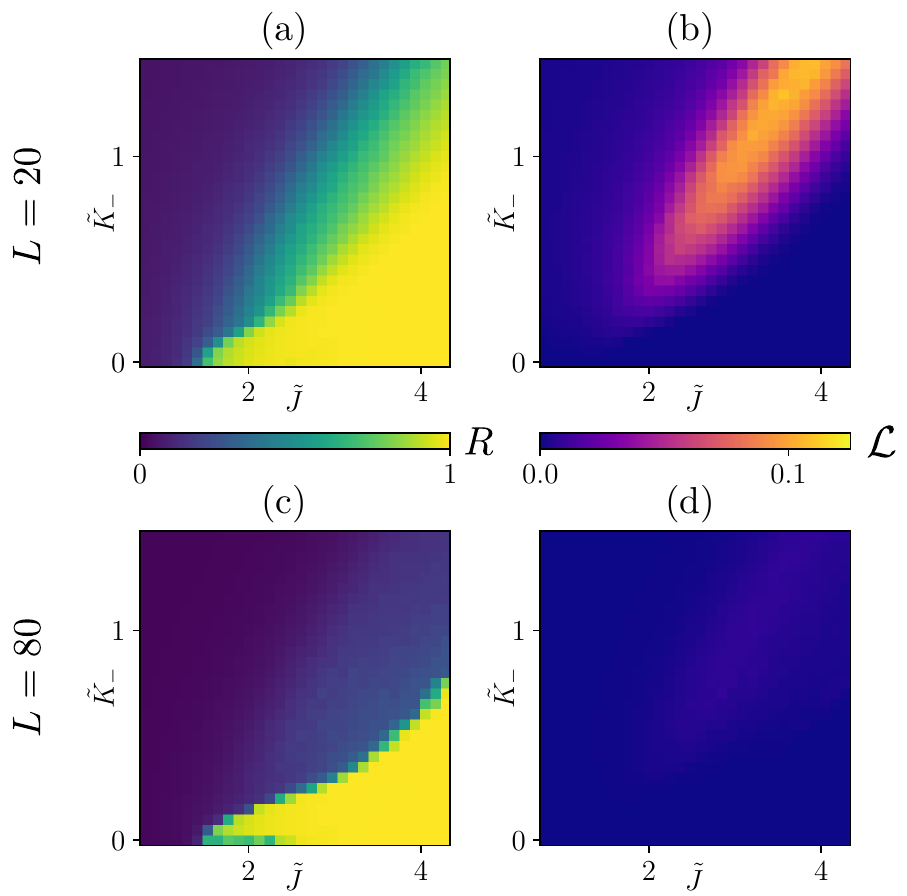}}
\caption{\textbf{Color maps of the order parameters in 2D in the general asymmetric case with fixed (non-zero) $\tilde{K}_+$}. (a) $R$ and (b) $\mathcal{L}$ color maps as a function of $\tilde{J}$ and $\tilde{K}_-$, with $\tilde{K}_+=0.1$ and $L=20$. (c-d) The same with $L=80$. Obtained from MC simulations.}
\label{2D_colormaps_2}
\end{figure}

Figure~\ref{2D_colormaps_2} shows color maps of $R$ and $\mathcal{L}$ as a function of $\tilde{J}$ and $\tilde{K}_-$ with a fixed non-zero value of $\tilde{K}_+$, for a 2D system with $L=20$ (panels a-b) and $L=80$ (panels c-d). For $L=20$ the phase diagram resembles the mean-field prediction (Fig.~\ref{MF_with_Kp}), which includes a swap phase. However, for $L=80$ both $R$ and $\mathcal{L}$ diminish and there is no apparent swap phase, supporting the conclusions of Sec.~\ref{swap_Kp}.
\begin{figure*}[ht]
\centering
{\includegraphics[width=1\textwidth,draft=false]{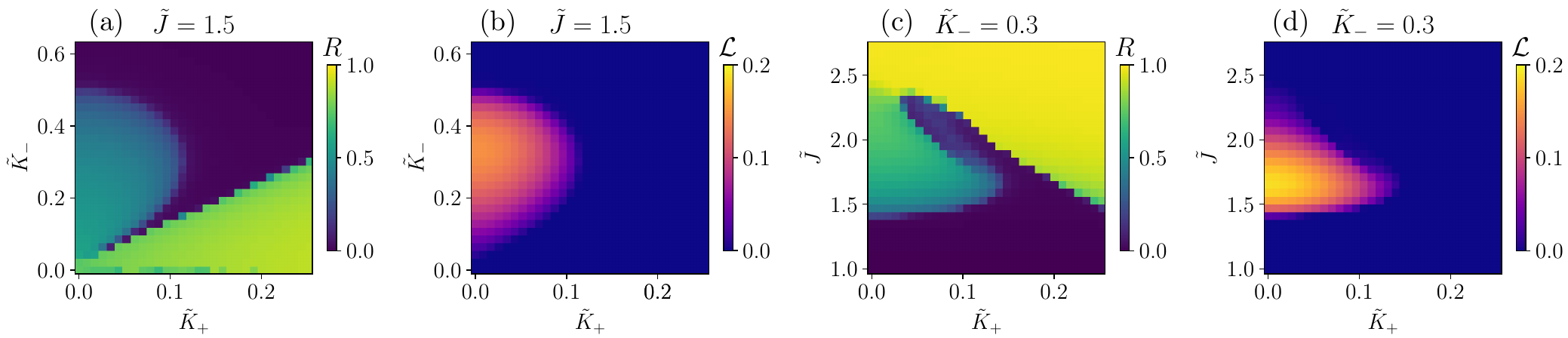}}
\caption{\textbf{Color maps of the order parameters in 3D in the general asymmetric case for a fixed value of $\tilde{J}$ and for a fixed value of $\tilde{K}_-$}. (a) $R$ and (b) $\mathcal{L}$ color maps as a function of $\tilde{K}_+$ and $\tilde{K}_-$ with $\tilde{J} = 1.5$ and $L=80$. (c) $R$ and (d) $\mathcal{L}$ color maps as a function $\tilde{K}_+$ and $\tilde{J}$ with $\tilde{K}_-=0.3$ and $L=80$. Obtained from MC simulations.}
\label{Phase_diagram_Kp_cuts}
\end{figure*}
%

\section{Color maps of the order parameters in 3D in the general asymmetric case} \label{3D_PD_cuts}
In the general asymmetric case, the phase diagram is three-dimensional as it depends on $\tilde{J}$, $\tilde{K}_+$, and $\tilde{K}_-$. In the main text the order parameters of the 3D system were calculated from MC simulations as a function of $\tilde{J}$ and $\tilde{K}_-$ for a fixed value of $\tilde{K}_+$ (Fig.~\ref{3D_PD_with_Kp}). Figure~\ref{Phase_diagram_Kp_cuts} shows the order parameters on additional cuts through the three-dimensional phase diagram, both for a fixed value of $\tilde{J}$ and for a fixed value of $\tilde{K}_-$, for a 3D system of linear size $L=80$. In both cuts, the three phases of disorder ($R=0$ and $\mathcal{L}=0$), swap ($R\neq0$ and $\mathcal{L}\neq0$) and static-order ($R\neq0$ and $\mathcal{L}=0$) are visible. However, the transition between the swap phase and the static-order phase is mediated by a disordered phase, as was also pointed out in Sec.~\ref{NO_SNIC}. The only region where a direct transition appears to occur is in the ${\tilde K}_+\to0$ limit, see the bottom-left part of Fig.~\ref{Phase_diagram_Kp_cuts}a-b and the upper-left part of Fig.~\ref{Phase_diagram_Kp_cuts}c-d. However, in this case, the static-order phase is predicted to be unstable due to droplet growth as discussed in Sec.~\ref{Sec_static}, and therefore this direct transition is only observed in finite systems.

\section{Supplementary movies}
Movie 1: Evolution in 2D from droplet-induced oscillations to disorder caused by spiral defects. Obtained from MC simulations with parameters: $L=150$, $\tilde{J}=2.8$, and
$\tilde{K}=0.3$.

Movie 2: 3D stable swap phase in steady state. Obtained from MC simulations with parameters: $L=80$, $\tilde{J}=1.5$, and
$\tilde{K}=0.1$.

Movie 3: 3D system coarsening from a fully
disordered state to scroll waves, and finally to global oscillations. Obtained from MC simulations with parameters: $L=80$, $\tilde{J}=2.1$, and $\tilde{K}=0.3$.
Line defects of scroll waves are found by coarse-graining the microscopic
system as explained in Sec.~\ref{coarsening} with $l=3$ and $s_{\rm max}=0.2$.

Movie 4: Destruction of the static-order phase in 2D by droplet growth, shown as a function of system size. Obtained from MC simulations with parameters: $\tilde{J}=2.8$ and $\tilde{K}=0.3$.

Movie 5: Droplet-capture mechanism in a 2D system when inter-species interactions are not fully anti-symmetric. Obtained from MC simulations with parameters: $L=150$, $\tilde{J}=3.3$,
$\tilde{K}_+=0.1$, and $\tilde{K}_-=0.3$.

%

\end{document}